\documentclass[twocolumn]{aastex631}

\usepackage{subcaption}
\usepackage{soul, xcolor}

\usepackage{mathtools,amssymb,amsmath}
\begin{document}

\setstcolor{red}

\title{The Exoplanet Radius Valley from Gas-driven Planet Migration and Breaking of Resonant Chains}

\author[0000-0003-1878-0634]{André Izidoro}

\affiliation{Department of Physics and Astronomy  6100 MS 550, Rice University, Houston, TX 77005, USA}
\affiliation{Department of Earth, Environmental and Planetary Sciences, 6100 MS 126, Rice
University, Houston, TX 77005, USA}

\author[0000-0002-0298-8089]{Hilke E. Schlichting}
\affiliation{Department of Earth, Planetary, and Space Sciences, The University of California, Los Angeles, Charles E. Young Drive East, Los Angeles, CA, USA}


\author[0000-0001-8061-2207]{Andrea Isella}
\affiliation{Department of Physics and Astronomy  6100 MS 550, Rice University, Houston, TX 77005, USA}

\author[0000-0001-5392-415X]{Rajdeep Dasgupta}
\affiliation{Department of Earth, Environmental and Planetary Sciences, 6100 MS 126, Rice
University, Houston, TX 77005, USA}

\author{Christian Zimmermann}
\affiliation{Max-Planck-Institut für Astronomie Königstuhl 17, 69117 Heidelberg, Germany}

\author[0000-0002-8868-7649]{Bertram Bitsch}
\affiliation{Max-Planck-Institut für Astronomie Königstuhl 17, 69117 Heidelberg, Germany}



\begin{abstract}

The size frequency distribution of exoplanet radii between 1 and 4$R_{\oplus}$ is bimodal with peaks  at $\sim$1.4~$R_{\oplus}$ and $\sim$2.4~$R_{\oplus}$, and a valley at $\sim$1.8$R_{\oplus}$. This radius valley separates two classes of planets -- usually referred to as ``super-Earths''  and ``mini-Neptunes'' -- and its origin remains debated. One model proposes that super-Earths are the outcome of photo-evaporation or core-powered mass-loss  stripping the primordial atmospheres of the mini-Neptunes. A contrasting model interprets the radius valley as a dichotomy in the bulk compositions, where super-Earths are  rocky planets and mini-Neptunes are water-ice rich worlds. In this work, we test whether the migration model is consistent with the radius valley and how it distinguishes these views. In the migration model, planets migrate towards the disk inner edge forming a chain of planets locked in resonant configurations. After the gas disk dispersal, orbital instabilities  ``break the chains'' and promote  late collisions. This model broadly matches the  period-ratio and planet-multiplicity distributions of Kepler planets, and  accounts for resonant chains such as TRAPPIST-1, Kepler-223, and TOI-178. Here, by combining the outcome of planet formation simulations with compositional mass-radius relationships, and assuming complete loss of primordial H-rich atmospheres in late giant-impacts, we show that the migration model accounts for the exoplanet radius valley and the intra-system uniformity (``peas-in-a-pod'') of Kepler planets. Our results suggest that planets with sizes of $\sim$1.4~$R_{\oplus}$ are mostly rocky, whereas  those with sizes of $\sim$2.4~$R_{\oplus}$ are mostly water-ice rich worlds. Our results do not support an exclusively rocky composition for the cores of mini-Neptunes.

\end{abstract}



\section{Introduction} \label{sec:intro}

Kepler transit observations have shown that planets with sizes between those of Earth (1$R{_\oplus}$) and Neptune ($\sim$4$R{_\oplus}$ ) are extremely common~\citep{lissauertal11a,batalhaetal13,howardetal13,fressinetal13,marcyetal14,fabryckyetal14}. Demographics analysis suggest that at least 30-55\% of the sun-like stars host one or more planets within this size range and with orbital periods shorter than 100 days \citep{mayoretal11,howardetal12,fressinetal13,petiguraetal13,zhuetal18,mulders18,muldersetal18,heetal19,heetal2021}. Uncertainties in stellar radius estimates from photometric Kepler observations prevented a detailed assessment of the intrinsic planet size distribution~\citep{fultonetal17,petiguraetal17}. Specific trends in planet-sizes only started to emerge from the data with more precise determination of stellar radii by follow-up surveys (California Kepler Survey; CKS)  and the use of Gaia improved parallaxes~\citep{johnsonetal17,vaneylenetal18b,petiguraetal2022}.  These studies showed that the size frequency distribution of planets  between $\sim$1 and $\sim$4$R{_\oplus}$ is bimodal with peaks at $\sim$1.4$R{_\oplus}$ and $\sim$2.4$R{_\oplus}$, and a valley at $\sim$1.8$R{_\oplus}$ (\cite{fultonetal17,fultonpetigura18,petigura20}). The best characterized planets with sizes of about $\sim$1.4$R{_\oplus}$ are consistent with rocky composition, as constrained by their estimated bulk densities~\citep{fortneyetal07,adamsseageretal08,lopezfortney14,weissmarcy14,dornetal15,wolfgangetal16,chenkipping17,bashietal17,otegietal20,zengetal16,zengetal19}. These planets are usually referred to as ``super-Earths''. Planets with sizes of about $\sim$2.4$R{_\oplus}$ are consistent with the presence of volatiles -- which could reflect either rocky-cores with H-He rich atmospheres or ice/water rich planets~\citep{kuchner03,rogersseager10,lopezfortney14,zengetal19,otegietal20,mousisetal20}. These planets are commonly referred to as ``mini-Neptunes''. For a detailed discussion, see reviews by~\cite{beanetal21} and \cite{weissetal2022}, and references therein.

 The planet size distribution also reveals an intra-system uniformity in planet radii, where planets in the same systems tend to have similar sizes~\cite[$R_{\rm j+1}/R_{\rm j}\approx1$, where $R_{\rm j+1}$ and $R_j$ are the radius of two adjacent planet-pairs; ][]{weissetal18}. This is popularly known as the ``peas-in-a-pod'' feature of exoplanets.

The planet-size distribution strongly constrains planet formation and evolution models. Different mechanisms have been proposed to explain Kepler-planets' bimodal distribution, including atmospheric loss via photo-evaporation~\citep{lopezfortney12,lopezfortney13,owenyanqin13,kurosakietal13,lugeretal15,mordasini20} and core-powered effects~\citep{ginzburgetal18,rogersetal21,guptaetal19,guptaetal22}. The photo-evaporation model suggests that super-Earths are  the photo-evaporated rocky cores of mini-Neptunes~\citep{owenyanqin13,mordasini20,owencampos20,zhangetal22}.  Atmospheric loss via core-powered effects also supports a predominantly rocky composition for super-Earths and the cores of mini-Neptunes~\citep{ginzburgetal18,guptaetal19,guptaetal22}. However, the bimodal size distribution of planets smaller than $4R_{\oplus}$ has been also interpreted as reflecting distinct compositions between rocky super-Earths and the ice/water rich mini-Neptunes~\citep{izidoroetal21a,zengetal19,venturinietal20}.

 The goal of this work is to investigate whether the observed radius distribution of super-Earths and mini-Neptunes is consistent with planet formation models, and, in particular, with the migration model \citep[][see \cite{raymondetal20} and \cite{beanetal21} for a detailed discussion of other plausible formation models]{izidoroetal17,izidoroetal21a}. Some of the key questions we will address in the following are: Can the size difference between super-Earths and mini-Naptunes be explained by planet formation models? What is the role and relevance of atmospheric loss processes in the context of formation and early evolution of planetary systems?

The migration scenario proposes that super-Earths and mini-Neptunes formed during the gas disk phase and experienced gas-driven planet migration~\citep[e.g.][]{terquempapaloizou07,idalin08,idalin10,mcneilnelson10,hellarynelson12,cossouetal14,colemannelson14,colemannelson16,izidoroetal17,ogiharaetal18,raymondetal18,carreraetal18,carreraetal19}. This model suggests that  convergent migration promotes the formation of chains of planets anchored at the disk inner edge and locked in first order mean motion resonances with each other, the so-called resonant chains. In a first order mean motion resonance, the ratio of orbital periods of two resonant planets is a ratio of integers with difference of one (e.g. $P_2/P_1=  2/1$, but it also requires libration of associated resonant angles). After gas  disk dispersal a large fraction of the resonant chains become dynamically unstable~\citep{izidoroetal17,izidoroetal21a,lambrechtsetal19,estevesetal20,estevesetal2022}. This instability phase leads to orbital crossing among planets and giant impacts, producing planets spaced by Hill radii in agreement with Kepler observations~\citep{puwu15,izidoroetal17}. Giant impacts are expected to erode  primordial atmospheres~\citep{liuetal15,inamdarschlichting16,bierstekeretal19,chanceetal22} and change planetary architectures. It remains elusive, however, if  this scenario is consistent with the size distribution of exoplanets. In this paper, we revisit the migration model to test if it is consistent with the observed peaks in the size distribution of planets at $\sim$1.4$R{_\oplus}$ and $\sim$2.4$R{_\oplus}$, and a valley at $\sim$1.8$R{_\oplus}$. We will also test how this model matches the so called ``peas-in-a-pod''  feature of exoplanets~\citep{weissetal18,millhollandwinn21}. 

\section{Methods}

\subsection{Planet formation simulations}\label{sec:simulations}

Our analysis is based on the numerical simulations presented in \cite{izidoroetal21a} and a small sample of new simulations. These simulations model the formation of  super-Earths and mini-Neptunes ($1<R<4R{_\oplus}$; $1<M<20M{_\oplus}$) by following the evolution of Moon-mass planetary seeds as they growth  inside a circumstellar disk. Our model accounts for several physical processes such as gas assisted pebble accretion~\cite[e.g.][]{ormelklahr10,lambrechtsjohansen12,johansenlambrechts17}, gas driven planet migration~\cite[e.g.][]{baruteauetal14}, gas tidal damping of orbital eccentricity and inclination~\cite[e.g.][]{cresswellnelson06,
cresswellnelson08}, and mutual gravitational interaction of planetary embryos. 

From \cite{izidoroetal21a}, we selected three models that differ in terms of starting location of planetary seeds in the gaseous disk ($w$), the flux of pebbles available for the seeds to grow ($S_{\rm peb}$), the starting time of the simulations relative to the disk age ($t_{\rm start}$), and size of pebbles inside the disk water snowline\footnote{Location in the gaseous disk where water condenses as ice} ($R_{\rm peb}$; see Table 1 in \cite{izidoroetal21a}). 
More specifically, the three models have the following initial setup: 
\begin{itemize}
    \item Model-III with $w = 0.2 - 2$ au, $t_{\rm start}$= 0.5~Myr, $R_{\rm peb}$=1~cm, and  $S_{\rm peb}$=5, thereafter referred as model A;
    \item Model-II with $w = 0.7-20$ au, $t_{\rm start}$= 3~Myr, $R_{\rm peb}$=1~mm, $S_{\rm peb}$=5, thereafter referred as Model B;
    \item Model-III with $w = 0.2 -2$ au, $t_{\rm start}$= 0.5~Myr, $R_{\rm peb}$=1~mm, and $S_{\rm peb}$=10; thereafter referred as model C. 
\end{itemize}
The selected models produce: i) planetary systems dominated by planets with rocky composition (model A); ii)  planetary systems dominated by water-rich worlds (model B); and iii) planetary systems with mixed populations of water-rich and rocky worlds (model C).

The breaking the chains scenario broadly matches the observed period ratio distribution of exoplanets with radii smaller than $\sim4R{_\oplus}$ if the great majority of the resonant chains become dynamically unstable after gas disk dispersal~\citep{izidoroetal17,izidoroetal21a}. A good match to observations requires an instability rate of more than 90-95\%, but not 100\%. The breaking the chains model suggest that the Kepler sample is well matched by mixing $>$90\% of  unstable systems with less than  $<$10\% of stable systems. This is because observations also show that resonant chains exist, so not all chains should become dynamically unstable. Iconic examples of resonant chains are TRAPPIST-1~\citep{gillonetal17,lugeretal17}, Kepler-223~\citep{millsetal16}, and TOI-178~\citep{leleuetal21} systems.

For each model of \cite{izidoroetal21a}  we have $\sim$50 simulations available that slightly differ in the initial distribution of planetary seeds and masses. For model A and B, about 90\%, and 76\% of the planetary systems become naturally unstable after the gas disk dispersal, respectively. For model C, the fraction of unstable systems is about 50\%. It is possible that all these fractions could be higher if we had integrated our simulations for longer timescales (e.g. $>$1~Gyr, instead of $\sim$50~Myr;~\cite{izidoroetal21a}), but this is  computationally too demanding. It is also possible that our simulations miss some important physics that may help triggering dynamical instabilities after gas disk dispersal. These include planetesimal scattering effects~\citep{chatterjeeford15,raymondetal22}, interactions of planet chains with distant external perturbers~\citep[][Bitsch \& Izidoro in prep]{laipu16,bitschetal20}, planet-star tidal interaction effects~\citep{bolmontmathis16},  and spin-orbit misalignment effects~\citep{spaldingbatygin16}.

Whereas model A and B provide 45 and 38 unstable systems, respectively, model C produces only 25 unstable systems \citep[see Figure 20 of ][]{izidoroetal21a}. To overcome potential problems caused by the smaller number of unstable systems in model C (e.g. misleading results due to small number statistics), we have increased their number  by artificially triggering dynamical instabilities in the 25 systems that remain stable after 50 Myr \footnote{We have  verified that the main results presented in this paper would not change qualitatively if we had used the original sample of \cite{izidoroetal21a}.}. This was done by changing the pericenter of the orbit of the two innermost planets by a random amount varying between -5\% and +5\% of their respective values. A similar approach  was adopted in simulating dynamical instabilities within the solar system~\citep[e.g.][]{levisonetal11,nesvornymorbidelli12}. Artificial triggering of dynamical instabilities, this time via an artificial reduction of the planet mass, has also been  used in studies  of the dynamical architecture of super-Earth and mini-Neptune systems~\citep{matsumotoogihara20,goldbergbatygin22}. After triggering instability, we extend the numerical integration of these 25  system for another $\sim$50~Myr. 

Finally, we expanded our model C-sample with 15 new simulations based on the Model~III of \cite{izidoroetal21a} that use slightly different pebble fluxes  ($S_{\rm peb}$=5 and $S_{\rm peb}$=2.5; $t_{\rm start} = 0.5 {\rm Myr}$)  and/or initial distribution of seeds (inside 1~au, instead of 2~au as in the nominal case). Out of these 15 new simulations, only one system  represents a stable system. All these simulations were numerically integrated for 50~Myr.

In summary,  model A, B,  and C are composed of  47 (2 stable systems), 41 (3 stable systems) and 65 (1 stable system) simulations, respectively. The ratio between stable and unstable systems broadly satisfies the fact that the migration model typically requires  less than  10\% of stable systems in order to match observations in terms of period-ratios and planet multiplicity distributions~\citep{izidoroetal17,izidoroetal21a}.

\subsection{Accretion of water/ices}

In \cite{izidoroetal21a}, planetary seeds grow via pebble accretion and mutual collisions. Pebbles beyond the snowline are assumed to have 50\% of their masses in water ice. Pebbles that drift inwards and  cross the water snowline are assumed to sublimate, losing their water component and releasing their silicate counterpart in the form of small silicate grains~\citep{morbidellietal15b}. Planets growing by pebble accretion beyond the snowline tend to become water-rich  whereas those growing inside the snowline tend to be rocky~\citep{izidoroetal21a,bitschetal19b}.  We model collisions as perfect merging events that conserve mass (including water) and linear momentum. Statistically speaking,  collisional fragmentation has a negligible effect on the final dynamical architecture of planetary systems produced in the breaking the chains model (\cite{estevesetal2022}; see also \cite{poonetal20}).

\subsection{Converting Planetary Mass to Planetary Radius}\label{sec:convert}

Our planet formation simulations provide the mass and composition of planets but not their size/radius. In order to compare our model to the exoplanet radius valley and the peas-in-a-pod trend,  we use compositional mass-radius relationships~\citep{zengetal16,zengetal19}. To this end, we use the  mass-radius-relationship (MRR) fits from \cite{zengetal19},\footnote{Available at \url{https://lweb.cfa.harvard.edu/~lzeng/planetmodels.html}}
 who modeled planets with the following compositions:
\begin{itemize}
    \item Earth-like rocky  composition: 32.5\%~Fe~+~67.5\% ~${\rm MgSiO_3}$. We refer to these planets simply as rocky planets. 
    \item Earth-like rocky  composition with ${\rm H_2}$ atmosphere: 99.7\%~Earth-like composition~+~0.3\%~${\rm H_2}$~envelope by mass. We refer to these planets as rocky planets with H-rich/primordial atmospheres.
    \item Water-rich composition: 50\%~Earth-like rocky core~+~50\%~${\rm H_2O}$ layer by mass. We refer to these planets as water-rich planets.
     \item Water-rich composition  with ${\rm H_2}$ atmosphere: 49.85\%~Earth-like composition + 49.85\% ${\rm H_2O}$ layer + 0.3\% ${\rm H_2}$ envelope by mass. We refer to these planets as water-rich planets with H-rich/primordial atmospheres.
\end{itemize} 
The  water mass fraction of our synthetic planets is an outcome of the simulations that tracks the composition of the material accreted by each planet. In general, these compositions vary between 0\% and 50\% of mass in water. However, because mass radius relationships for intermediate compositions (e.g. water-mass fraction of 25\%) are not publicly available, we assume that planets with water-mass fractions larger than 10\% are water rich worlds and those with lower water-mass fractions are rocky.  
We argue that this simplification does not degrade the quality of our study because most of our final planets have either 0\% or $>$20\% of water content. Nevertheless, in order to account for uncertainties coming from our model-simplifications (e.g.  we  neglect  water/mass loss via impacts; see ~\cite{marcusetal10,leinhardtstewart12,bierstekeretal20,estevesetal2022}) and the few planets with intermediate water content, we calculated planet radii assuming  1-$\sigma$ uncertainties of 7\%. This uncertainty is motivated by  the typical difference in size of planets (with no atmosphere) with 25\% water-mass fraction and those with 50\% ~\cite[see][]{zengetal16,zengetal19}. In the Appendix of the paper, we also test our results against the empirical mass radius relationship of ~\cite{otegietal20}, and we show that our main conclusions do not qualitatively change.
 
\subsection{Atmospheric loss via giant impacts}

In simulations of \cite{izidoroetal21a}, gas accretion onto growing planets was not taken into account~\citep[see papers by][focused on the formation of giant planets]{bitschetal19,bitschetal20}. This is a reasonable approximation because state-of-the-art 3D hydrodynamical simulations show that the atmosphere of planets with masses smaller than 10-15$M_{\oplus}$ should correspond to only a few percent of the core mass. In this mass range and below, recycling between the planetary atmosphere and the circumstellar disk is an efficient process, limiting atmospheric mass ~\citep{lambrechtslega17,cimermanetal17,bethunerafikov19,moldenhaueretal22}.

In all simulations presented here, the  masses of planets at the end of the gas disk dispersal are systematically smaller than 10$M_{\oplus}$. Both hydrodynamical simulations and analytic calculations show that their putative atmospheric masses would be limited to a few percent of their total masses ~\citep[e.g.][]{leechiang15,ginzburgetal16}. Following previous studies ~\citep[e.g.][]{owenwu17,ginzburgetal18,guptaetal19}, we assume  that the atmosphere-to-core  mass ratio of our planets at the end of the gas disk phase are  0.3\%  in our nominal simulations, but we also test cases with 0.1\%, 1\% and 5\% (these cases are presented in the Appendix).

We also assume that giant impacts ($M_{\rm p}/M_{\rm t}> 0.1$; where $M_{\rm p}$, $M_{\rm t}$ are the projectile and target masses, respectively) that take place after the gas disk dispersal, completely strip primordial atmospheres ~\citep{liuetal15,bierstekeretal19}, either leaving behind bare rocky, or bare water-rich cores. Following this definition, $\gtrsim$80-90\% of the late impacts in our simulations are  flagged as giant impacts. We do not model the formation of secondary/outgassed atmospheres in this work.

\subsection{Atmospheric loss via photo-evaporation}

The model of \cite{izidoroetal21a} also does not include photo-evaporation or core-powered mass-loss of planetary atmospheres. As discussed before, we do not model gas accretion, but the existence of gaseous atmospheres are assumed during our data analysis. 

Recent studies discovered that photo-evaporation and/or core-powered mass-loss might explain the exoplanet radius valley~\citep{owenwu17,guptaetal19,jinmordasinietal18,guptaetal20}. To investigate the robustness of our results to the possibility of subsequent atmospheric loss after the giant impact phase has concluded, we analyze our simulations including atmospheric mass loss by photo-evaporation after the giant impact phase. 

In order to test the impact of photo-evaporation in our model, we follow a simple energy-limited escape prescription to estimate  planet's atmosphere stability when subject to stellar X-ray and ultraviolet (XUV) radiation \citep[e.g.][]{owenwu17}. We follow the criterion by~\cite{misenerschlichting21}, which compares the atmospheric binding energy to the energy the planet receives from 100~Myr to 1 Gyr. If the ratio between these two quantities $\Phi$ is $\lesssim1$, then sufficient energy is received by the planet to have its atmosphere photo-evaporated. $\Phi$ is given by
\begin{equation}
    \begin{split}
          \Phi & \simeq  \dfrac{f}{3.3 \times 10^{-3}} \left( \frac{E_{\rm out}}{5.2\times10^{45} {\rm erg}} \right) ^{-1} \left( \dfrac{\eta}{0.1} \right)^{-1}\\ 
      &  \times \left( \dfrac{M_{\rm c}}{3 M_{\oplus}} \right)^{5/4} \left( \frac{T_{\rm eq}}{1000~K} \right)^{-4} \left( \dfrac{R_p}{R_c} \right)^{-2},
    \end{split}
\end{equation}
where $f$ is the atmosphere-to-core mass ratio, that in our simulations varies from 0.1\% to 5\%, $M_{\rm c}$ is the planetary core mass, $T_{\rm eq}$ is the planet equilibrium temperature, $R_{\rm p}$ is the planet radius, and $R_{c}$ is the core radius. The integrated stellar energy output is set to $E_{\rm out} =  5.2\times10^{45}$~erg, while the dimensionless efficiency parameter describing the amount of energy  available for driving mass-loss is set as $\eta=0.1$~ \citep{owenwu17}. 

For every planet in our simulations, $M_c$ is provided directly by our planet formation simulation, while $R_p$ and $R_c$ are calculated from \cite{zengetal19} assuming 
planets with a fixed and identical atmosphere-to-core mass ratio ($f$) and bare planets, respectively. The planet equilibrium temperature is calculated as 
\begin{equation}
    T_{\rm eq } = 279~\left( \dfrac{a_{\rm p}}{1~{\rm au}} \right)^{-1/2}~{\rm K},
\end{equation}
where $a_p$ is the planet's orbital semi-major axis.
For all the planets with an atmosphere, i.e., those that did not experience a late giant impact, we calculate $\Phi$, and, if $\Phi < 1$, we  
remove the planet atmosphere and assign to that planet a radius equal to the core radius $R_c$. Vice versa, if $\Phi > 1$, we assume that photoevaporation 
is inefficient and assign to that planet the original radius $R_p$.  We  verified that our simplified treatment of photoevaporation is qualitatively consistent with more sophisticated photoevaporation models of the literature~\citep[e.g.][]{owenwu17}.

\section{Results}

\begin{figure*}
\centering
	\includegraphics[scale=.52]{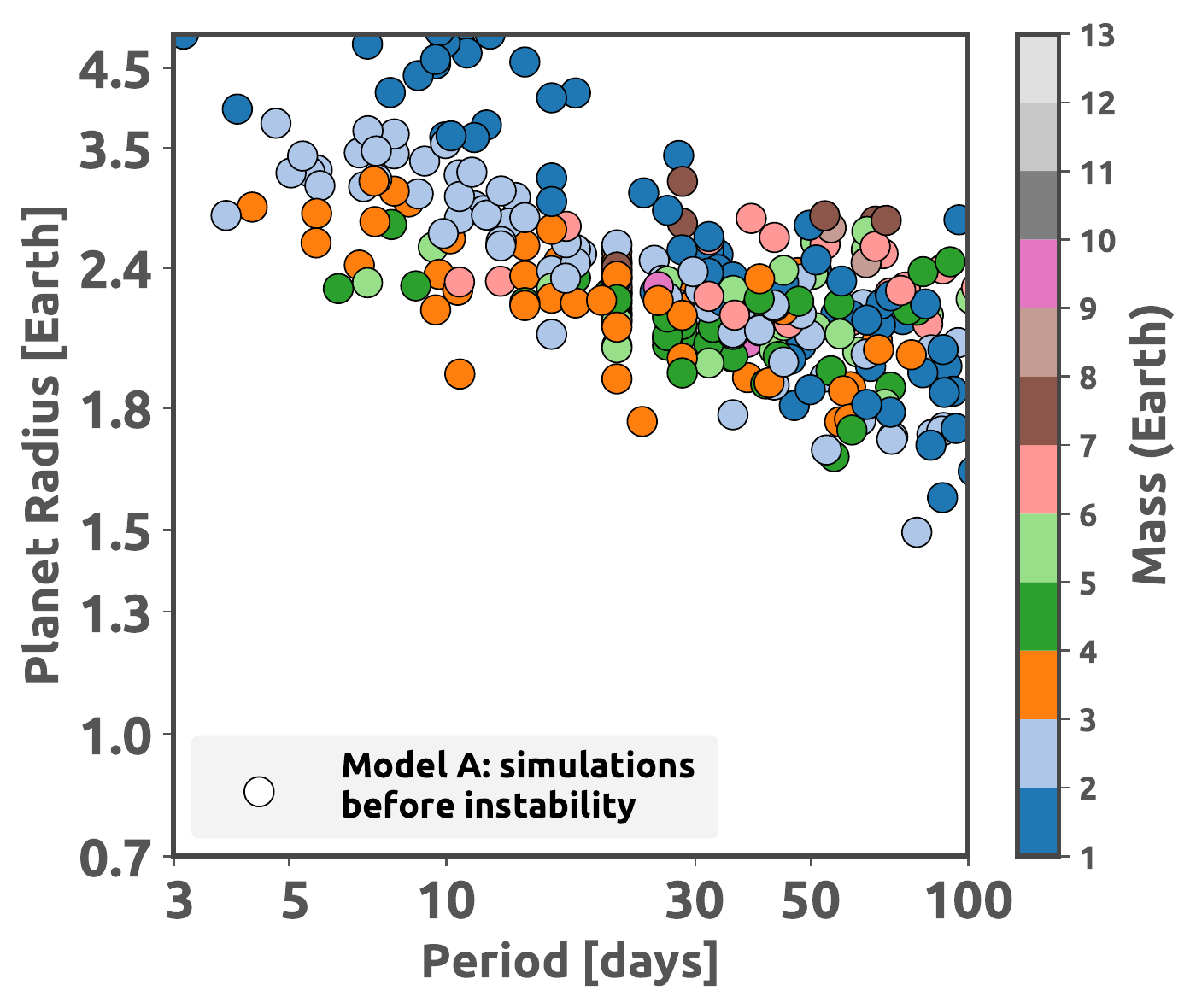}
    \includegraphics[scale=.52]{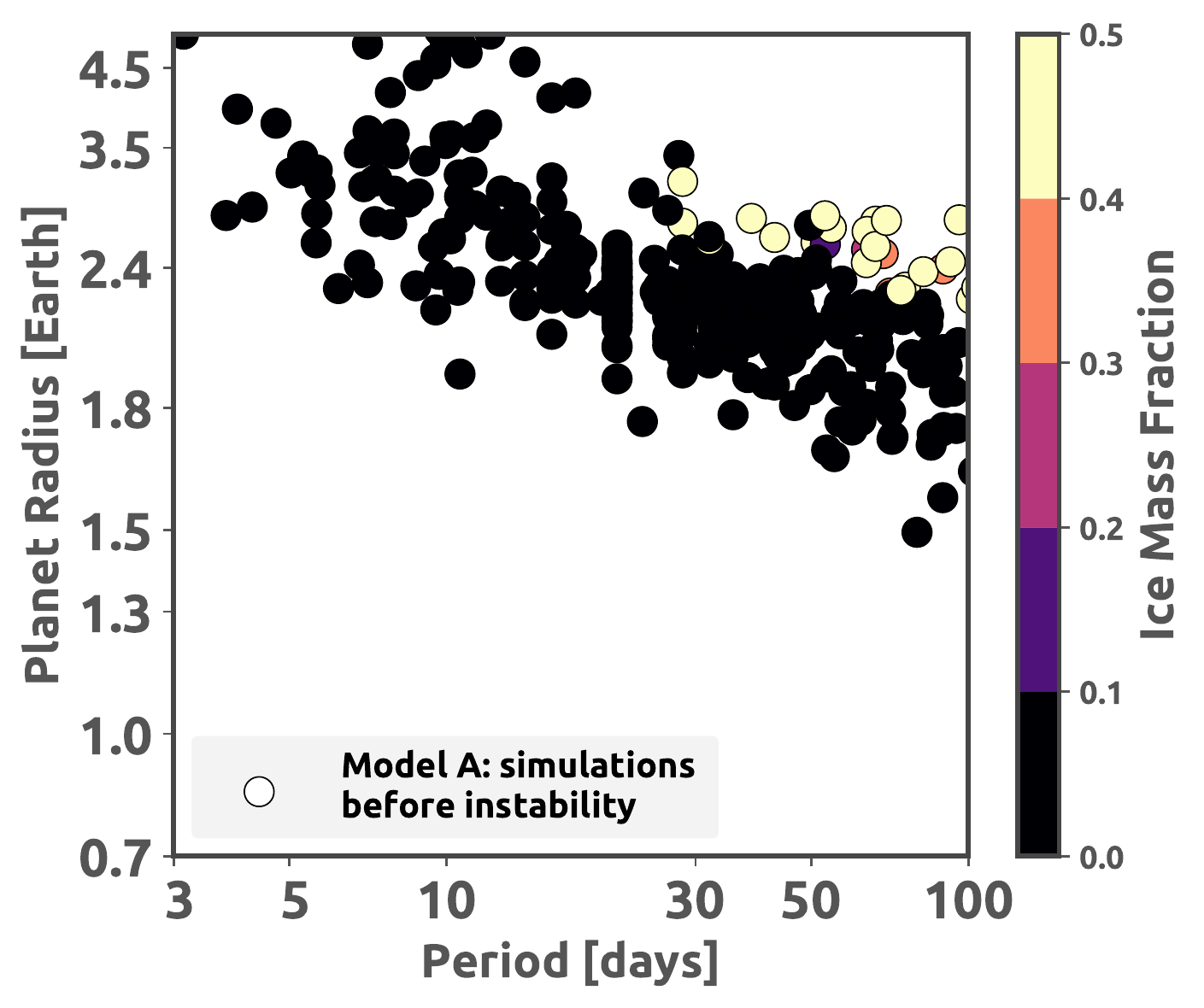}
    
    \includegraphics[scale=.52]{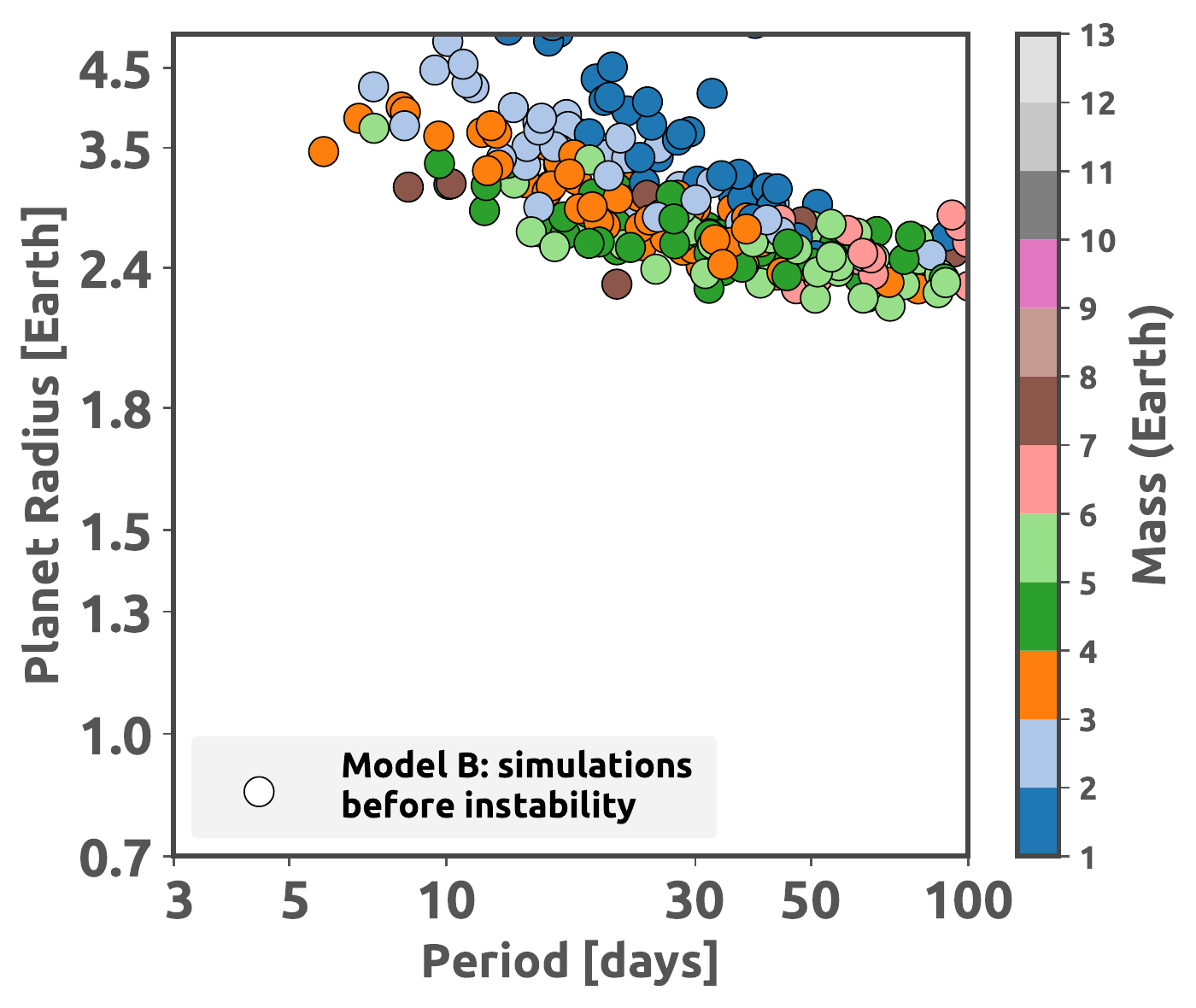}
    \includegraphics[scale=.52]{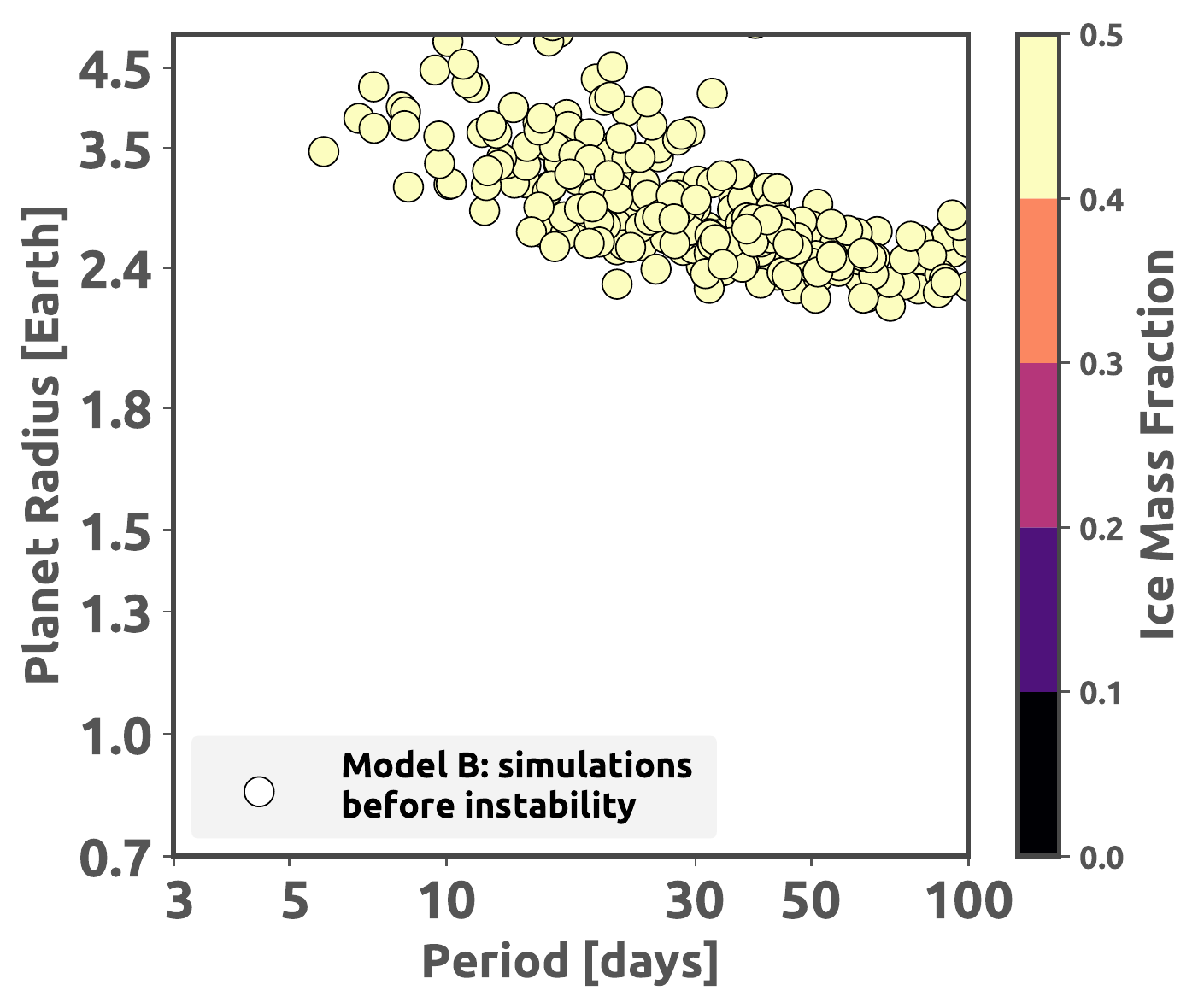}
    
    \includegraphics[scale=.52]{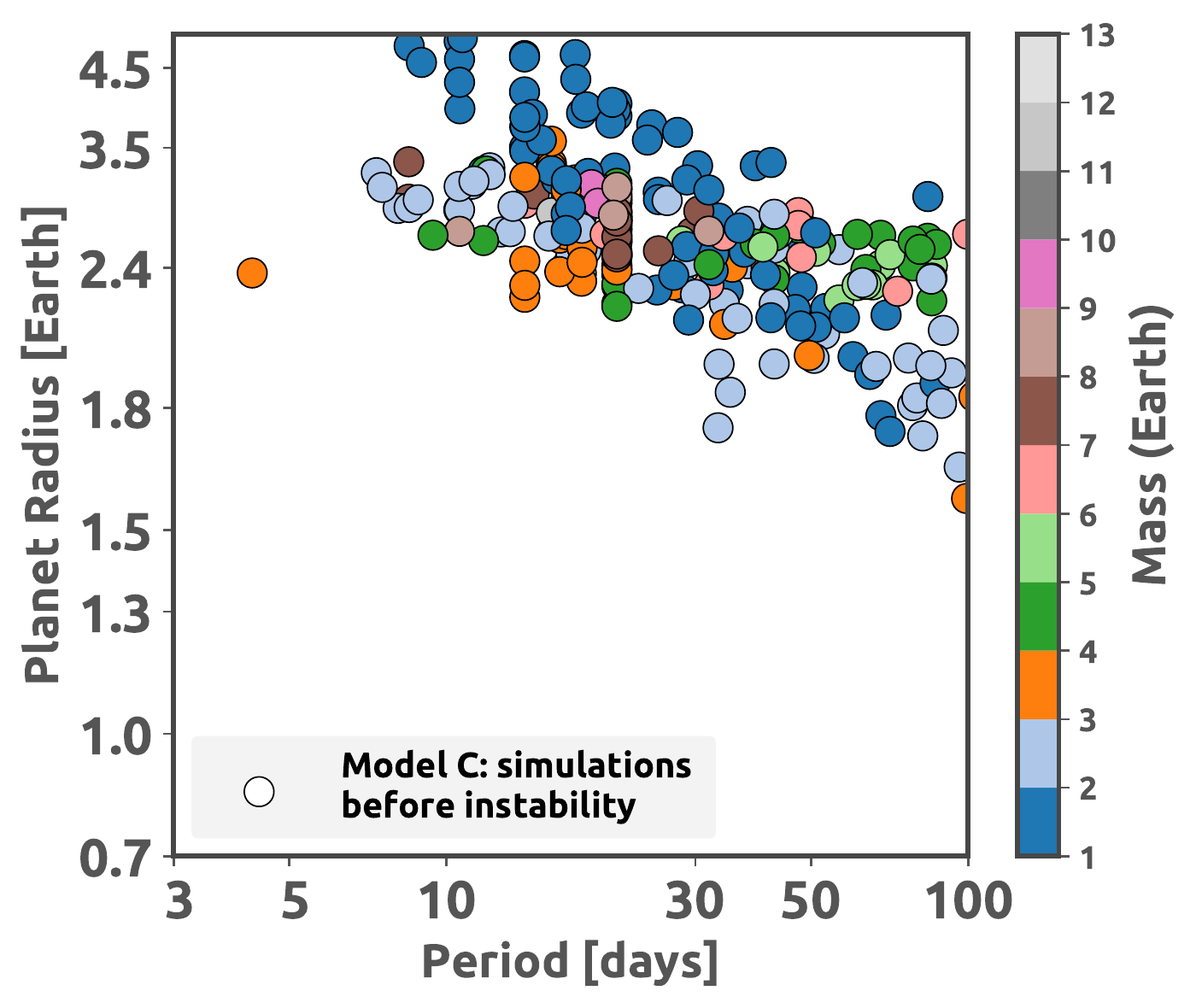}
    \includegraphics[scale=.52]{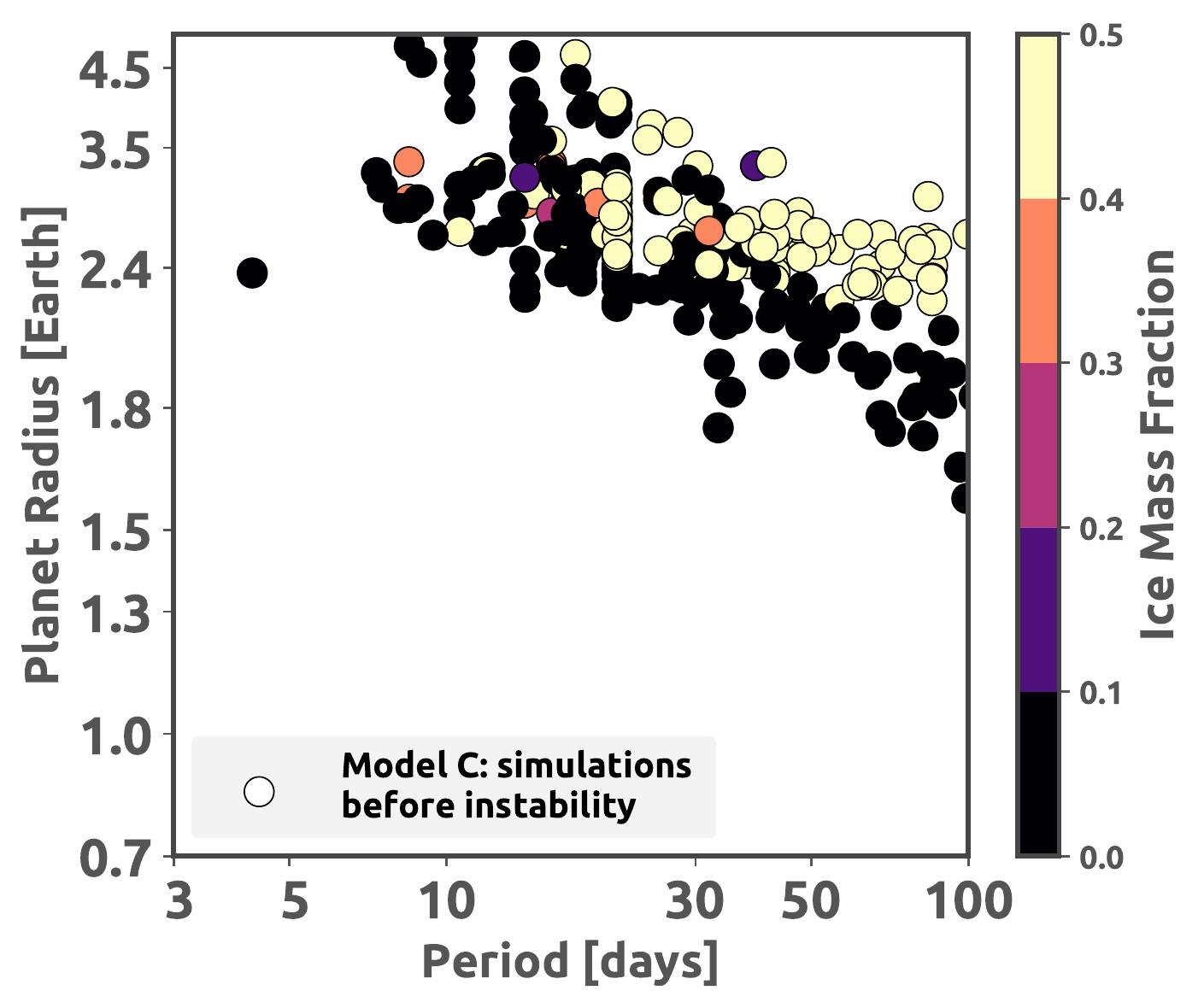}        
   \caption{Architecture of planetary systems at the end of the gas disk dispersal phase, i.e., before the onset of dynamical instabilities and the breaking of resonant chains. Planet sizes are calculated assuming primordial atmosphere-to-core mass ratios of 0.3\%. The horizontal axis shows orbital period and the y-axis show planetary radius as calculate from the mass radius relationship from \cite{zengetal19}. Planets are shown as individual dots. Panels show the results of all simulations and all planets (P$<$ 100 days) produced in each model.  From top to bottom, it shows model A, B, and C, respectively. Individual planets are color-coded according to the their masses (left-column) and  ice/water-mass fractions (right-column panels).}
    \label{fig:systemsbeforeinstability}
\end{figure*}

Figure \ref{fig:systemsbeforeinstability} shows the orbital period versus planet radius for every planet in our simulations at the end of the gas disk phase, i.e. before orbital dynamical instabilities take place.  From top-to-bottom, we show the results of model A, B, and C. The panels on the right show planets color-coded by their water/ice mass fraction, while the panels of the left show planets color-coded by mass.  At this stage, all the planets have an hydrogen atmosphere-to-core mass ratio of 0.3\% and most of the planets are locked in resonant chains. The simulated planets are typically less massive than 10~${\rm M_{\oplus}}$~\citep{izidoroetal21a} and, due to the presence of primordial atmospheres, are larger than $\sim$1.5~$R_{\oplus}$. Model A, B and C are dominated by rocky (black),  water/ice rich (yellow), and mixed composition planets, respectively.

\begin{figure*}
\includegraphics[scale=.52,trim = 0cm 1.55cm 1.88cm 0.1cm,clip]{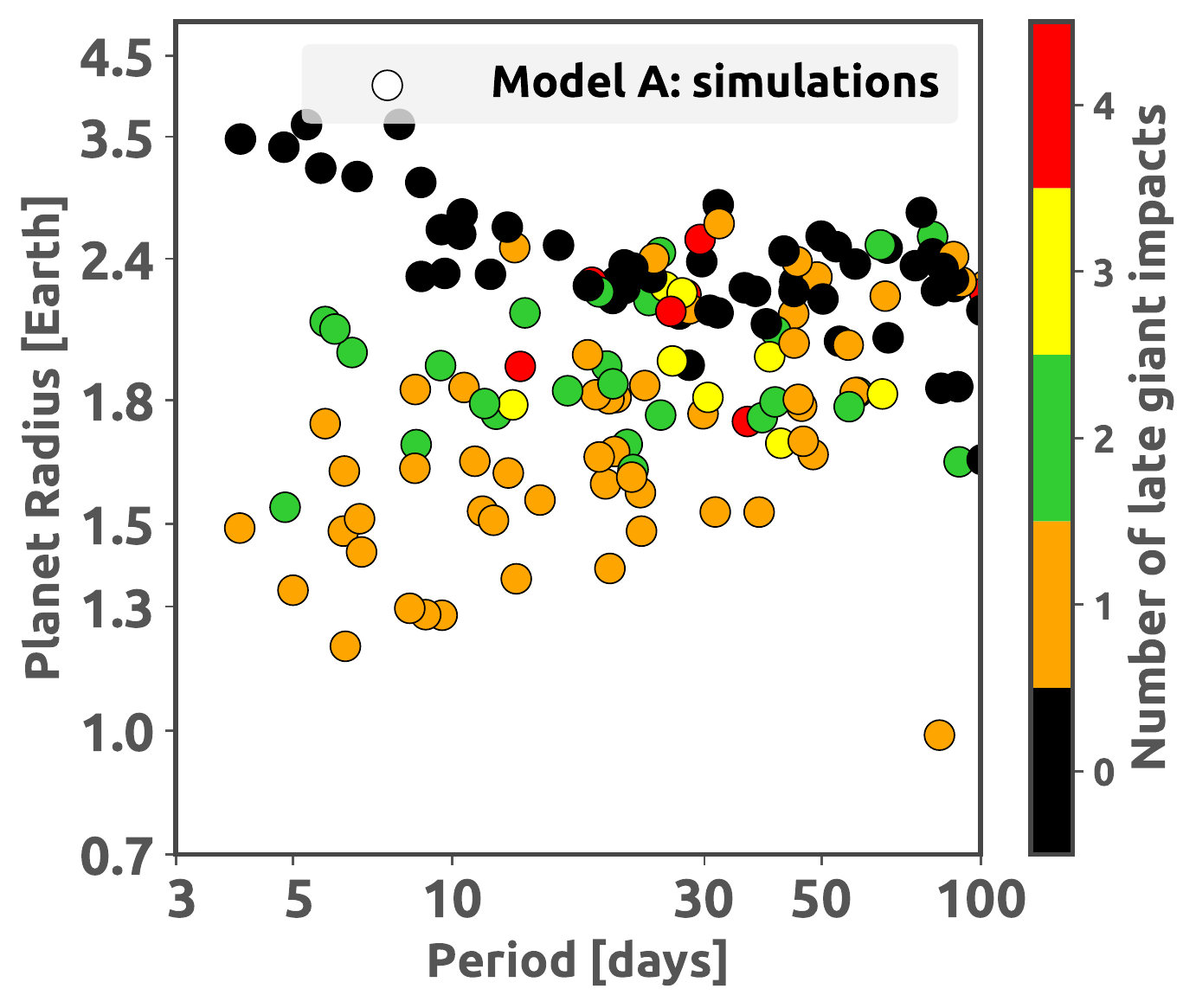}
\hspace{-0.295cm}
\includegraphics[scale=.52,trim = 1.9cm 1.55cm 1.88cm 0.1cm, clip]{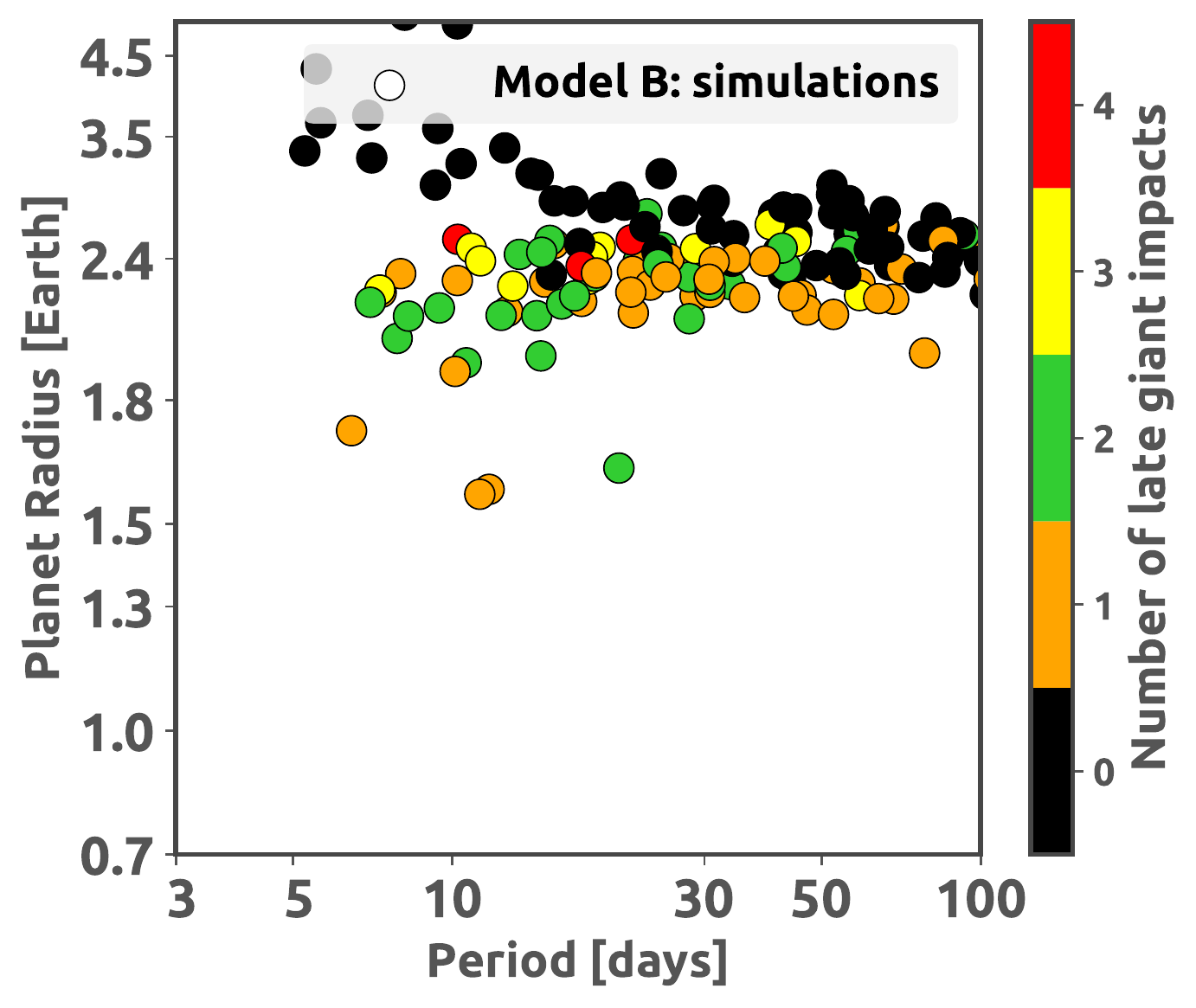}
\hspace{-0.295cm}
\includegraphics[scale=.52,trim = 1.9cm 1.55cm 0cm 0.1cm,clip]{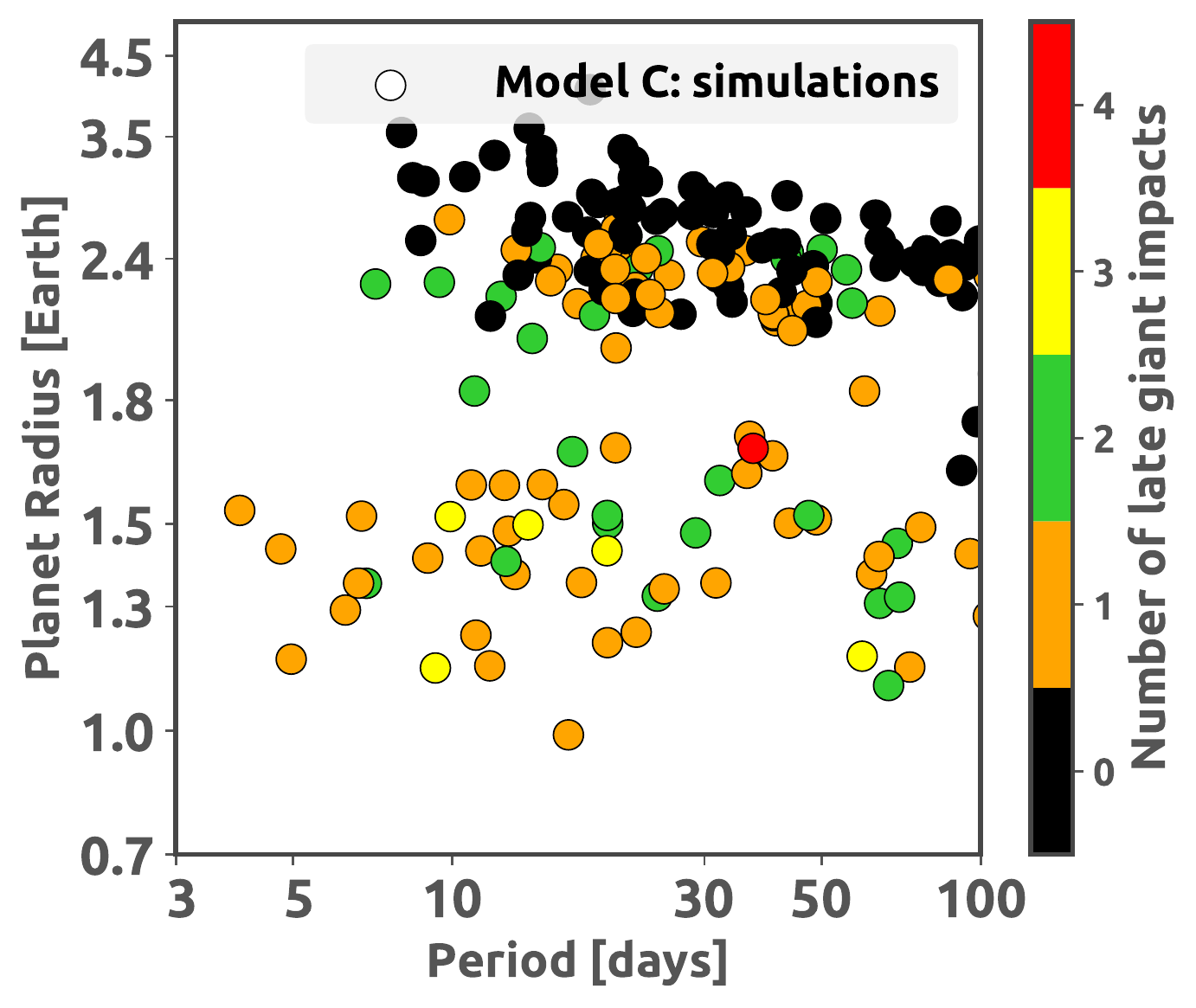}

\includegraphics[scale=.52,trim = 0cm 1.55cm 1.88cm 0.1cm,clip]{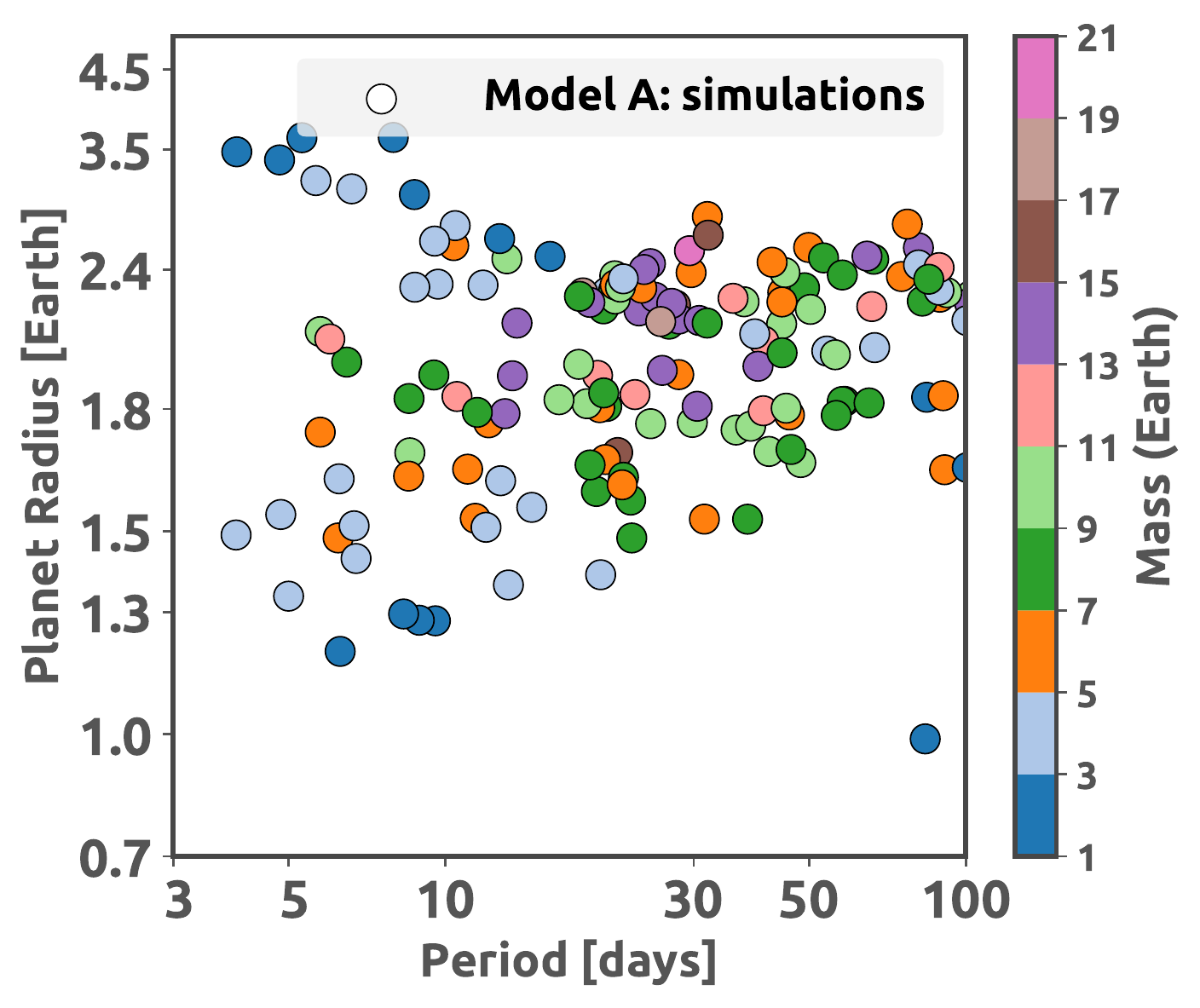}
\hspace{-0.44cm}	
\includegraphics[scale=.52,trim = 1.90cm 1.55cm 1.88cm 0.1cm, clip]{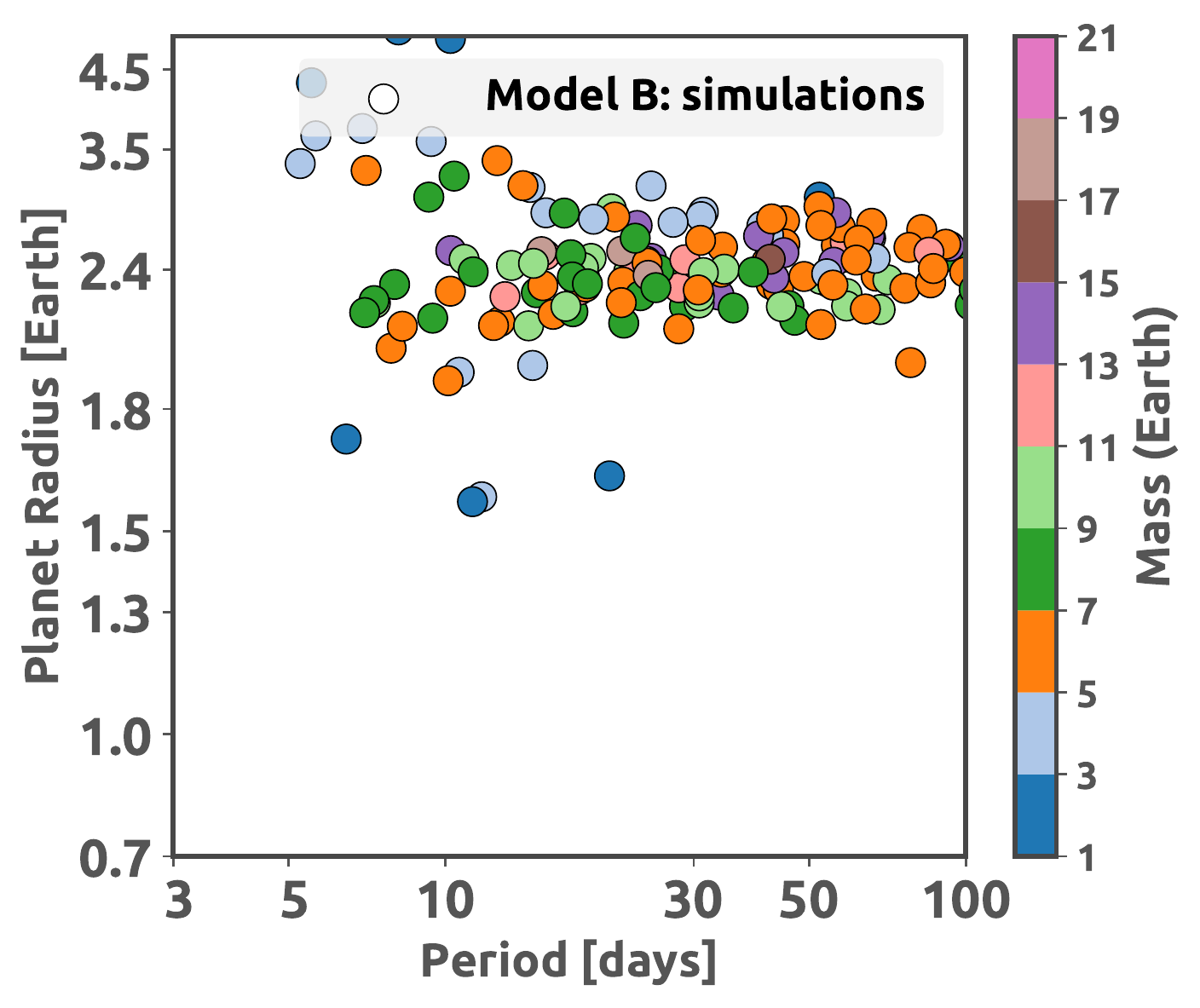}
\hspace{-0.44cm}	
\includegraphics[scale=.52,trim = 1.9cm 1.55cm 0cm 0.1cm,clip]{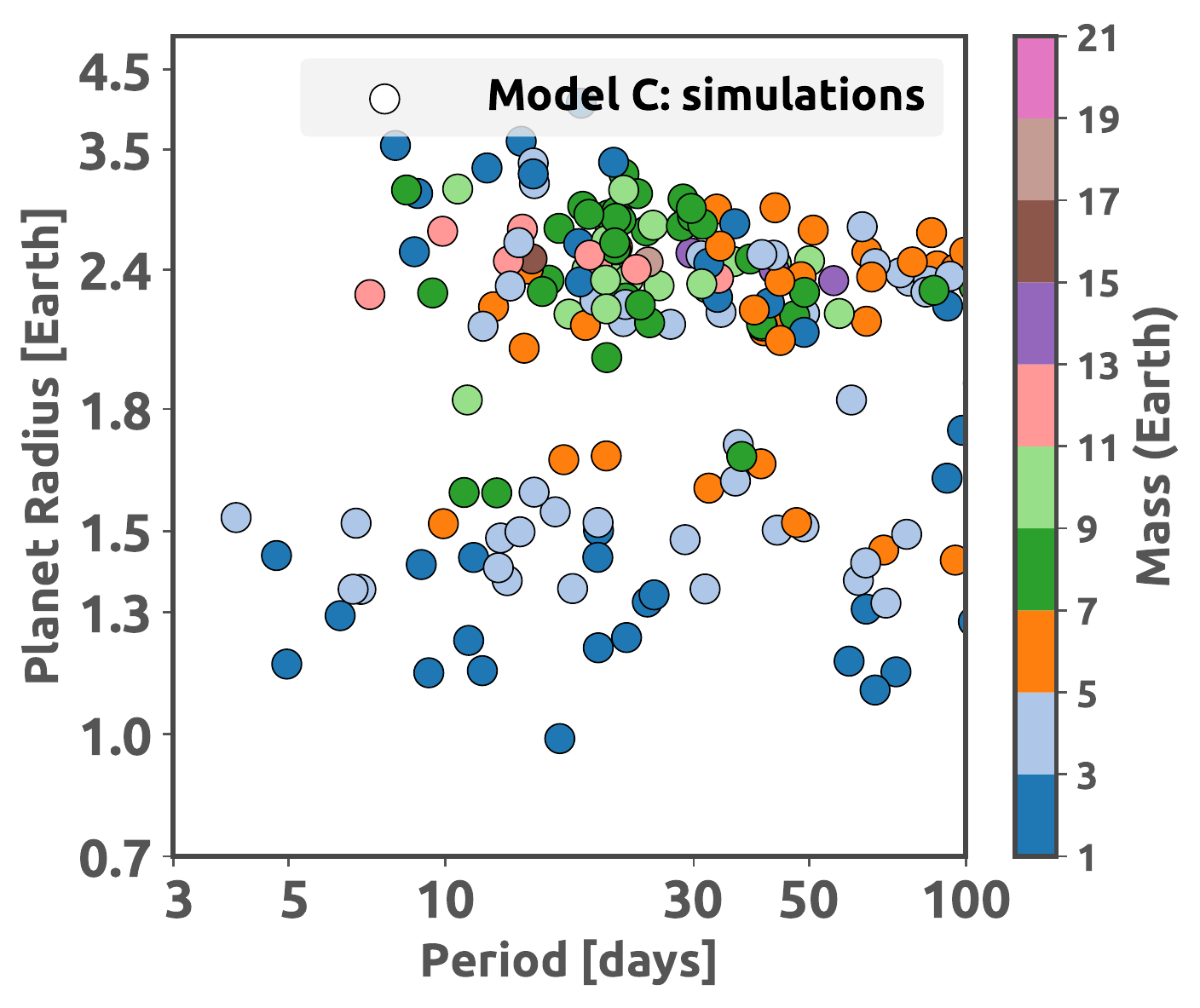}

\includegraphics[scale=.52,trim = 0cm 0cm 1.88cm 0.1cm,clip]{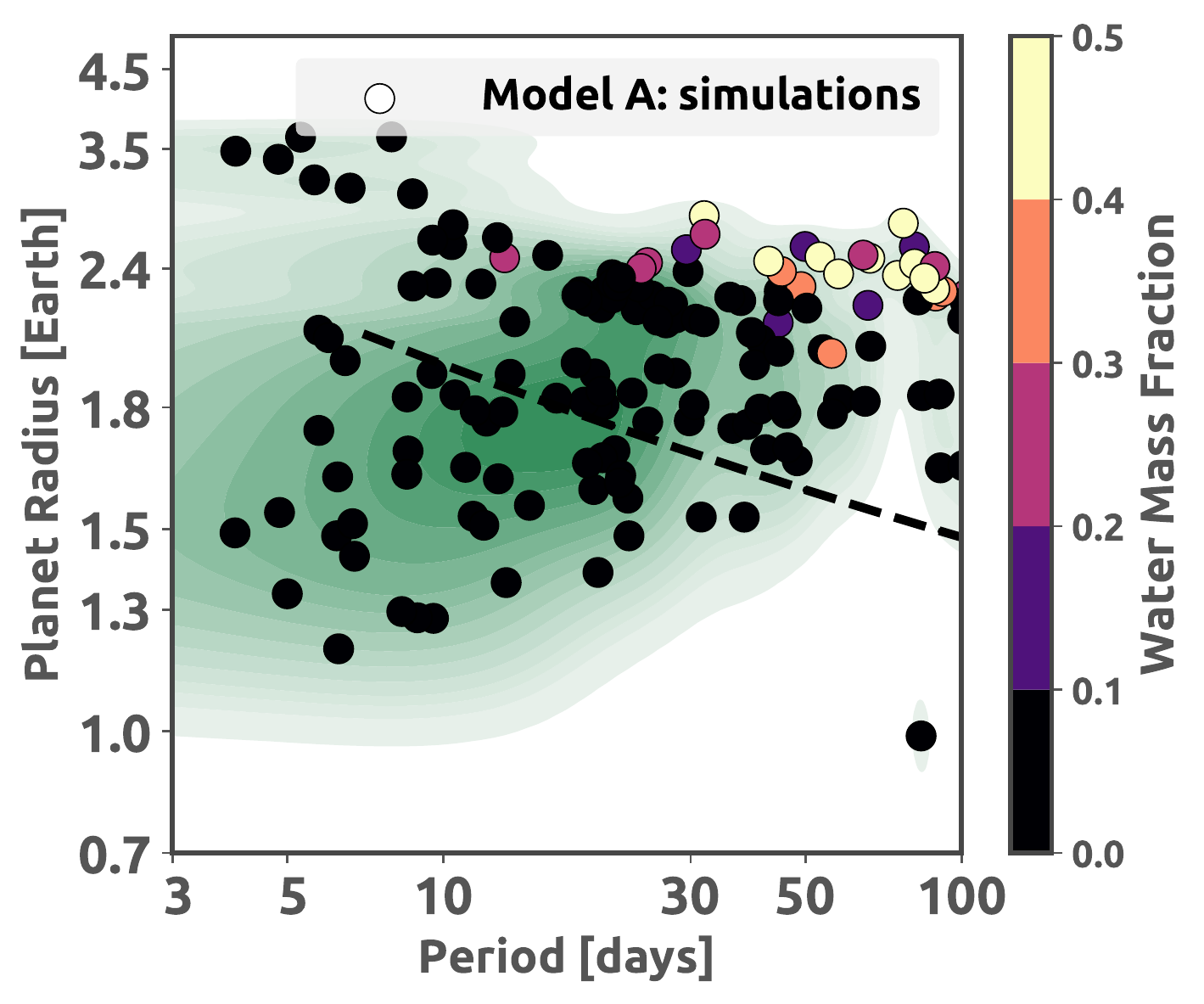}
\hspace{-0.49cm}	
\includegraphics[scale=.52,trim = 1.9cm 0cm 1.88cm 0.1cm, clip]{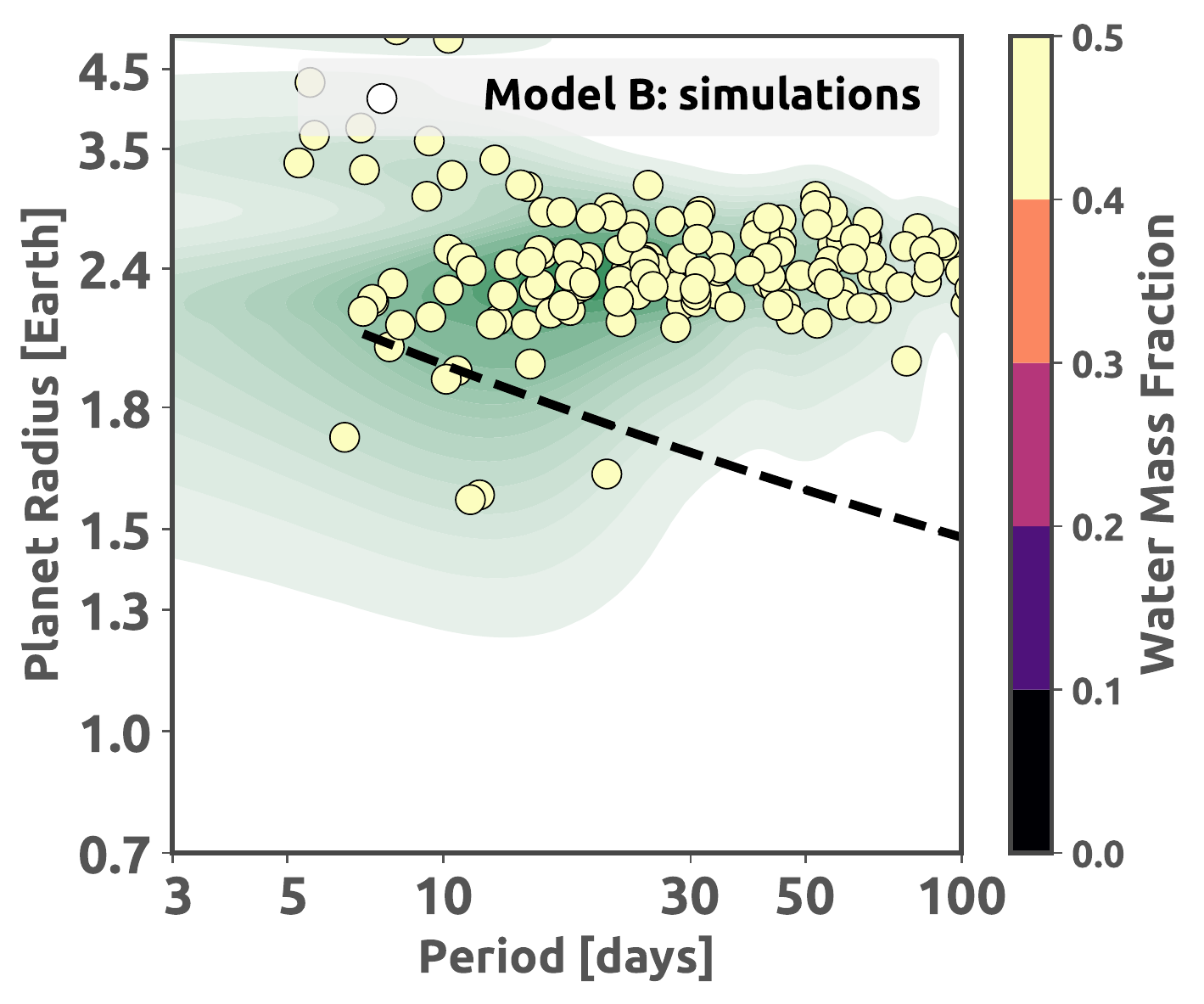} 
\hspace{-0.49cm}	
\includegraphics[scale=.52,trim = 1.9cm 0cm 0cm 0.1cm,clip]{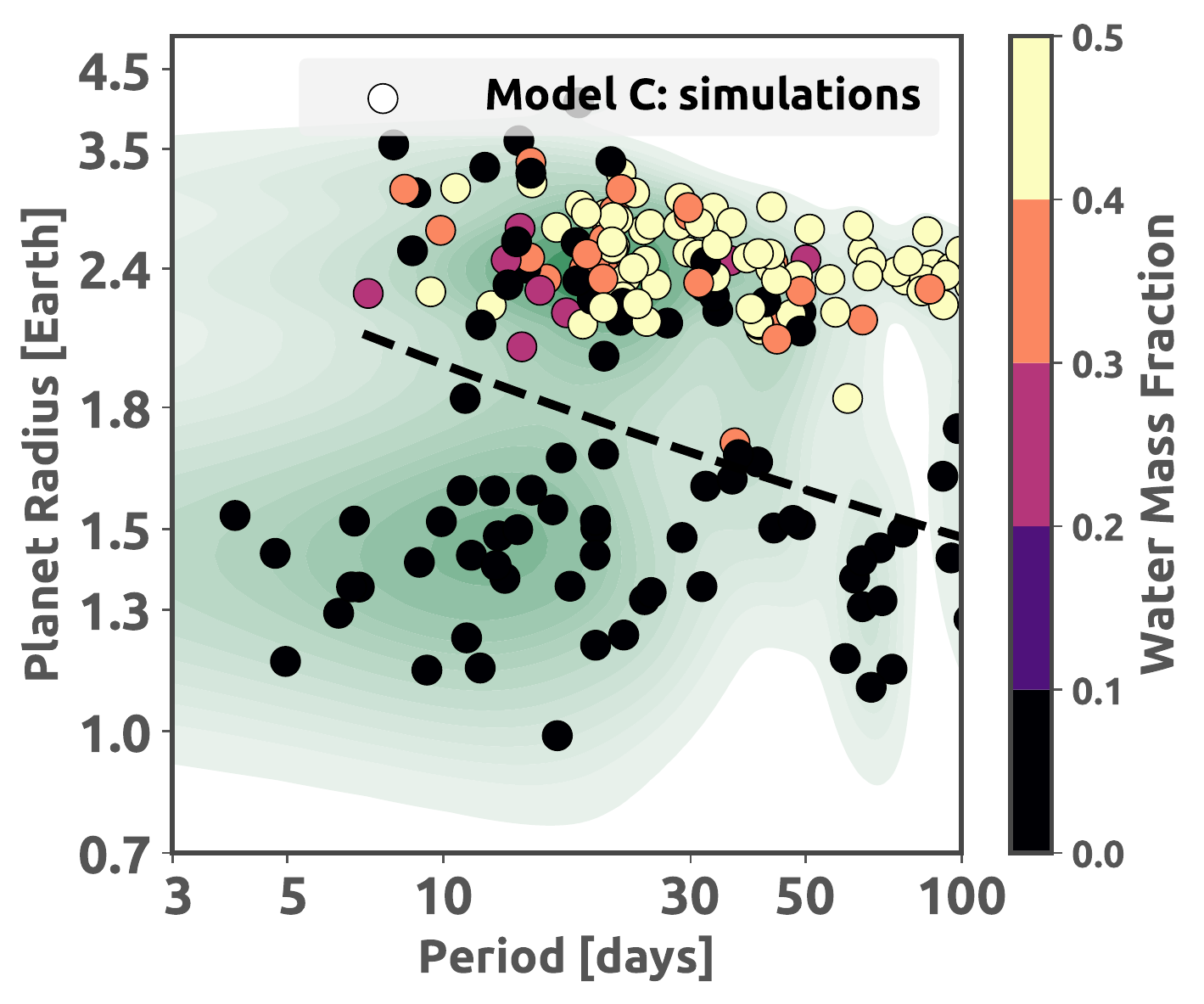}
\caption{Architecture of planetary systems at the end of our simulations, i.e., after dynamical instabilities have taken place. The horizontal axis shows orbital period, and the y-axis planetary radius as calculate from the mass radius relationships of \cite{zengetal19}.  Planets are shown as individual dots. We only show planets with orbital periods shorter than 100 days. In all these models, more than $\sim$95\% of the resonant chains became unstable after gas disk dispersal. Models A, B, and C are shown from left to right. Points are color-coded according to the planets'
number of late giant impacts (top row of panels), masses (middle row of panels), and ice-mass fractions (right column). We neglect the effects of photo-evaporation and/or other subsequent atmospheric mass-loss in all these cases (see Appendix for simulations where the effects of photoevaporation are included). The dashed lines in the bottom panels show the exoplanet radius valley slope, calculated  as $R =10^{-0.11{\rm log}_{10}(P)+0.4}$~\citep{vaneylenetal18,guptaetal19}. The green background-contours show the kernel density distribution of the planets in our simulations (dots of the figure).}
\label{fig:after_instabilities}
\end{figure*}

Figure \ref{fig:after_instabilities} shows the final architecture of our planetary systems after dynamical instabilities have taken place (and without the effects of photoevporation; see Appendix). In addition to showing planet masses and water mass fractions via color-coding (as in Figure \ref{fig:systemsbeforeinstability}), we also show the number of late giant impacts that each planet experienced during the final stage of our simulations between 5-100~Myr. Panels in the left, central, and right columns show model A, B and C, respectively. 
During the dynamical instability phase, planets experienced up to 4 giant impacts, with most of the planets experiencing one or two giant impacts. About 26\% of the planets in model A, 36\% of the planets in model B, and 43\% of planets in model C did not experience any giant impact after gas disk dispersal. Model C exhibits the lowest number of late giant impacts because its planetary systems are less crowded at the beginning of the dynamical instability phase as compared to model A and B. 

The green shaded regions in the bottom panels of Figure \ref{fig:after_instabilities} shows the density distribution of the simulated planets, with darker green indicating higher density. For comparison purposes, we also show with a black dashed line the location of the center of the exoplanet planet radius valley as a derived by previous studies~\citep{vaneylenetal18,guptaetal19}.
In model A, the planet density distribution has a single peak that covers a broad range of planet radii from roughly 1.5 to 3 $R_{\oplus}$. This distribution is not consistent with exoplanet observations.  Model B also shows a single-peaked distribution centered at about 2.4~$R_{\oplus}$, but, in contrast to model A, the peak covers a narrow range of planet radii, from roughly 2 to 3 $R_{\oplus}$. However, exoplanets show a second peak in the radius distribution at about 1.4~$R_{\oplus}$, which is not accounted for by model B. Finally, model C shows two peaks at $\sim$1.4~$R_{\oplus}$ and 2.4~$R_{\oplus}$ and a deficit of planets at $\sim$1.8~$R_{\oplus}$ which coincides fairly well with the location of the exoplanet planet radius valley~\citep[see black-dashed line;][]{fultonpetigura18,vaneylenetal18}. Model C is therefore the model that best matches the demographics of exoplanetary systems.

In order to understand the origin of the radius valley in model C, it is important to recall that this model produces a dichotomy in composition, with planets larger than 2$R_{\oplus}$ being water-rich  and planets under 1.5$R_{\oplus}$ being mostly rocky. The  top-right panel of Figure \ref{fig:after_instabilities} shows that all planets smaller than 1.5$R_{\oplus}$ experienced at least one late giant impact. To understand how the radius valley emerges, let's take as an example a rocky planet that before the instability phase has an orbital period of 23 days, a mass of about 3$M_{\oplus}$, and  an atmosphere-to-core mass ratio of 0.3\%. The radius of this planet, as calculated using the mass radius relationship of~\cite{zengetal19}, is roughly 2.1$R_{\oplus}$. If this planet collides with  an equal mass planet of similar composition during the instability phase it will lose its primordial atmosphere~\citep{bierstekeretal19}, and its final mass and radius will be roughly 6$M_{\oplus}$ and 1.6$R_{\oplus}$, respectively. Collisions result in atmospheric loss, which reduces the radius of rocky planets significantly, moving them from above to below the radius valley. 

Note, however, this is not the case for water-rich planets. To illustrate this contrasting scenario, let's take a second planet with the same orbital period of 23 days, same mass and atmosphere-to-core mass ratio,  but with water-rich composition (water-mass fraction of 50\%) at the end of the gas disk phase. In this case, the planet radius before the instability phase is about 2.9$R_{\oplus}$. If the planet collides with an equal mass planet of similar composition, its mass will double and its radius will be roughly 2.1$R_{\oplus}$, which is still above the radius valley. This shows that the origin of the radius valley in our model is associated to a dichotomy in (core) composition. In addition, water-rich planets/cores are typically more massive at the end of the gas disk phase than rocky ones, due to the higher efficiency of pebble accretion and larger pebble isolation mass beyond the snowline than in the inner disk~\citep{lambrechtsetal14,bitschetal18b,bitsch19}. This reduces the likelihood that water-rich planets move below or fill the radius valley. If dynamical instabilities after the disk dispersal, and the resulting planetary collisions, are a  common process of planet formation, our models indicate that the radius valley does not form if planets/cores have similar compositions and masses, as in model A and B (see also Section \ref{sec:discussion}).

\begin{figure*}
\centering
    \includegraphics[scale=.5]{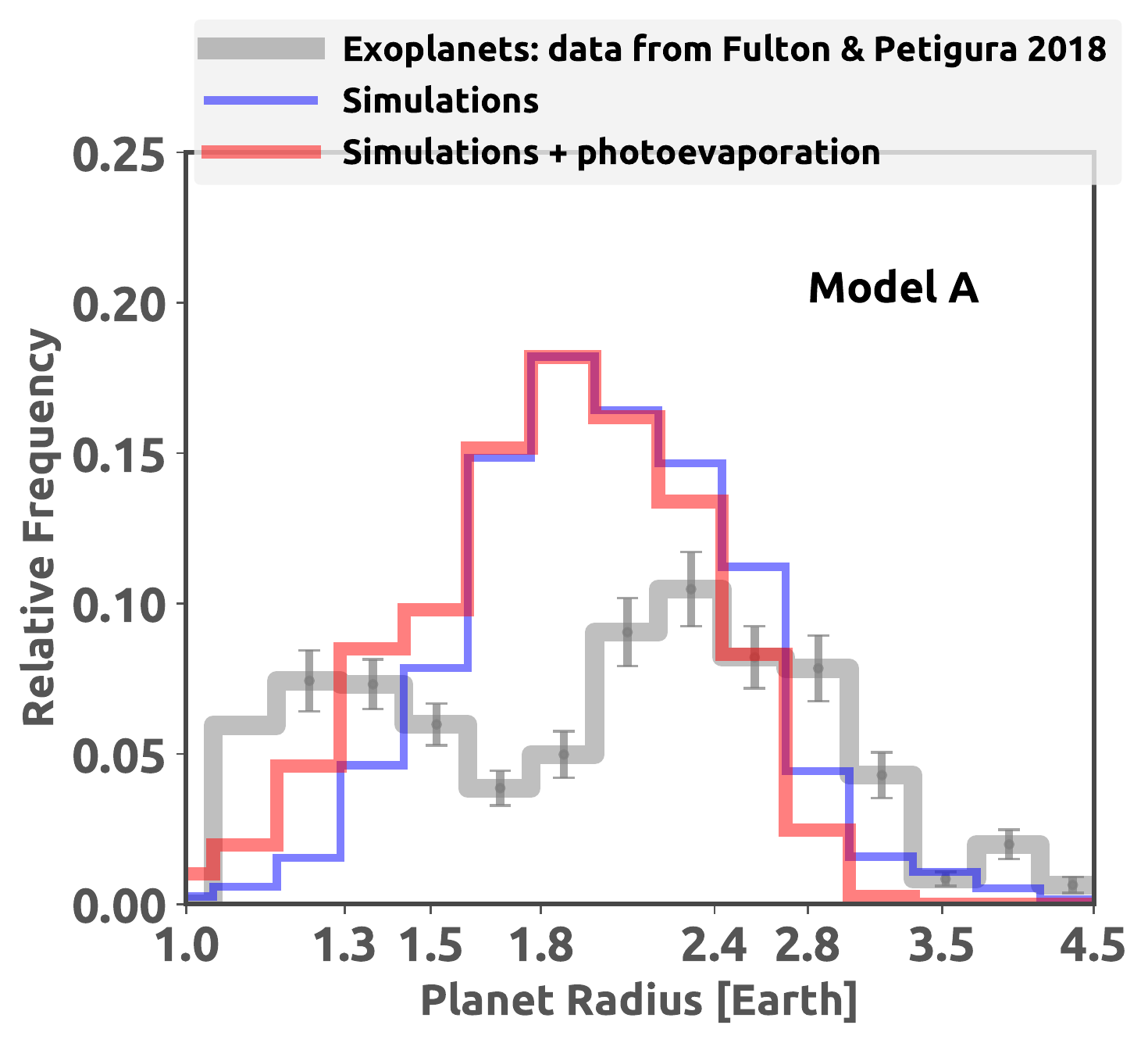}
        \includegraphics[scale=.5]{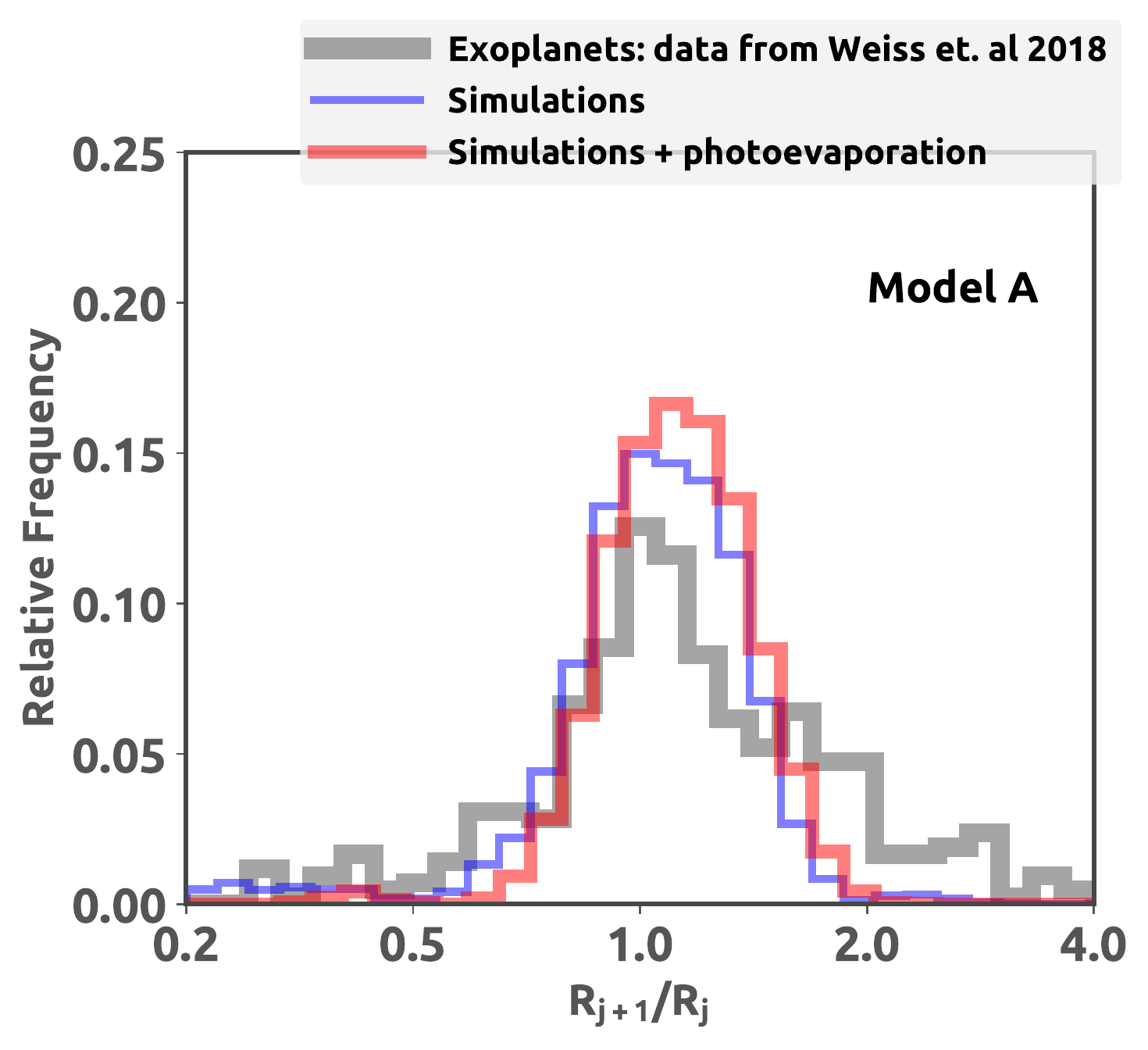}
    \includegraphics[scale=.5]{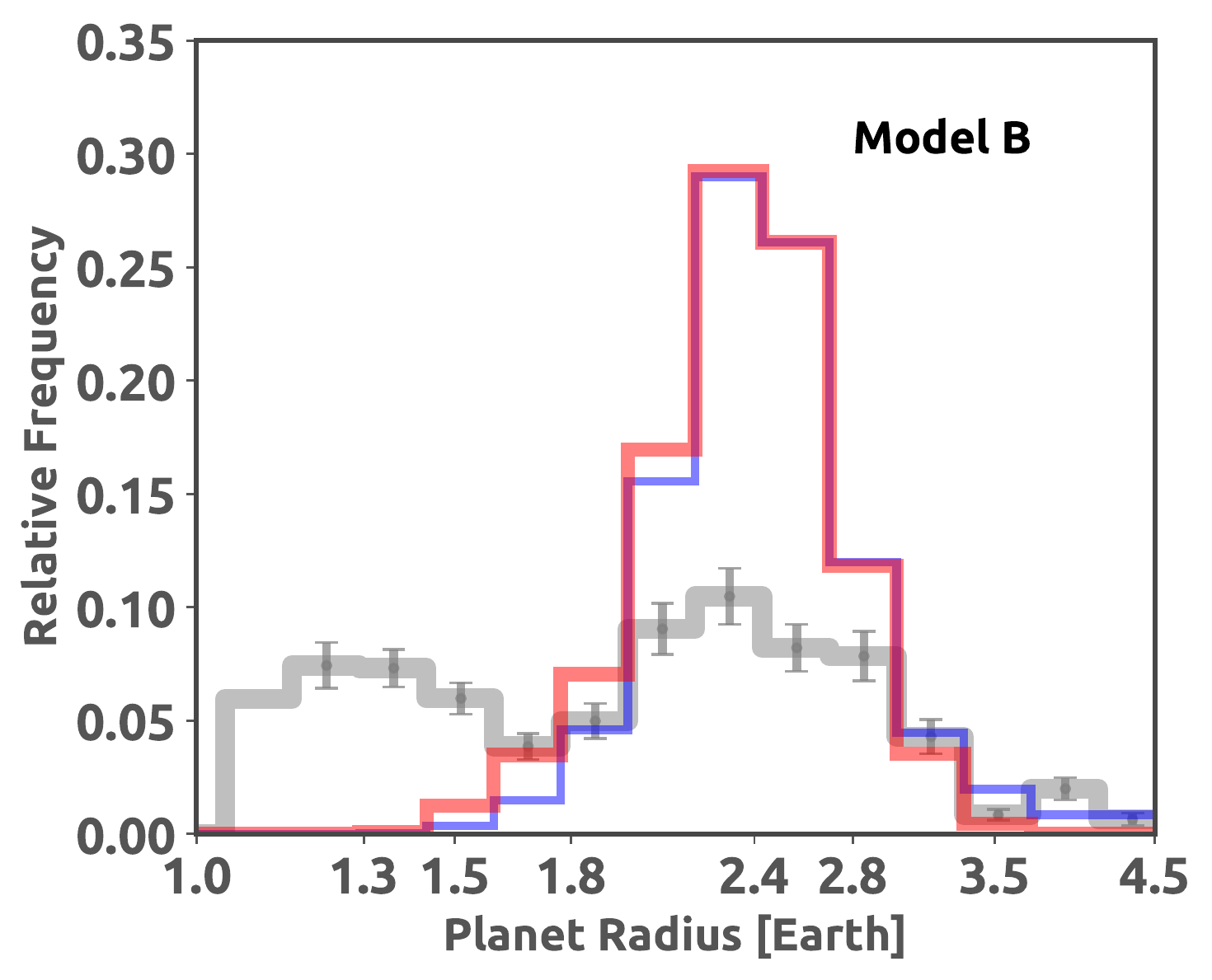}
        \includegraphics[scale=.5]{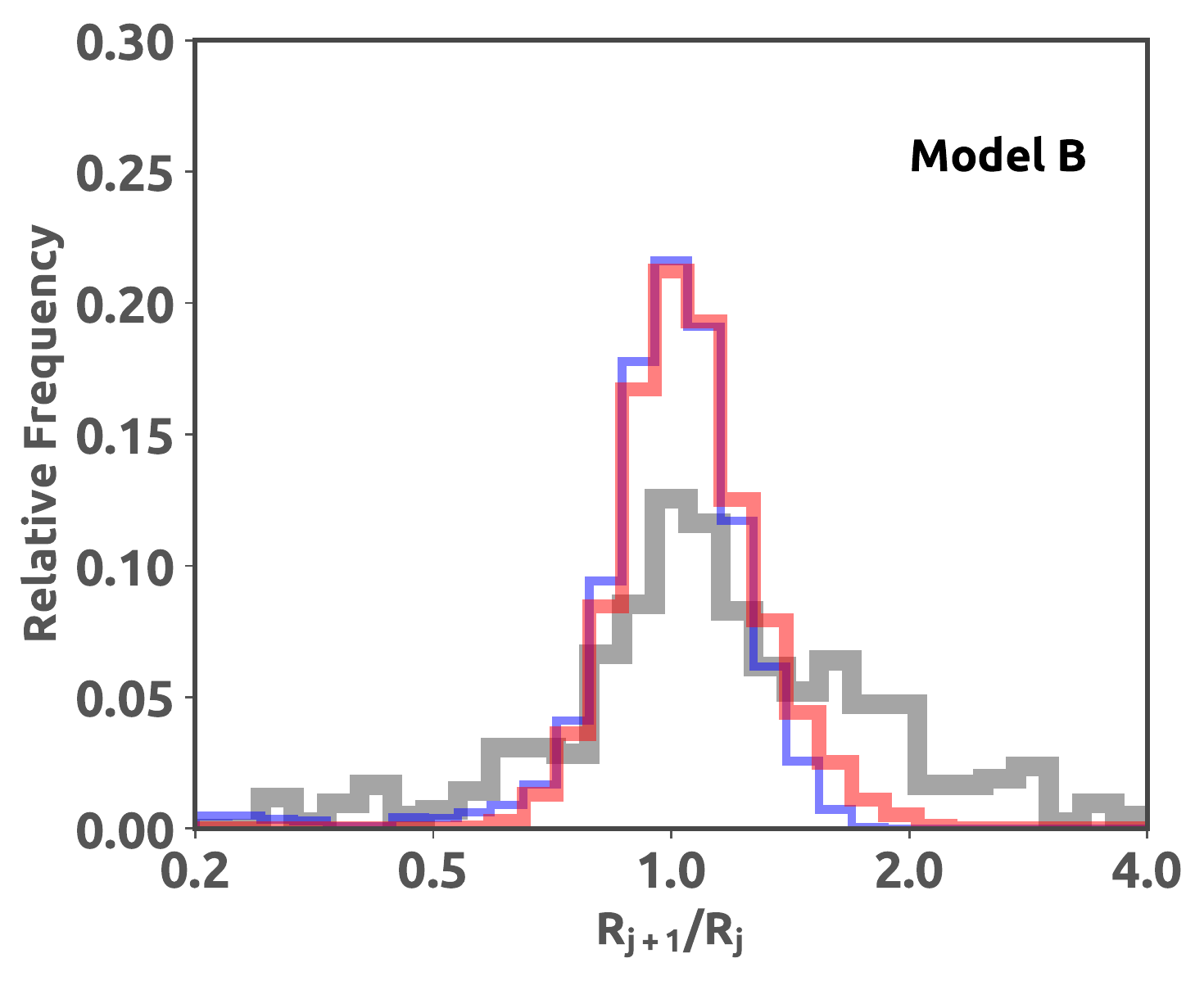}
     \includegraphics[scale=.5]{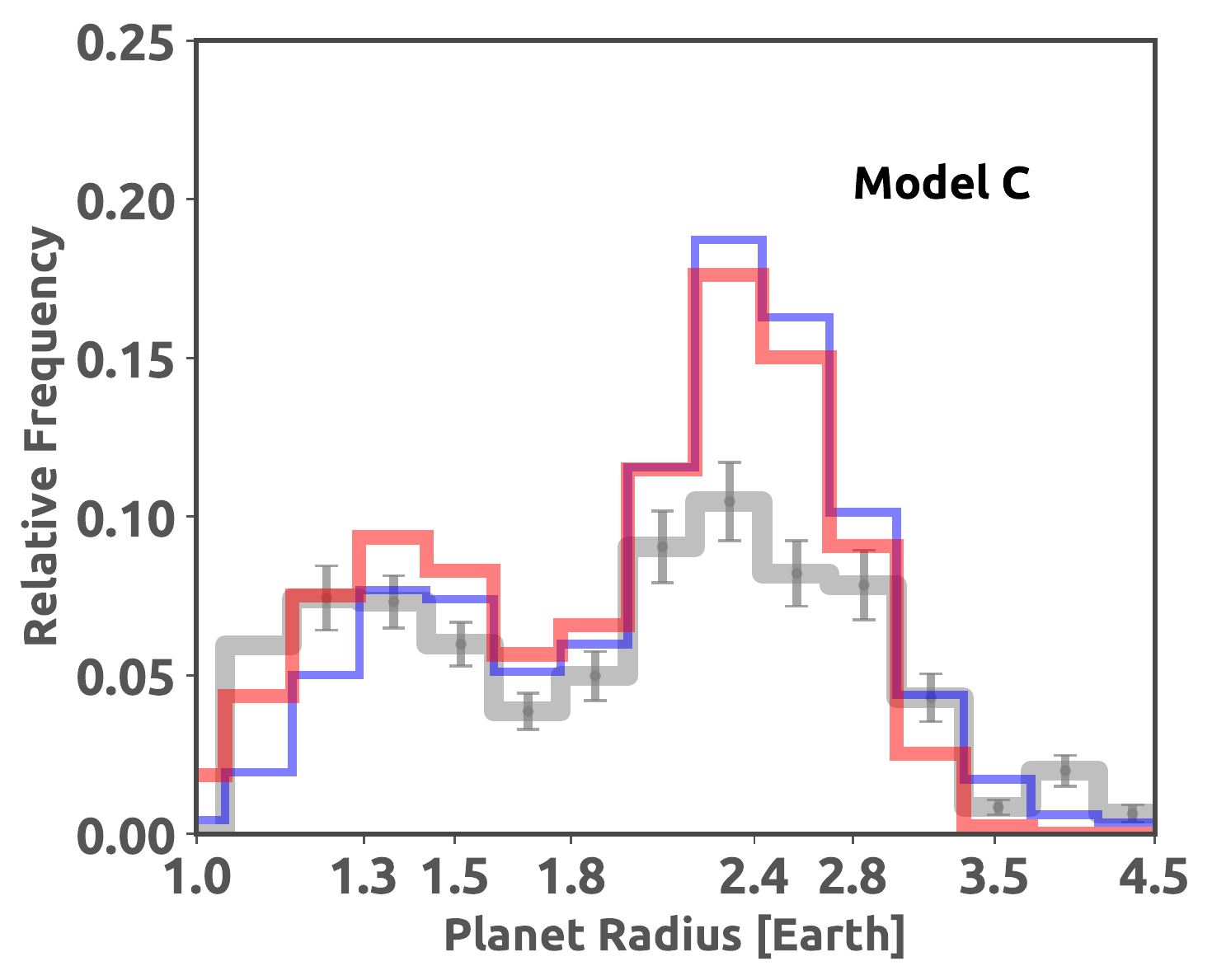}
        \includegraphics[scale=.5]{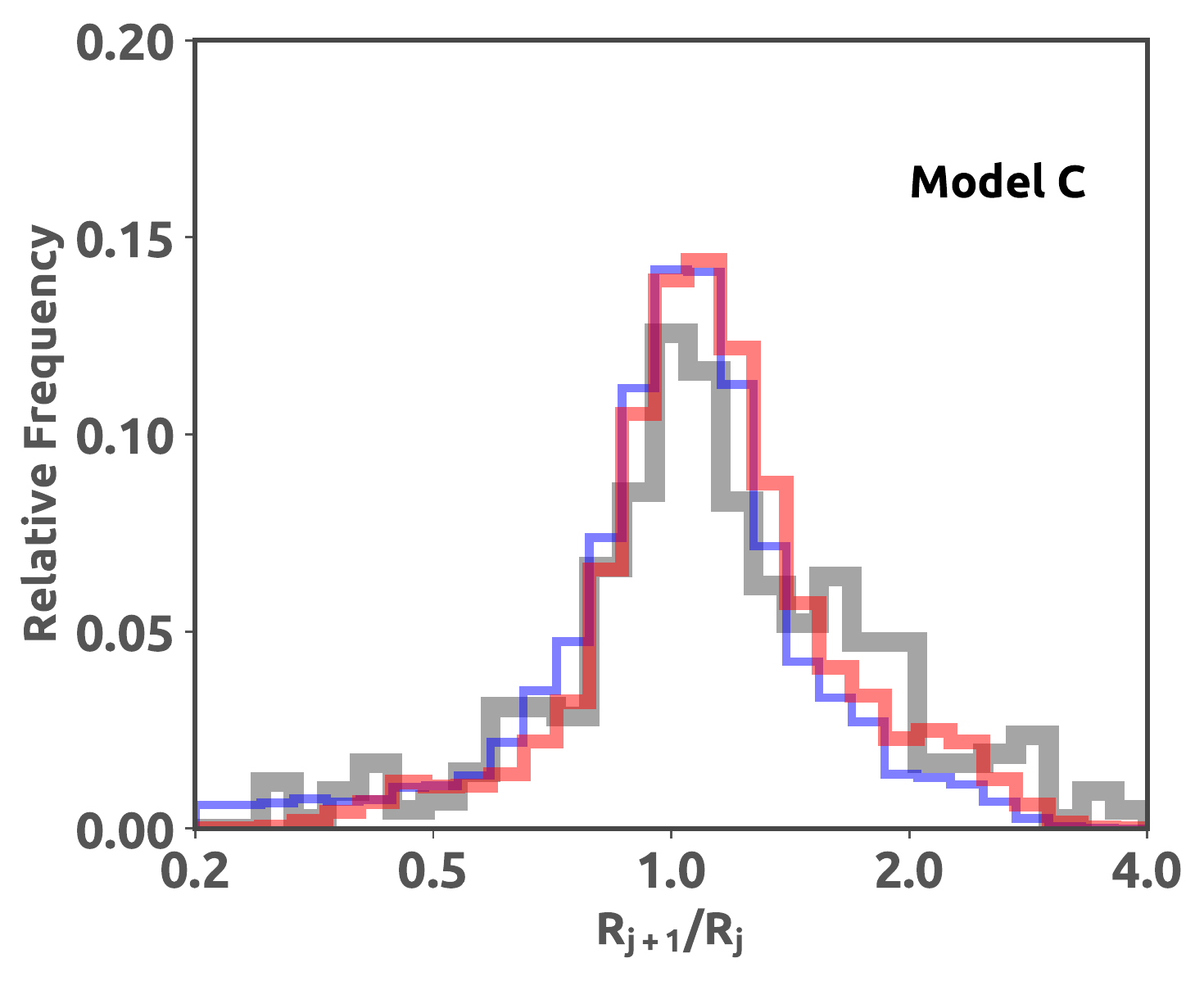} 
    \caption{{\bf Left:}  Planet radius distribution. {\bf Right:} Planet size-ratio  distribution of adjacent planet pairs. These distributions are computed after dynamical instabilities and the breaking of the resonant chains. Exoplanets are shown in gray. Blue shows the outcome of our planet formation simulations. Red shows the outcome of our planet formation simulations including the effects of photo-evaporation. From top-to-bottom, the panel-rows show model A,  B, and  C. For all models, we use the mass-radius relationship of~\cite{zengetal19} assuming an uncertainty of 7\% in size and an initial atmosphere-to-core mass ratio of 0.3\%. We build these histograms accounting for uncertainties in radius by generating the possible radius of each planet 100 times.}
    \label{fig:distributions}
\end{figure*}

Supporting the results of Figure \ref{fig:after_instabilities}, Figure \ref{fig:distributions} shows that the only model showing a valley in the planet-radius  distribution is model C. The peaks at 1.4-1.5$R_{\oplus}$ and 2.4$R_{\oplus}$ are also pronounced, matching fairly well the location of the peaks in the CKS data. Our simulations, however, produce a peak at 2.4$R_{\oplus}$, higher than that of exoplanets after completeness corrections~\citep{fultonpetigura18}. Assuming that the relative sizes of the observed peaks are truly representative of reality -- which may not be the case, for instance, due to observational bias -- one could imagine several ways to reconcile our model with observations. Possible solutions could be achieved via an increase in  the number of systems with rocky planets,  an increase in the efficiency of photo evaporation or the inclusion of additional atmospheric mass-loss mechanisms, or a reduction in the efficiency of formation of water-rich planets (see also Appendix, where we test a different mass-radius relationship and different atmosphere-to-core mass ratios). 

The panels on the right column of Figure \ref{fig:distributions} show the size ratio distribution of adjacent planets. The exoplanet sample in this case comes from~\cite{weissetal18}, but still from the CKS sample. All our models produce size-ratio  distributions that are as narrow as the exoplanet sample and broadly match the peak at $R_{j+1}/R_j\approx1$. However, model C provides the best match to the peak and shape of the frequency distribution. These results demonstrate that our model,  which self-consistently accounts for planets' compositions,  dynamical evolution and late giant impacts,  provides a natural explanation for both the observed distribution of  planetary radii (i.e, the radius valley feature) and size ratios (i.e., the peas-in-a-pod feature).

Additional results for cases including the effects of photoevaporation are shown in the Appendix.  Overall, we find that in our model photoevaporation has a negligible to small impact on the distribution of planets in the planet radius vs orbital period space. Planets populating the radius valley in model A do not have atmospheres as consequence of giant impacts, and therefore, they cannot be stripped via photoevaporation In conclusion, if late giant impacts are common, photoevaporation does not  seem to affect the general result that the radius valley requires a dichotomy in composition of planetary cores. We have also included the effects of observational bias in our simulations by performing synthetic transit observations of our planetary systems, and our results also do not qualitatively change (see Appendix).

In Figure \ref{fig:planettypes} we show the fraction of planets of different types produced in our simulations of model C. If we neglect the effects of photo-evaporation or any other sub-subsequent atmospheric mass-loss (Figure \ref{fig:A1a}), 27.8\%  of our planets are rocky, 17.2\% are rocky with primordial atmospheres, 29.8\% are water-rich, and 26.3\% are water-rich with primordial atmospheres. The number of planets with atmospheres -- regardless of the composition -- drops when we included the effects of photo-evaporation, as expected. However, the effects of photo-evaporation  increases the fraction of bare rocky planets by only $\sim$10\% and that of water-rich planets by only $\sim$2\% (see Figure \ref{fig:planettypes}). The breaking the chains model and photo-evaporation, or possibly other atmospheric mass-loss mechanisms, are not mutually exclusive processes but our results shows that giant impacts play the dominant role sculpting the radius valley.

\begin{figure*}
\hspace{1cm}\subcaptionbox{Before photoevaporation\label{fig:A1a}}[13em]{\centering \includegraphics[scale=.5]{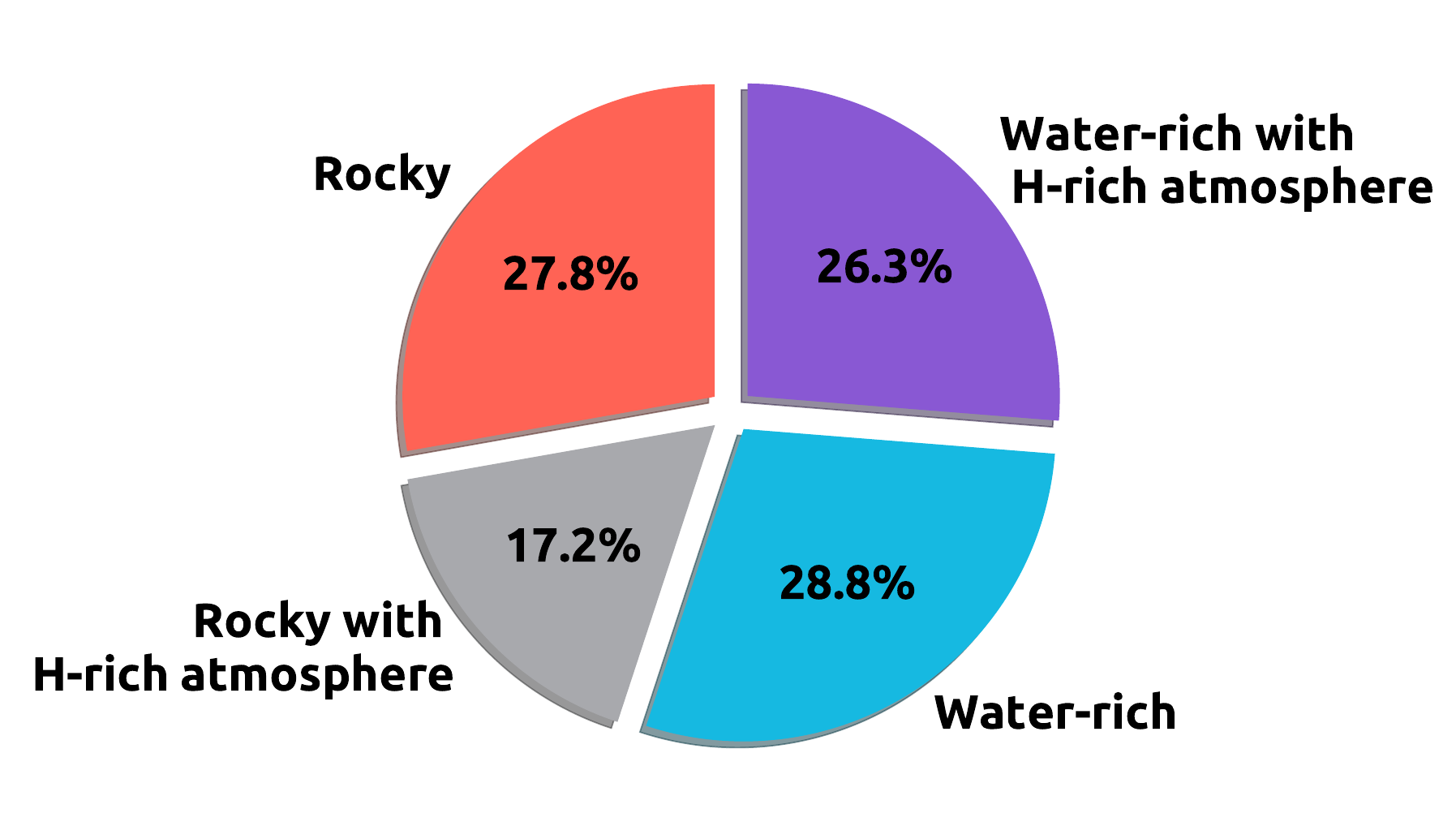}}
\hspace{5cm}\subcaptionbox{After photo-evaporation\label{fig:A1b}}[13em]{\centering \includegraphics[scale=.5]{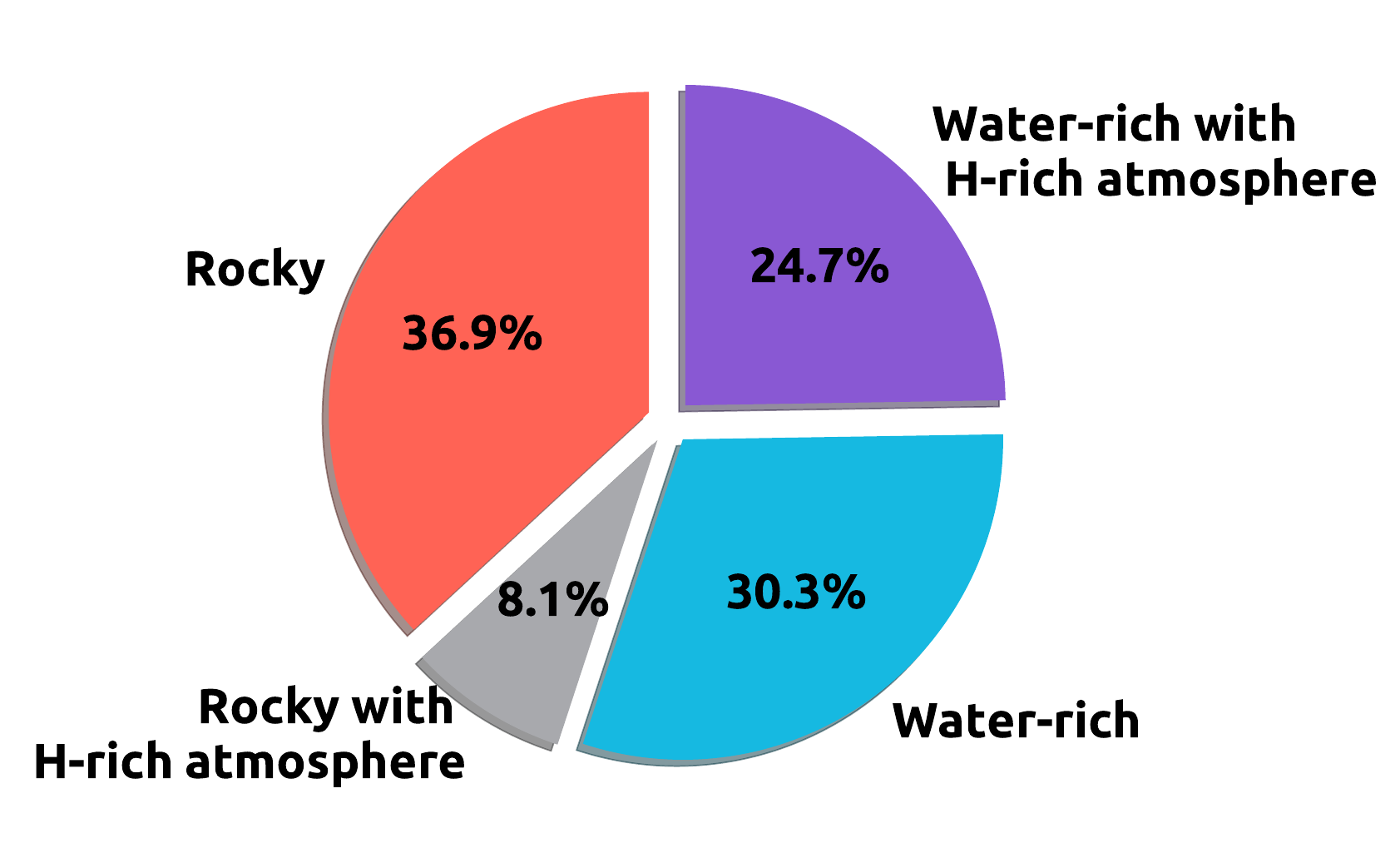}}
    \caption{Fraction of planets of different types produced in our simulations of model C, that match the radius valley and peas in a pod feature. The left-hand side pie chart shows the relative fractions when the effects of photo-evaporation are ignored. The right-hand side panel shows the relative fraction  when  we include the effects of photo-evaporation.} 
    \label{fig:planettypes}
\end{figure*}

\begin{figure*}
\centering
	\includegraphics[scale=.3]{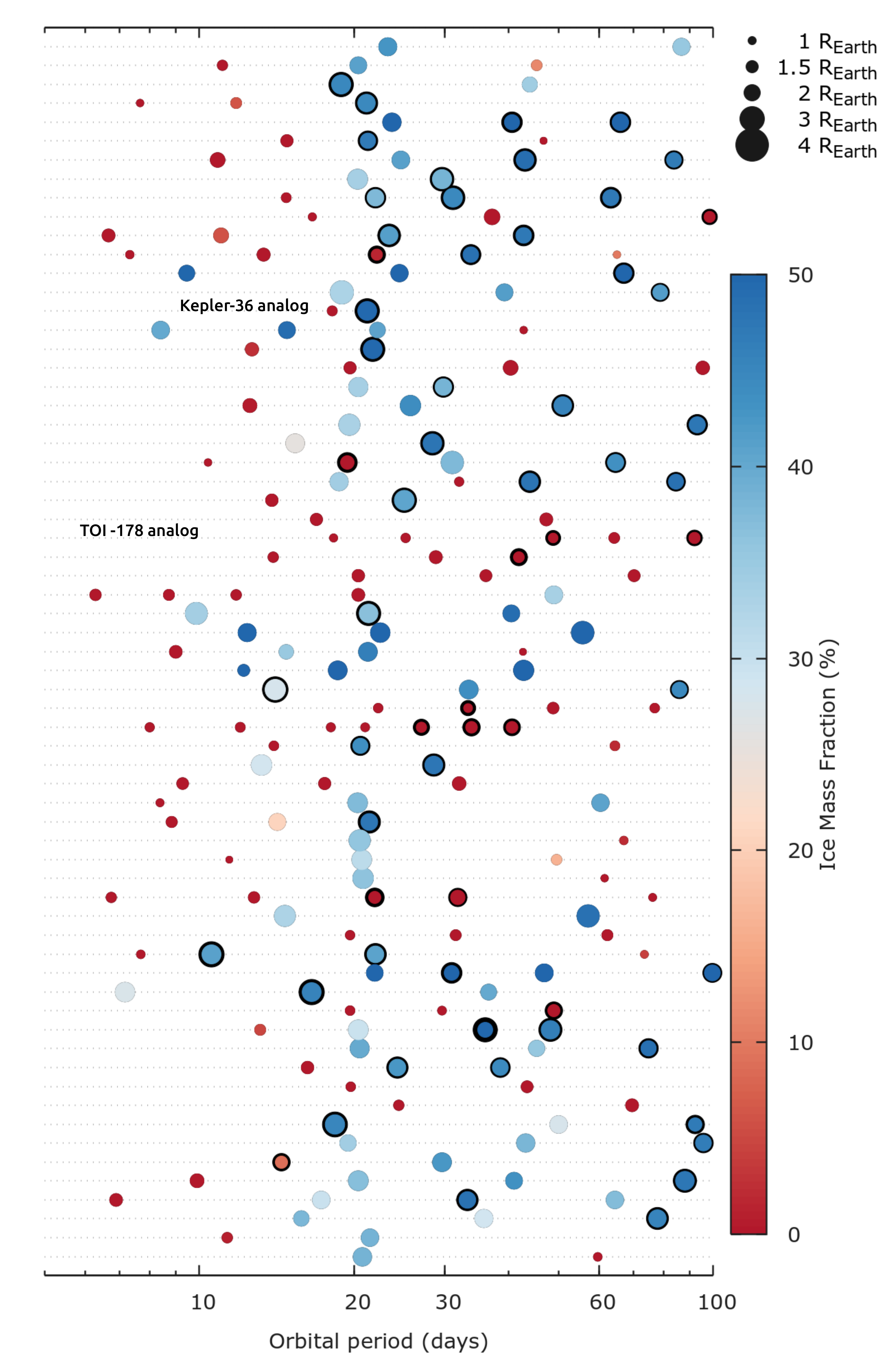}   
    \caption{Final planetary systems produced in Model-C.  Each line shows one planetary systems. The color-coding indicates the planet composition.  A black-ring around the dot indicates the presence of  atmosphere, and the full-size of the dot scales with its radius. When determining the final planet sizes, we include atmospheric stripping by photo-evaporation after the giant impact phase. Planets atmospheric masses, when present, are assumed to correspond to 0.3\% of the core mass.}
    \label{fig:final_systems}
\end{figure*}

Figure \ref{fig:final_systems} shows the final planetary systems produced in our simulations including the effects of photo-evaporation.  The innermost planets are typically rocky whereas water-rich mini-Neptunes tend to have relatively larger orbital periods~\citep{izidoroetal21a}. This orbital arrangement is broadly consistent with observations~\citep{millhollandwinn21}. Figure \ref{fig:final_systems} also shows that migration and dynamical instabilities promote a great diversity of planetary compositions. Some of systems  show, for instance, adjacent planets with distinct compositions as a rocky planet adjacent to a water-rich one or a rocky planet adjacent to a rocky planet with  a primordial atmosphere. These systems are consistent with the orbital architecture of intriguing  systems like  Kepler-36~\citep{carter12} and TOI-178 (\cite{leleuetal21}, see example systems in Figure \ref{fig:final_systems} and \cite{raymondetal18})  .

\section{Discussion}\label{sec:discussion}

Our model A, which was designed to produce planetary systems dominated by rocky planets/cores~\citep{izidoroetal21a} failed to reproduce the exoplanet radius valley. Model A was based on a specific set of initial conditions (see Section \ref{sec:simulations}) but systems dominated by rocky planets -- as Model A -- can be achieved using different initial parameters in the model of~\cite{izidoroetal21a}. So, can we conclude that the breaking the chains scenario is generally  inconsistent with the exoplanet radius valley when planetary systems are dominated by planets/cores with rocky composition?  
The best way to address this question would be to perform an extensive exploration of the parameter space that define our model. However, this is impossible due to the computational time required to perform the simulations. Nevertheless, we can try to solve this problem using a different approach, based on the following reasoning. 

Let's first assume that as postulated by our model all the super-Earths are the result of collisions between mini-Neptunes.
In this case, super-Earths should be more massive than mini-Neptunes, and this is in direct contrast with the observations (see \cite{chenkipping17,otegietal20}). For instance,
the ``super-Earth'' peak is observed at  $\sim$1.4~$R_{\oplus}$~\citep{fultonpetigura18}, which corresponds to a mass of about 3.5$M_{\oplus}$, assuming an Earth-like composition. The inner edge of the valley corresponds to a size of about 1.6$R_{\oplus}$ and translates to a mass of $\sim6M_{\oplus}$~\citep{zengetal19}, also for an Earth-like composition. In the context of the breaking the chains model, rocky planets less massive than $\sim$6$M_{\oplus}$ ($<$1.6$R_{\oplus}$)  are envisioned to be the outcomes of one or more giant impacts that stripped primordial planetary atmospheres of mini-Neptunes. As one or two collisions are a common outcome of dynamical instabilities, it implies that mini-Neptunes should have masses in the range of $\lesssim$1-3$M_{\oplus}$ before collisions.

 We have performed additional simulations
to further investigate the plausibility of this scenario. We refer to this additional set of simulations as model D. These simulations were designed to produce exclusively rocky planets with  masses relatively  lower than those produced in model A.  Before dynamical instabilities, the average mass of planets in this model is 1.9$M_{\oplus}$ (compare to 3.6$M_{\oplus}$ in model A). We  adjusted some of the free parameters of our model, in order to produce lower mass planets in model D. In model D, Moon-mass planetary seeds were initially distributed between 0.2 and $\sim$1~au, $t_{\rm start}$= 0.5~Myr, $R_{\rm peb}$=1~mm, and $S_{\rm peb}$=5. (compare with parameters of model A described in see Section \ref{sec:simulations}).

\begin{figure}
\centering
\includegraphics[scale=.52]{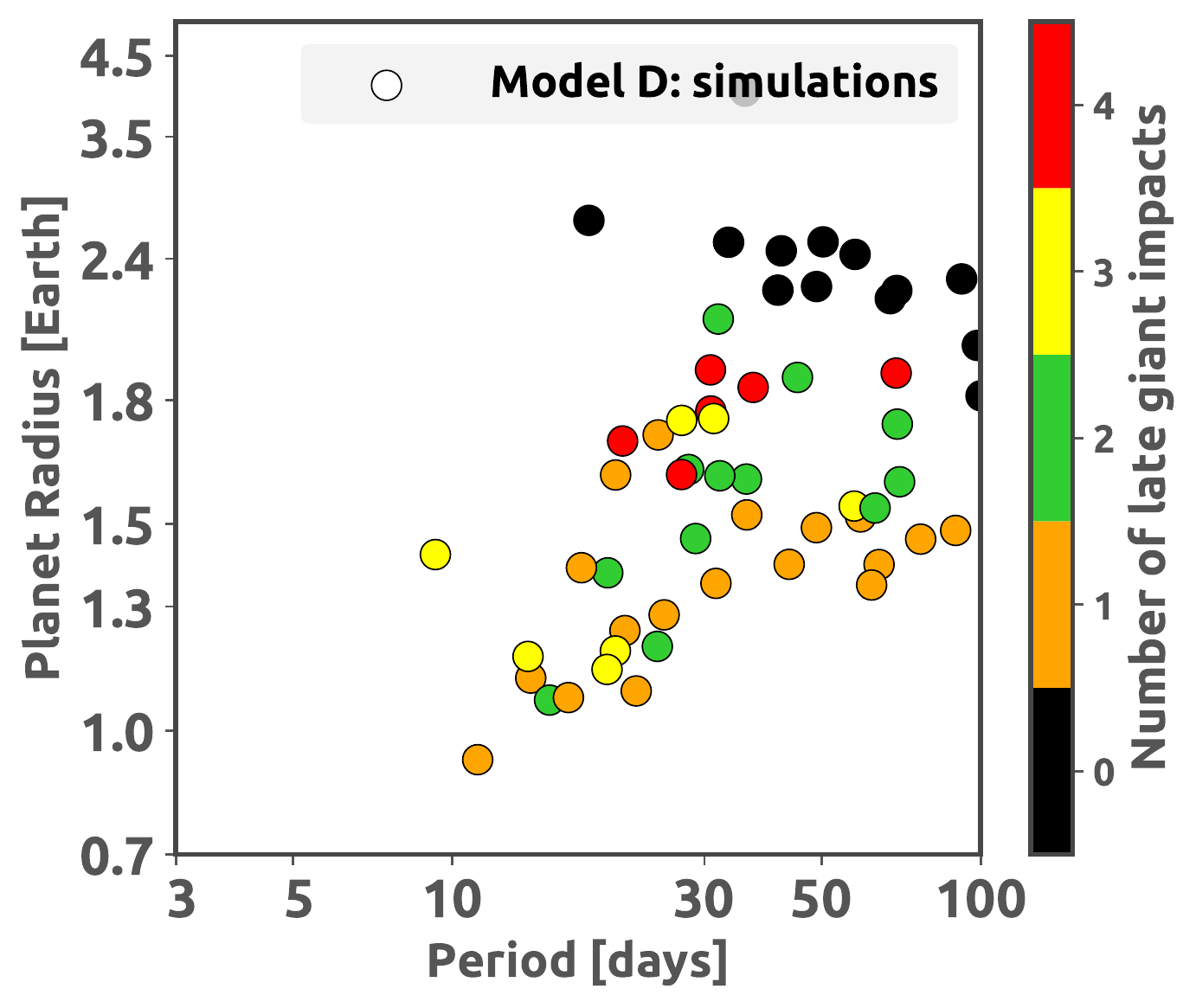}

\includegraphics[scale=.52]{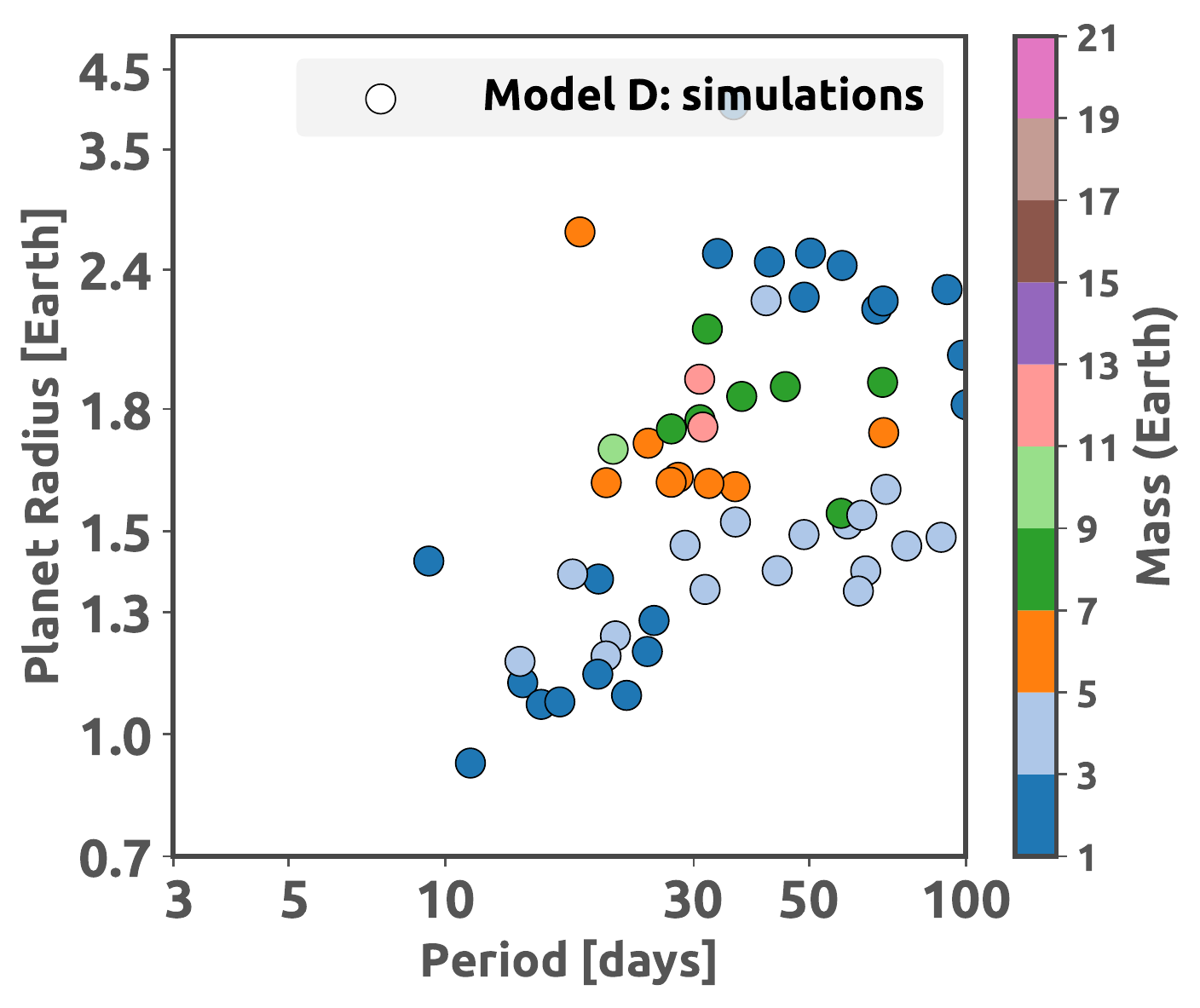}

\includegraphics[scale=.52]{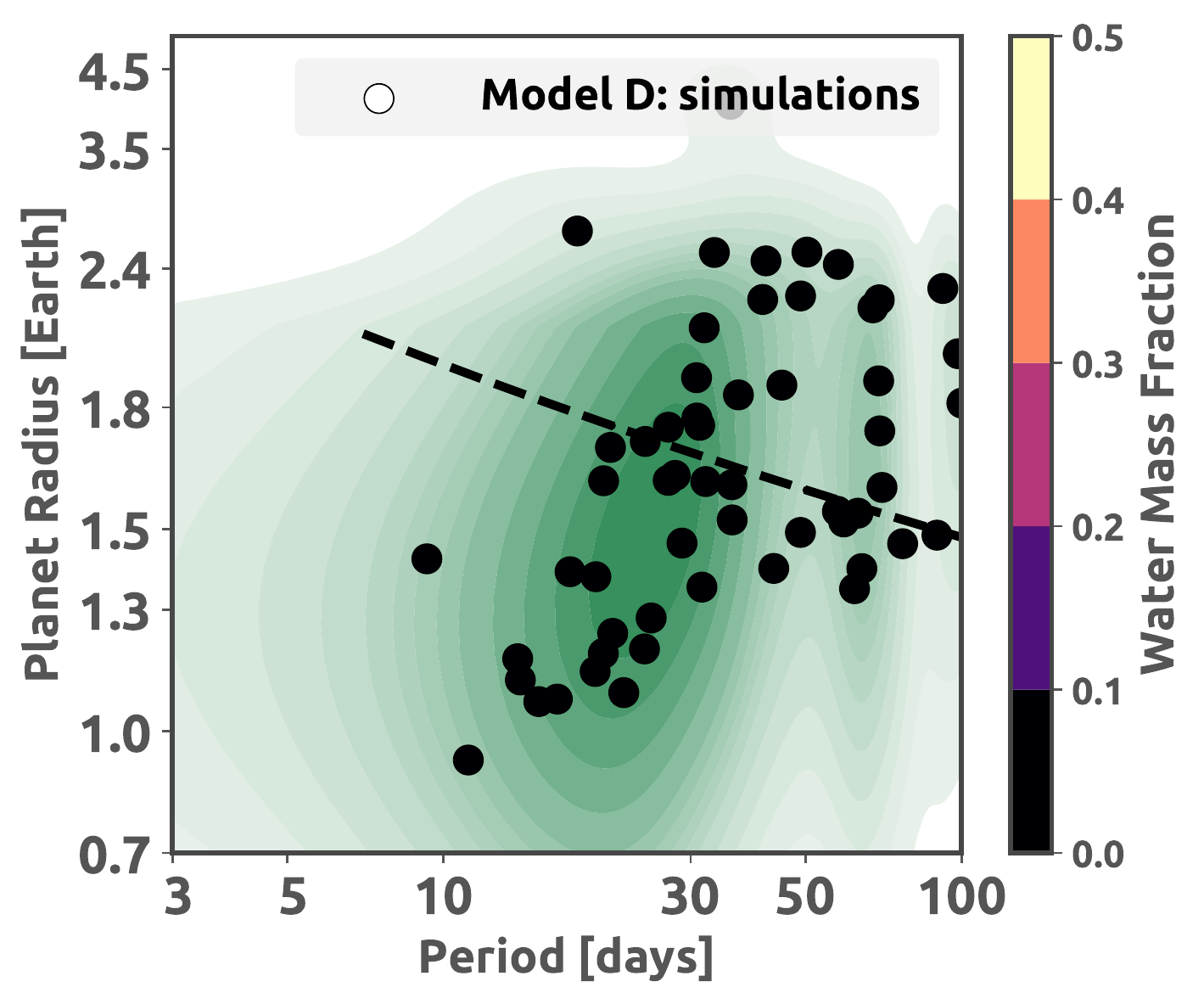}

\caption{The same as Figure 2 but for model D.}
    \label{fig:modelD}
\end{figure}

 Figure \ref{fig:modelD} shows the outcome of simulations of Model D after dynamical instabilities (as Figure \ref{fig:after_instabilities}). It shows that even when rocky planets are systematically less massive than those of model A, there is  no clear valley in the planet size distribution at $\sim$1.8$R_{\oplus}$. Interestingly, the final planet-mass distribution of model D is broadly consistent with that predicted by photo-evaporation models~\citep{owenwu17}. We have verified that -- after dynamical instabilities -- model's D planet-mass distribution broadly corresponds to a Rayleigh distribution with a mode of 3$M_{\oplus}$ (see \cite{owenwu17}). Yet, a valley in the radius distribution is not created  in model D because giant impacts play the dominant role stripping planetary atmospheres and filling/destroying the radius gap in our model. Note that this is different from standard photo-evaporation and core-powered atmospheric mass-loss models~\citep[e.g.][]{owenwu17,jinmordasinietal18,guptaetal19}. In these models, atmospheric accretion during planet formation is assumed to ``fill'' the radius valley before it can be sculpted by photo-evaporation or cored-powered mass-loss effects. In model D  (as in model A), planets  populating the radius valley have no atmospheres to be stripped via photo-evaporation or core-powered mass loss effects because they were stripped via late giant impacts. 

A last issue with model D is that most planets with radii larger than $\sim$2$R_{\oplus}$ (dark-blue dots in the middle panel of Figure \ref{fig:modelD}) have masses of $\sim1-3M_{\oplus}$. This mass range conflicts with the exoplanet data which suggests that mini-Neptunes typically have masses  larger than 5$M_{\oplus}$ ~\cite[e.g.][]{zengetal19,otegietal20}. Consequently, this scenario is not realistic on it's own.

Another potential mean of creating a radius valley with exclusively rocky planets in our model would require the existence of  two classes of  rocky planets with hydrogen-rich atmospheres. As before, the first class -- the progenitor of super-Earths  -- would consist of rocky planets with hydrogen-rich atmospheres and  masses of $\sim1-3M_{\oplus}$ planets. The second class of planets -- the progenitors of mini-Neptunes --  would consist of hydrogen-rich rocky planets with masses of $\gtrsim6-9M_{\oplus}$.
Before instabilities, both classes of planets have atmospheres and radii that put them about the radius valley.  
Collisions between $1-3M_{\oplus}$ planets would create bare rocky planets with masses between $2-6M_{\oplus}$ and radii below 1.8 $R_{\oplus}$. These planets would therefore populate the super-Earth side of the radius valley (as in our previous scenario). Vice versa, collisions involving $\gtrsim6-9M_{\oplus}$ planets would produce bare rocky planets with masses  $\gtrsim12-18M_{\oplus}$ and radii larger than 1.8~$R_{\oplus}$, which will be above the radius valley. In principle, one could assume that these two classes of planets could come from systems with two very distinct types of planetary system architectures. However, observations show that super-Earths and mini-Neptunes co-exist in the same planetary systems~\cite[e.g.][]{carreraetal18,millhollandwinn21,hawthornetal22} and have orbital configurations such that super-Earths are usually found closer-in as compared to mini-Neptunes ~\cite[e.g.][]{,millhollandwinn21,hawthornetal22}. This implies that these envisioned populations of progenitors should also coexist in the same system, instead of making two distinct classes of planetary systems.  None of the existing   planet formation models self-consistently accounting for migration predict a strong dichotomy in mass and orbital radii for rocky planets in a same system~\cite[e.g.][]{ogiharaetal15a,lambrechtsetal19}. Building planetary systems with such generic features -- if possible -- would require very specific and perhaps unrealistic  assumptions on the distribution of rocky material in the inner parts of  protoplanetary disks.

In light of these issues, we argue that,  if close-in  exoplanets formed via disk migration  and subsequent dynamical instabilities that lead to giant impacts -- the {\it breaking-the-chains evolution} -- then a large fraction of mini-Neptunes (and their cores) are water-rich planetary objects, as predicted by model C.

\section{Conclusion}
In this work we  have revisited the breaking the chains scenario for the formation of super-Earths and mini-Neptunes~\citep{izidoroetal17,izidoroetal21a} with the goal of testing if this model is consistent with the exoplanet radius valley~\citep{fultonetal17,fultonpetigura18} and peas-in-a-pod feature~\citep{weissetal18,weisspetigura20}. We model the formation and dynamical evolution of planetary systems. We used simulations from \cite{izidoroetal21a} that include the effects of disk evolution, pebble accretion, gas driven planet migration, eccentricity and inclination damping due to planet-gas tidal interactions, and mutual gravitational interactions of planetary embryos (see also ~\cite{lambrechtsetal19} and ~\cite{bitschetal19}). We selected  three different set ups from that study, that produced planetary systems dominated by rocky planets, water-rich planets, and planets of mixed compositions (water rich and rocky). We assume that at the end of the gas disk phase planets have atmospheres corresponding to 0.1\%, 0.3\%, 1\% or 5\% of their masses, as suggested by observations~\citep[e.g.][]{lopezfortney13,zengetal21}, numerical simulations~\citep[e.g.][]{moldenhaueretal22,lambrechtslega17} and analytical atmospheric accretion models models~\citep[e.g.][]{ginzburgetal16}.

The breaking the  chains model proposes that after gas disk dispersal more than 90-95\% of the planetary systems become dynamically unstable which leads to a phase of giant impacts.  We assume that late giant impacts strip primordial atmospheres of low-mass planets~\citep{bierstekeretal19}. By using mass radius relationship from ~\cite{zengetal16,zengetal19} for different compositions and equilibrium temperatures, we show that the breaking the chains model is consistent with the planet radius distribution, which has peaks at $\sim$1.4~$R{_\oplus}$ and $\sim$2.4~$R{_\oplus}$, and a valley at $\sim$1.8$R{_\oplus}$. Our model is also consistent with the peas-in-a-pod feature~\citep{weissetal18,weisspetigura20}.
We also tested our model using the empirical mass-radius relationship of ~\cite{otegietal20} and the results do not qualitatively change (see Appendix). 

  Our results do not support an exclusively rocky composition for the cores of mini-Neptunes. A rocky composition  -- for super-Earths and mini-Neptune cores -- is favoured by photoevaporation \citep{owenwu13,mordasini14} and core-powered atmospheric mass-loss models~\citep{guptaetal19}. Instead, we  predict that planets larger than $\sim$2~$R_{\oplus}$ (mini-Neptunes, or their cores) are mostly water-rich ($>$10\% water by mass) and planets smaller than $\sim$1.6~$R_{\oplus}$ (super-Earths) are mostly rocky. We also suggest that orbital instabilities and late giant impacts are the dominant processes sculpting the orbital architecture of super-Earths and mini-Neptunes. Photoevaporation or other subsequent atmospheric mass-loss processes play a minor role if  giant impacts after gas disk dispersal are common as suggested by our model (see Appendix).

Our results suggest that planet formation starts early (e.g.~$<$$0.5-1$~Myr) in the inner disk. This is consistent with models of dust coagulation and planetesimal formation in the solar system ~\cite[e.g.][]{izidoroetal21b,morbidellietal22}, and the estimated ages of meteorites~\cite[e.g.][]{Kruijeretal17}. 
Our results also suggest that planet formation beyond the snowline may not be as efficient as currently thought~\citep[e.g.][]{drazkowskaalibert17}, or large-range inward migration is slower or less efficient than usually considered~\cite[e.g.][]{paardekooperetal11}. Indeed, 
if planets beyond the snowline grow quickly, they migrate inward and destroy the inner rocky systems~\citep[see Model B and][]{izidoroetal14b}, or become gas giants~\citep{bitschetal19,bitschetal20}. This is inconsistent with observations. Finally, we predict that a fraction of planets larger than $\sim$2$R{_\oplus}$ should be water-rich and have a primordial H-rich atmosphere. This prediction may be tested by future  James Webb Space Telescope observations.

\begin{acknowledgements}
We are very thankful to the anonymous reviewer for reading our paper and providing constructive comments that helped to improve this manuscript. A.~Iz. thanks Sean Raymond for inspiring conversations. A.~Iz., H.~S., R.~D., and A.~Is. acknowledge NASA grant 80NSSC18K0828 for financial support during preparation and submission of the work. A.~Iz. and A.~Is. acknowledge support from The Welch Foundation grant No. C-2035-20200401.  B.~B. thanks the European Research Council (ERC Starting Grant 757448-PAMDORA) for their financial support.
\end{acknowledgements}

\clearpage
\appendix
\setcounter{figure}{0}
\renewcommand{\thefigure}{A\arabic{figure}}
\section{Simulated transit observations and  complementary results}

In order to further compare our results to observations, we conduct synthetic transit observations of our planetary systems by following the procedure discussed in ~\cite{izidoroetal21a}. In brief, each system is observed from different lines of sight characterized by inclination angles evenly spaced by 0.1\arcdeg\ from  -20\arcdeg\ to 20\arcdeg\ relative to the plane of the primordial gas disk. Viewing angles in the azimuthal direction are evenly spaced by 1\arcdeg\, from 0\arcdeg\ to 360\arcdeg.
For each line of sight, we check which planets transit in front of their host star assuming 3.5 yr long observations (as in the case of Kepler observations; ~\citealt{weisspetigura20}), and use these to create a list of detected planets. For example, let's assume that star A is orbited by planets b, c, and d, which have different orbital inclinations. If only planet b is observed to transit when the system is observed along a specific line of sight, then we add the system composed by A and b to our new list. Then, if planets c and d are observed to transit when the system is observed from a different point of view, we add the system composed by A, c, and d to the list of detected planets as well. In this way, one synthetic planetary system can result in many different observed planetary systems.  In order to be considered detected, the signal-to-noise ratio (SNR) of a transiting planet must be SNR$>$ 10.  We calculate the SNR as
\begin{equation}
    {\rm SNR} = \frac{(R_{\rm p}/R_{\odot})^2\sqrt{3.5 {\rm yr}/P}}{{\rm CDPP}_{\rm 6{\rm hr}}\sqrt{6~{\rm hr}/T}},
\end{equation}
where $T$ is the transit duration,  ${\rm CDPP_{\rm 6hr}}$  is the 6~hr Combined Differential Photometric Precision, a measurement of the stellar noise level~\citep{christiansenetal12}, $R_{\rm p}$ and P are the orbital physical radius and orbital period of the transiting planet. This equation takes into account the fact that most Kepler stars were continuously observed for 3.5 yr. This simulator algorithm is more sophisticated than that used in ~\cite{izidoroetal17,izidoroetal21a} where detection rely on geometric transit only.  

In our N-body numerical simulations, the central star is always a solar type star ($R_{\star}=R_{\odot}$ and $\rho_{\star}=\rho_{\odot}$), and we do not need to make any assumption about the stellar photometric noise. However, to perform our synthetic transit observations we do. We randomly pick ${\rm CDPP_{\rm 6hr}}$ from direct measurements of the photometric noise for the CKS stellar sample \citep{weissetal18}.

For every observed synthetic planetary system, we randomly assign one star from the CKS sample and take its corresponding  ${\rm CDPP_{\rm 6hr}}$ to compute the SNR for each transiting planet. Finally, the transit duration is calculated as
\begin{equation}
    T= 13 {\rm hr} \left(\frac{P}{1 {\rm yr}}\right)^{1/3}\left(\frac{\rho_{\star}}{\rho_{\odot}}\right)^{-1/3}\sqrt{1 - b^2},
\end{equation}
where $b$ is the impact parameter.

\begin{figure*}
\centering
    \includegraphics[scale=.45]{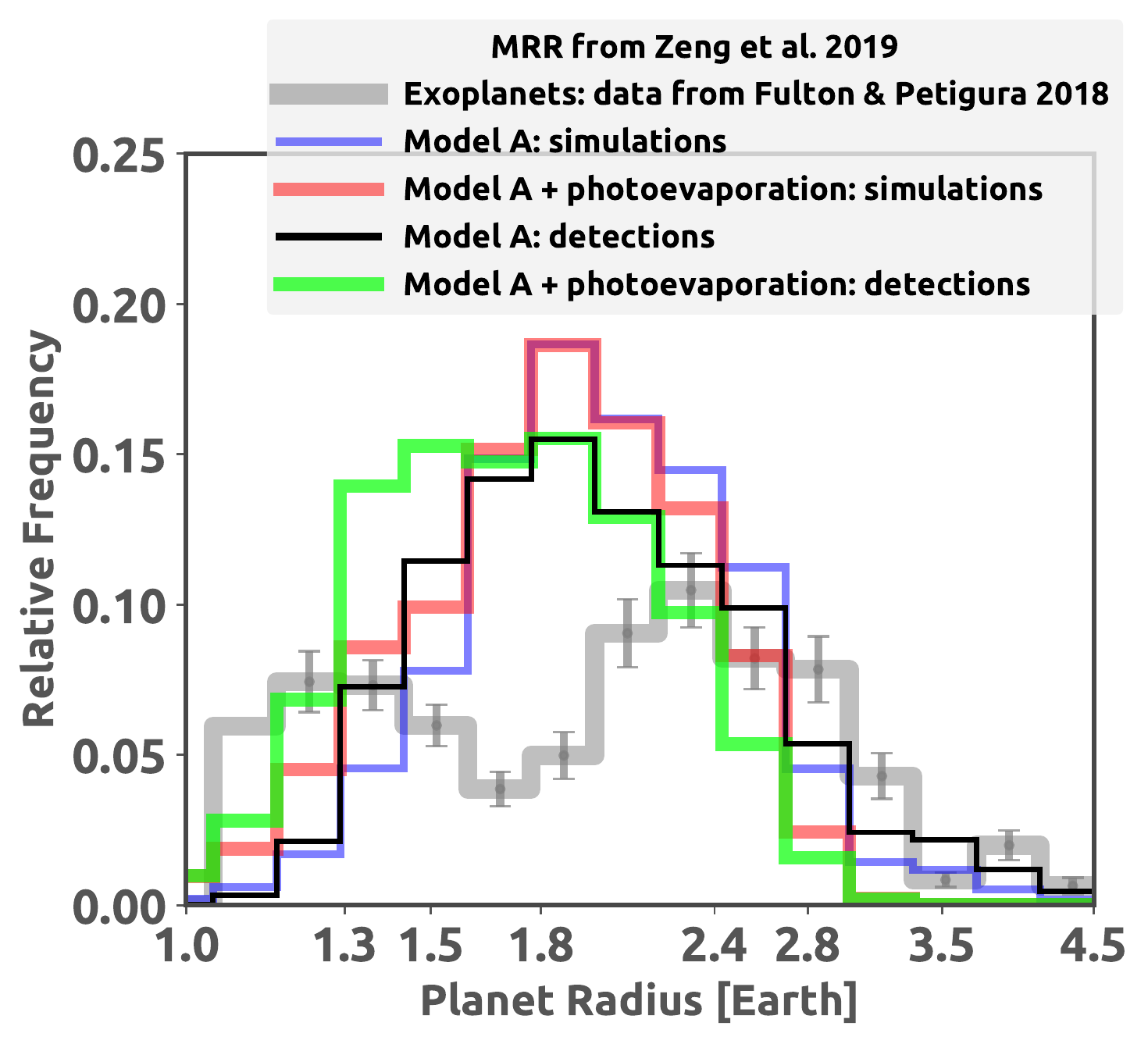}
        \includegraphics[scale=.45]{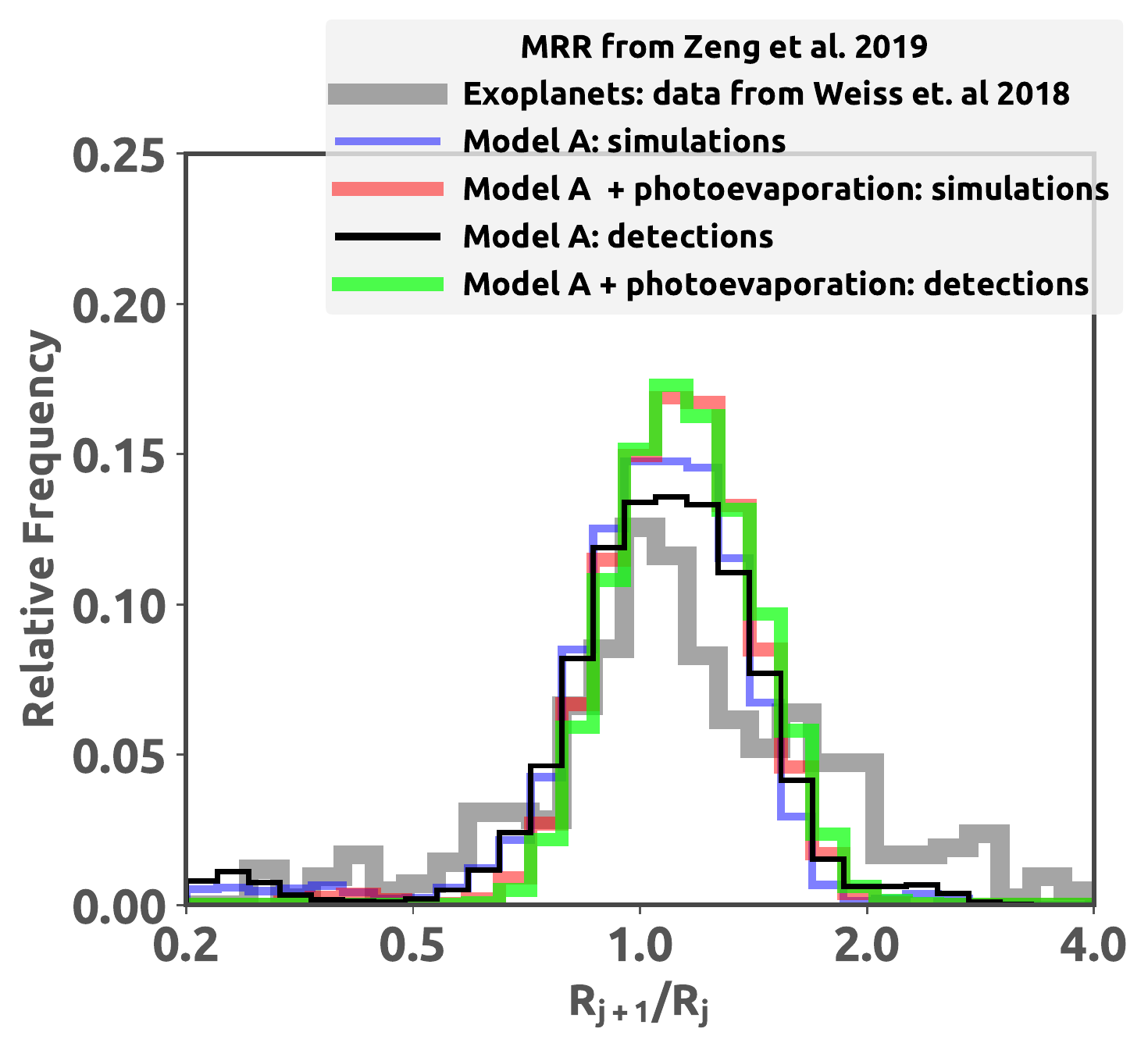}
    \includegraphics[scale=.45]{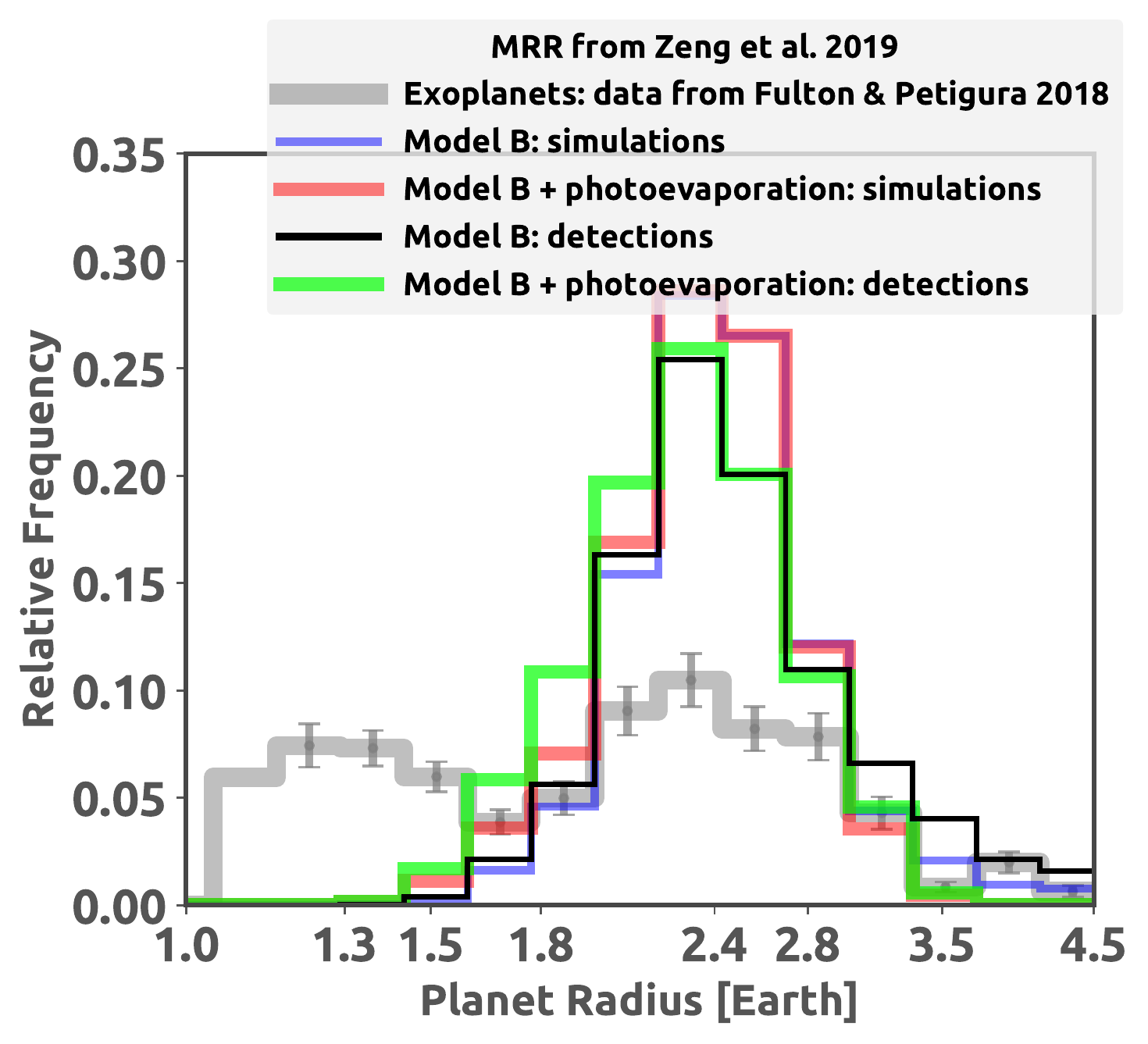}
        \includegraphics[scale=.45]{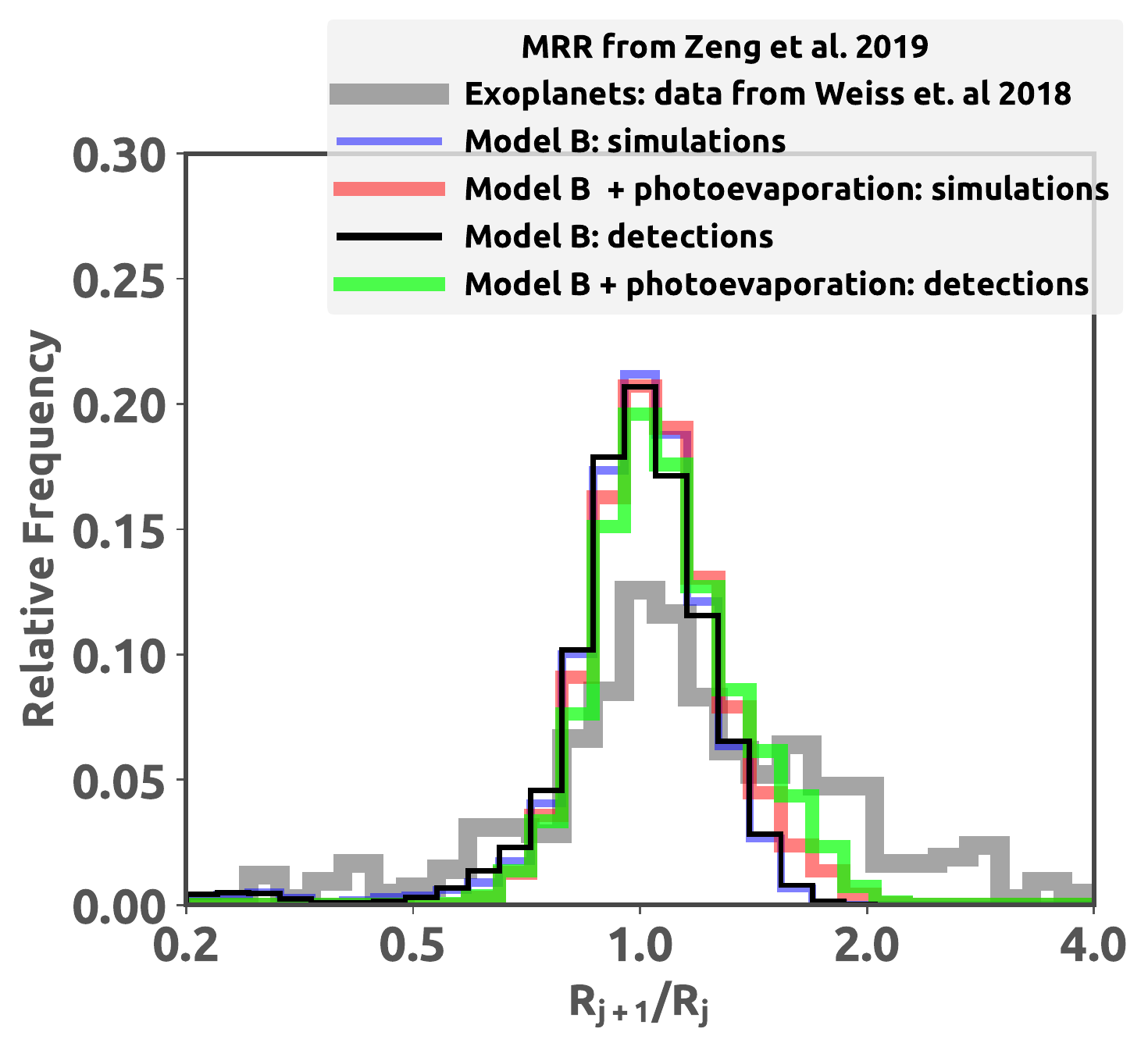}
     \includegraphics[scale=.45]{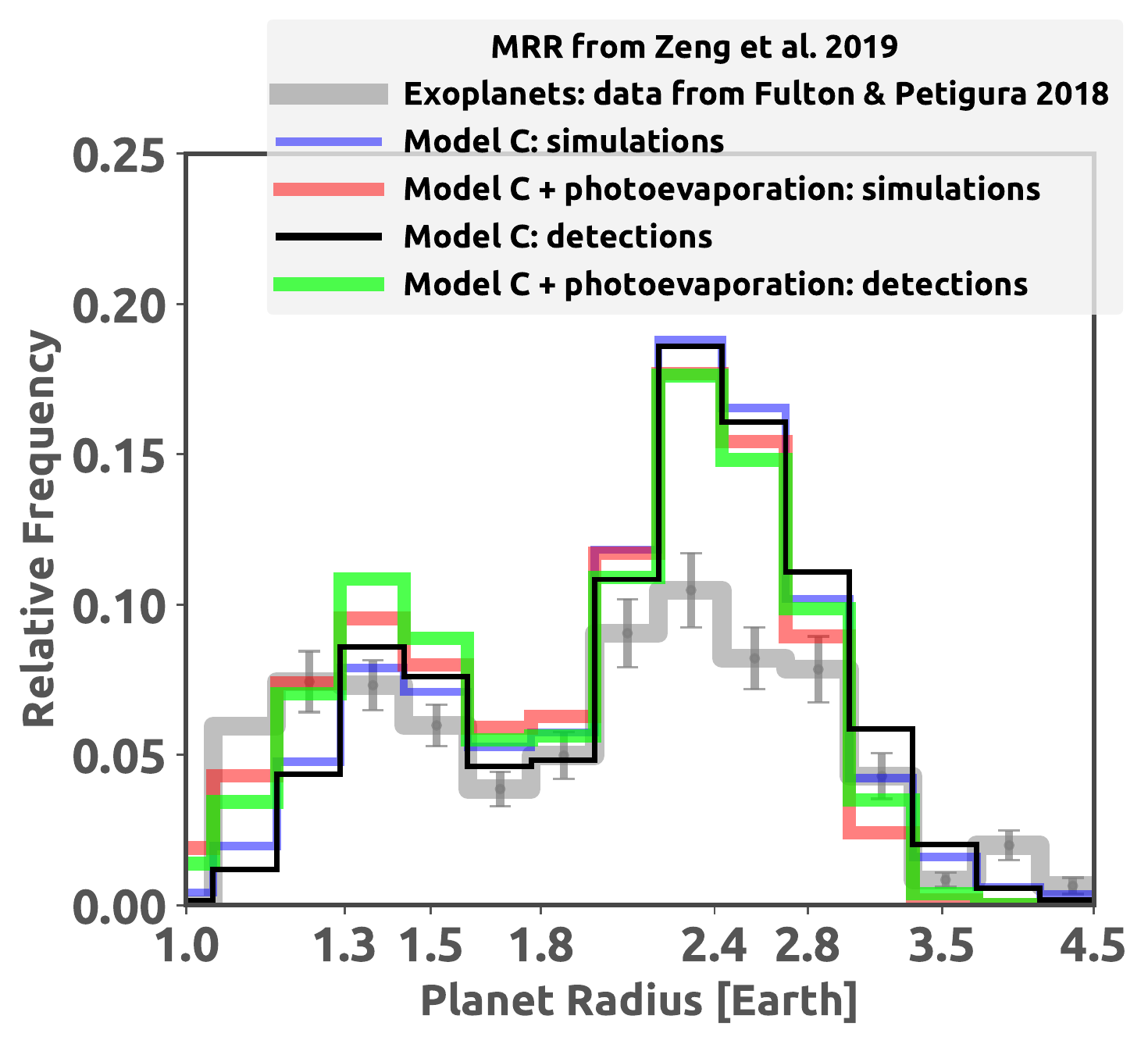}
        \includegraphics[scale=.45]{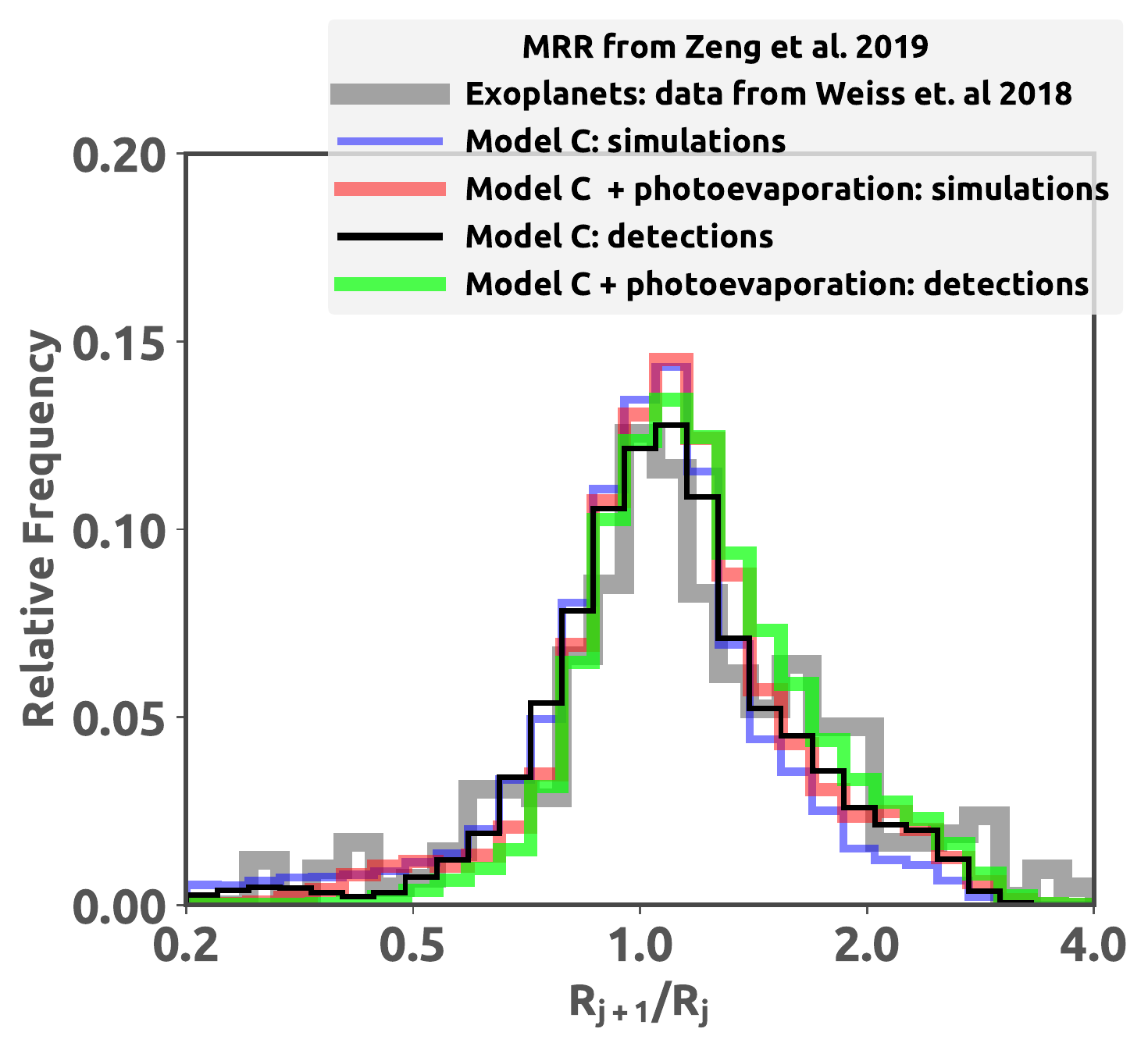} 
    \caption{{\bf Left:} Planet radius distribution. {\bf Right:} Planet size-ratio  distribution of adjacent planet pairs. Observations are shown in gray. Blue shows the outcome of our planet formation simulations. Red shows the outcome of our planet formation simulations modeling the effects of photo-evaporation. Black shows the synthetic transit observations of our simulations. Green shows the synthetic transit observations when we include the effects of photo-evaporation. From top-to-bottom, the panel-rows show model A,  B, and  C. For all models, we use the mass-radius relationship of~\cite{zengetal19} assuming an uncertainty of 7\% in size and an initial atmosphere-to-core mass ratio of 0.3\%.}
    \label{fig:distributions_with_observations}
\end{figure*}

Figure \ref{fig:distributions_with_observations} shows the planet radius distribution (left) and the planet size-ratio distribution for adjacent planets (right) of our simulations, synthetic detections, and exoplanets in the CKS sample \citep{fultonpetigura18,weissetal18}.  In order to construct these two histograms (blue and red),  we also account for uncertainties of 7\% in planet size (see Section \ref{sec:convert}).

As also shown in the main paper, model A and B fail to reproduce the bimodal distribution of planet radii, even when observational biases are taken into account. In model A, simulated detections that include the effect of photo-evaporation (green histogram) have a flat-top distribution of planetary radii across the planet valley region (between 1.3 and 1.8 $R_{\oplus}$) but over-predicts the relative frequency of these planets. In model B and C, we find that the inclusion of photo-evaporation has a small effect on the distribution of planet radii.

\begin{figure}
\centering
    \includegraphics[scale=.55]{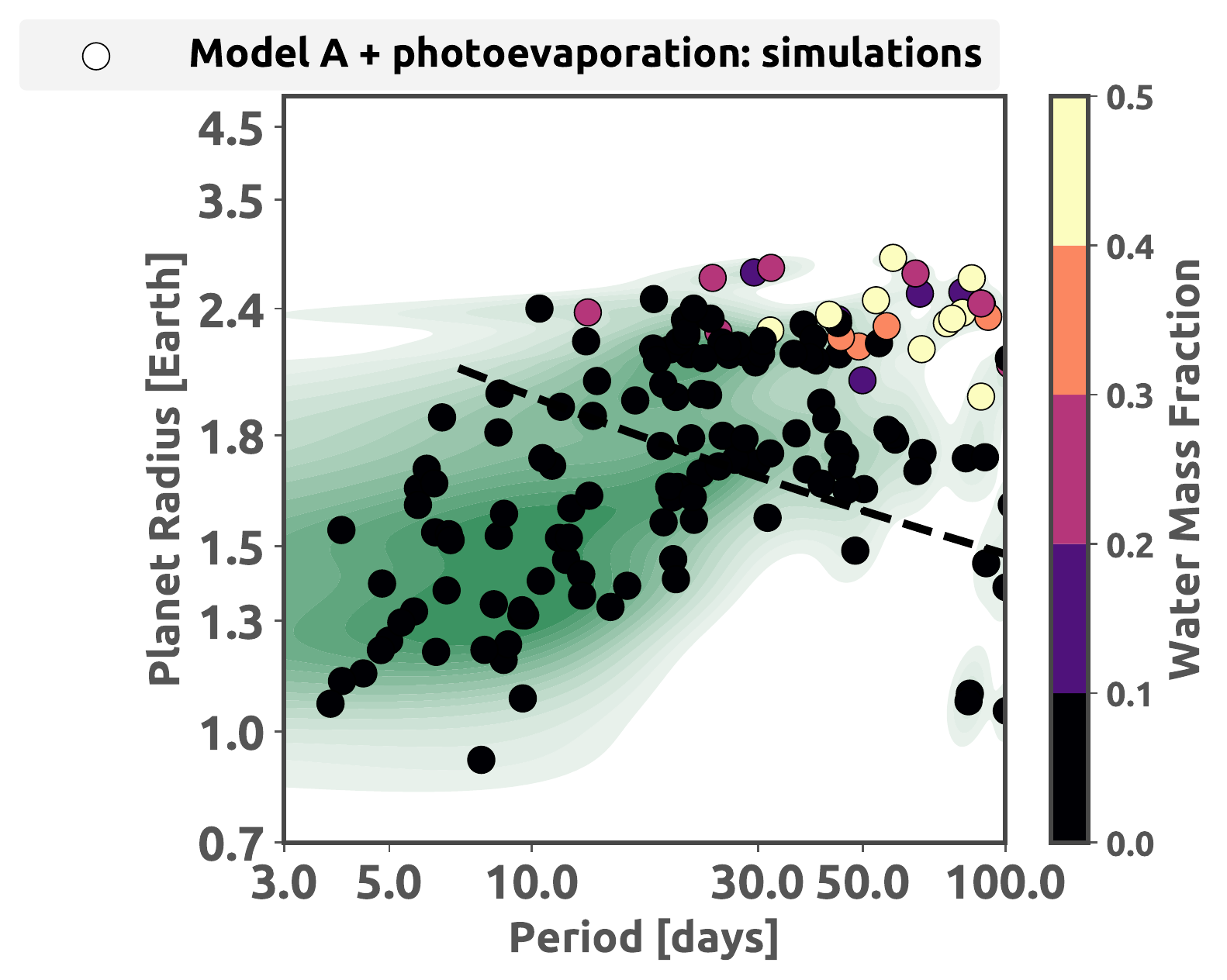}
        \includegraphics[scale=.55]{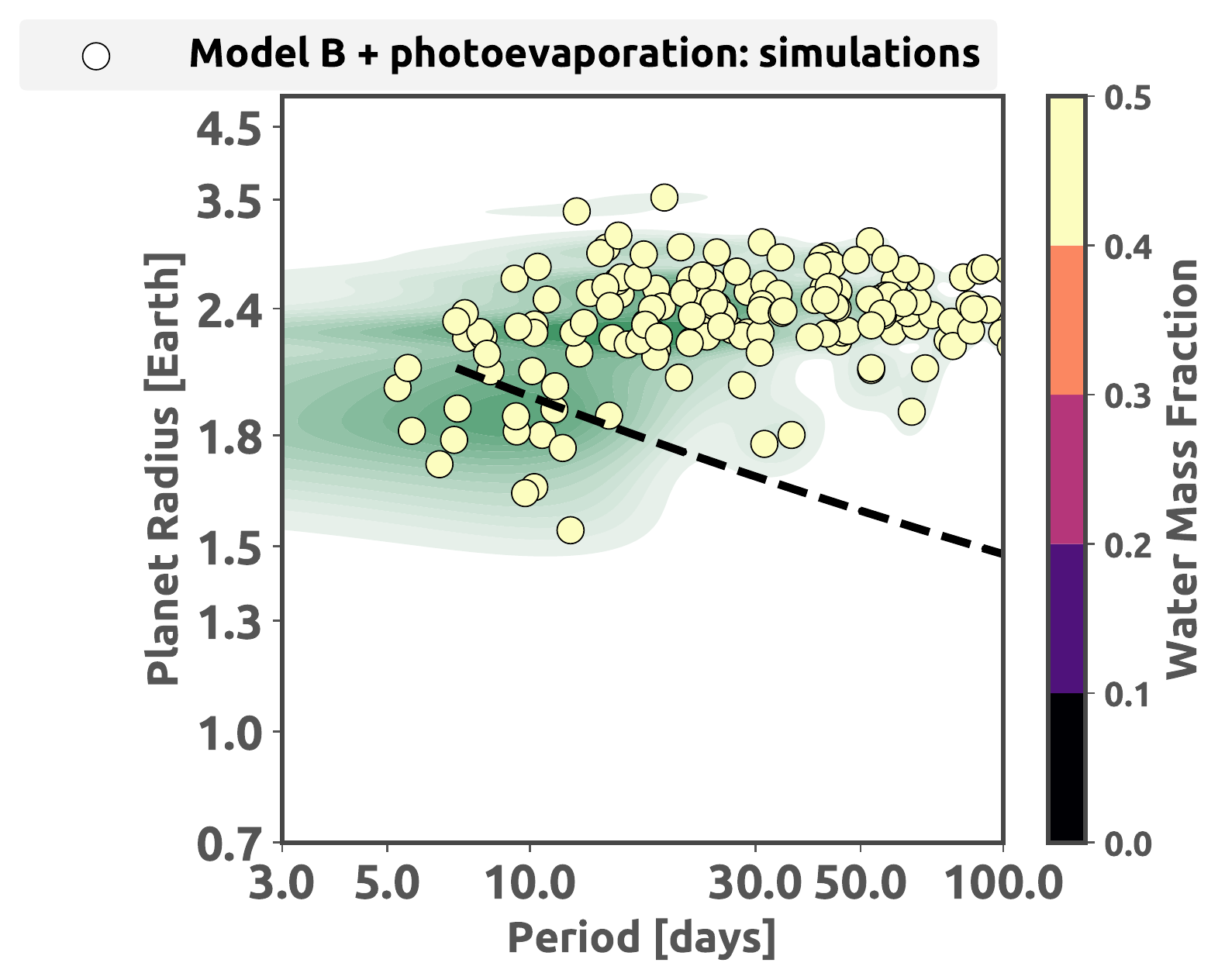}
        \includegraphics[scale=.55]{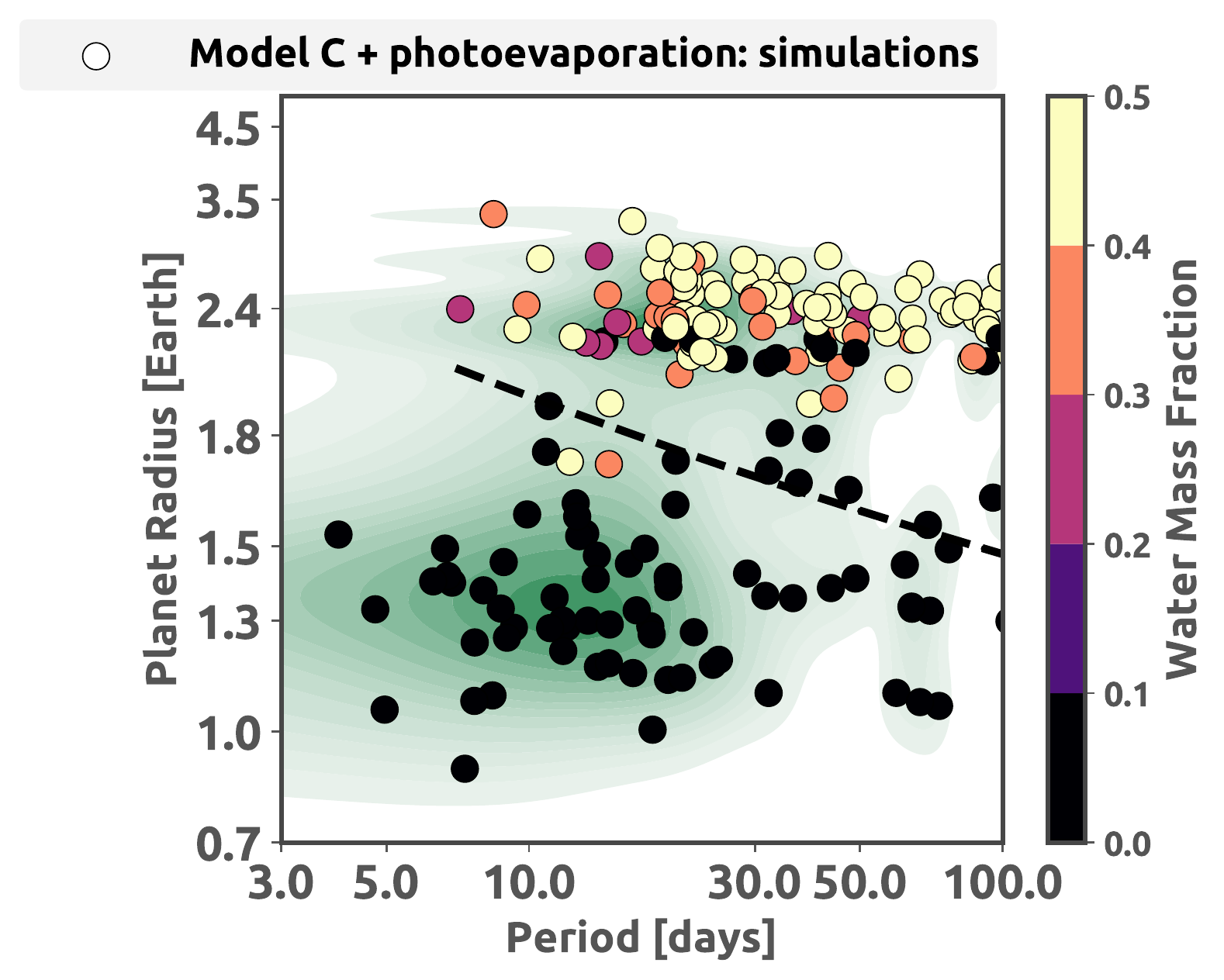} 
    \caption{Planetary architecture of our systems at the end of the simulations when including the effects of photo-evaporation. The assumed atmosphere-to-core mass fraction is 0.3\%. As in our nominal analysis, the mass-radius relationship used to convert masses to planet radii comes from ~\cite{zengetal19}.} 
        \label{fig:systemswithphoto}
\end{figure}

Figure \ref{fig:systemswithphoto} shows the final planetary architecture of our systems when we  include the effects of photoevporation. As one can see, the trends shown in Figure \ref{fig:after_instabilities} -- where we do not include the effects of photo-evaporation  -- do not change qualitatively.

\begin{figure}
\centering
    \includegraphics[scale=.45]{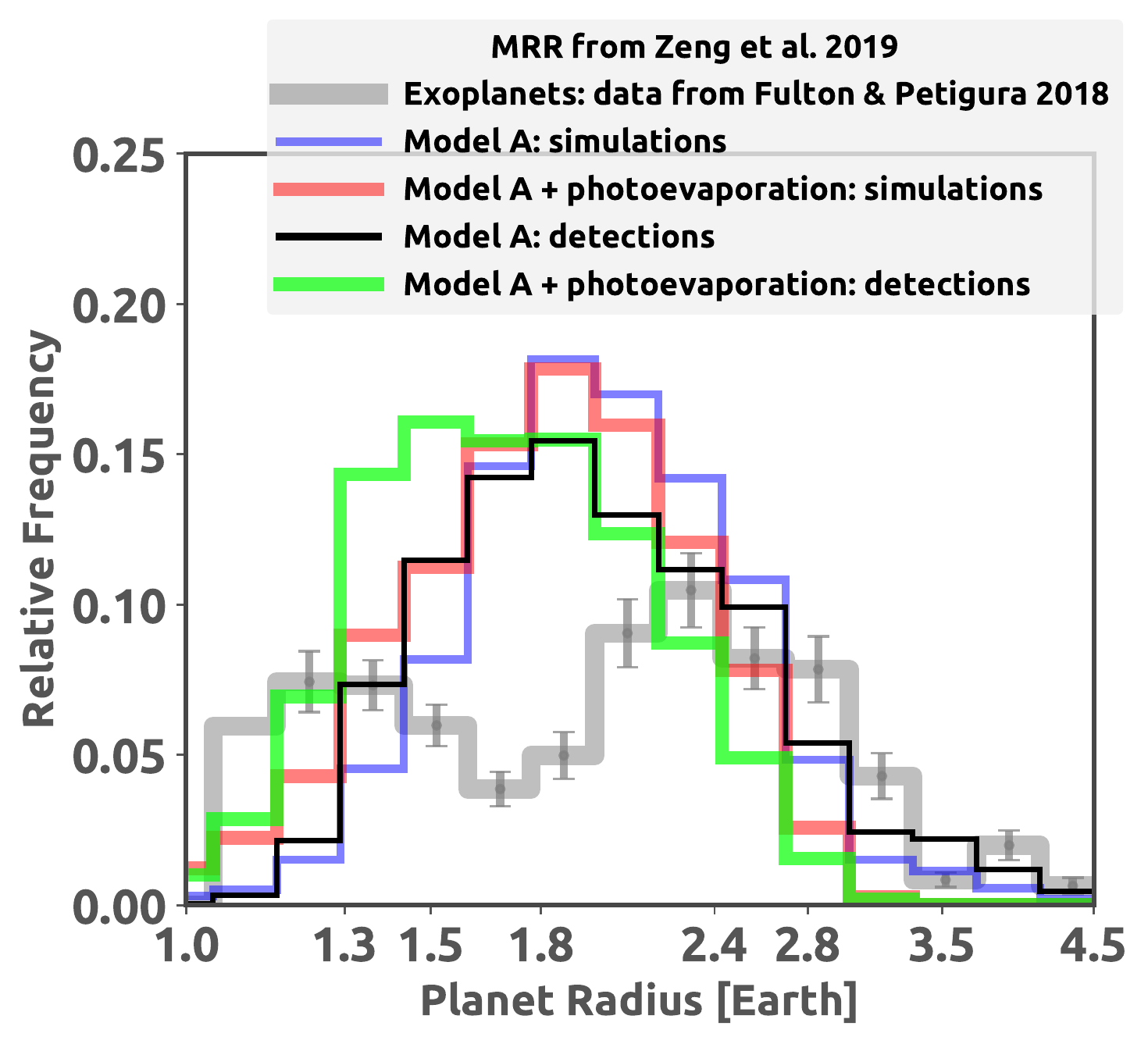}
        \includegraphics[scale=.45]{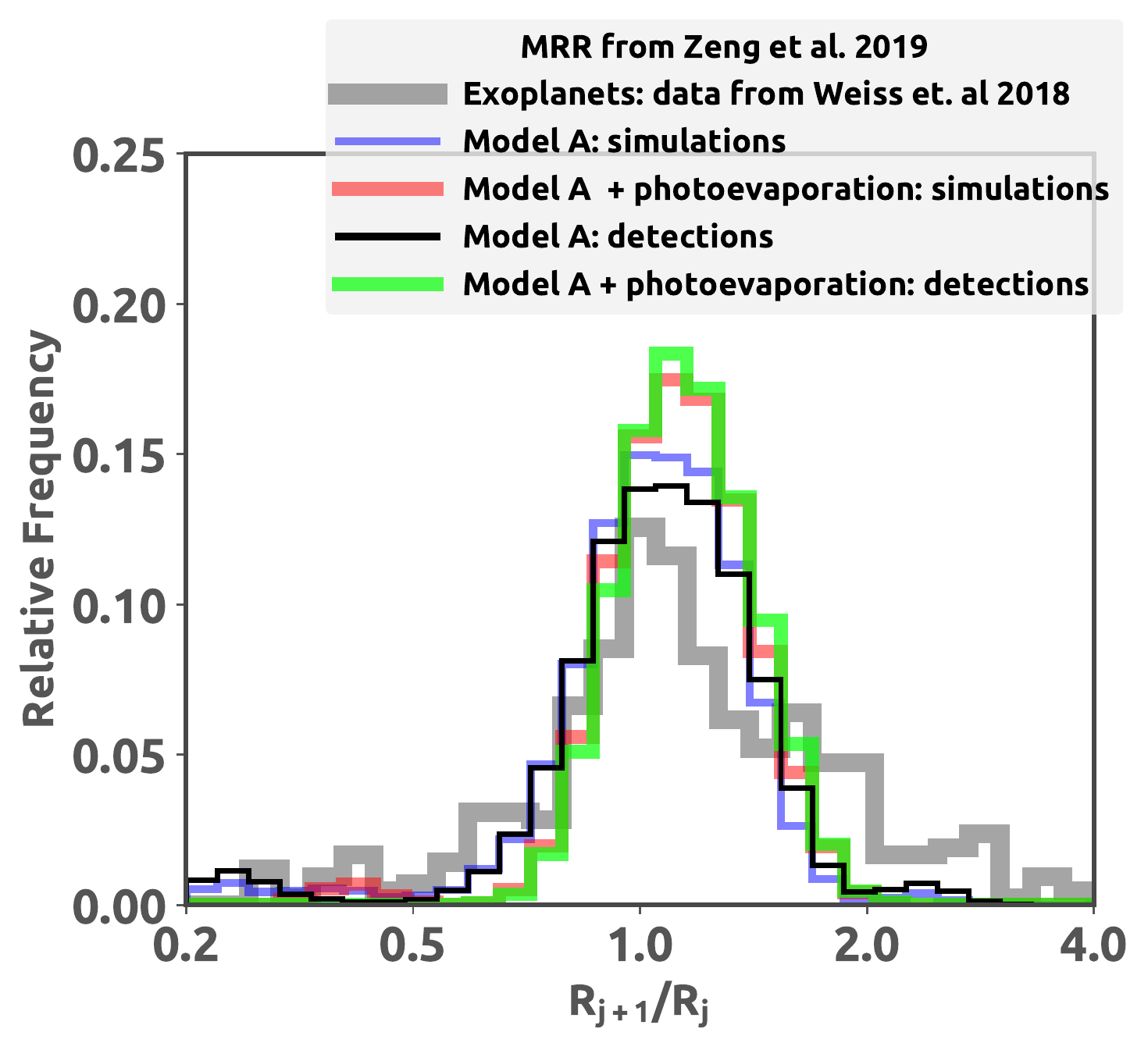}
    \includegraphics[scale=.45]{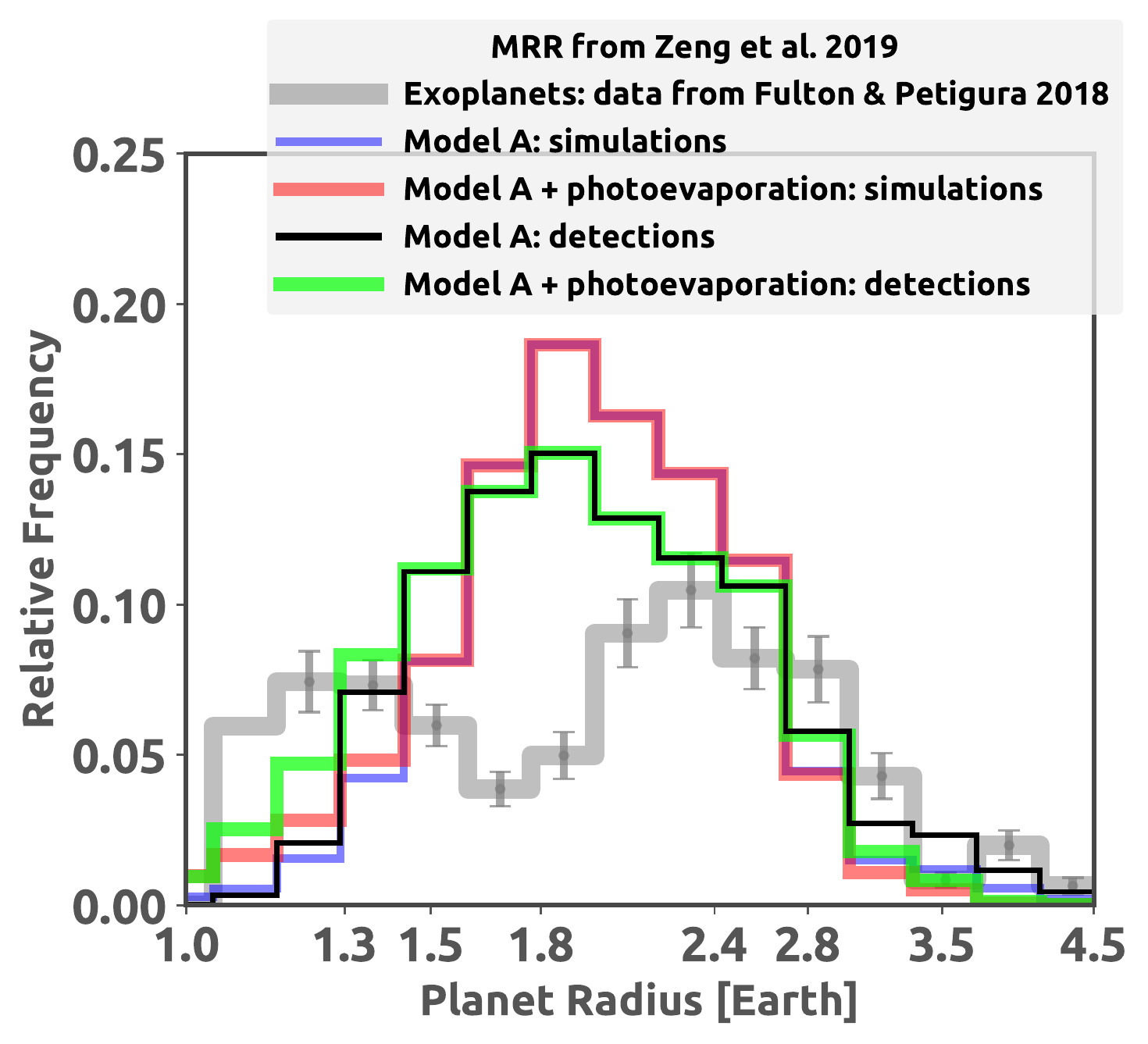}
        \includegraphics[scale=.45]{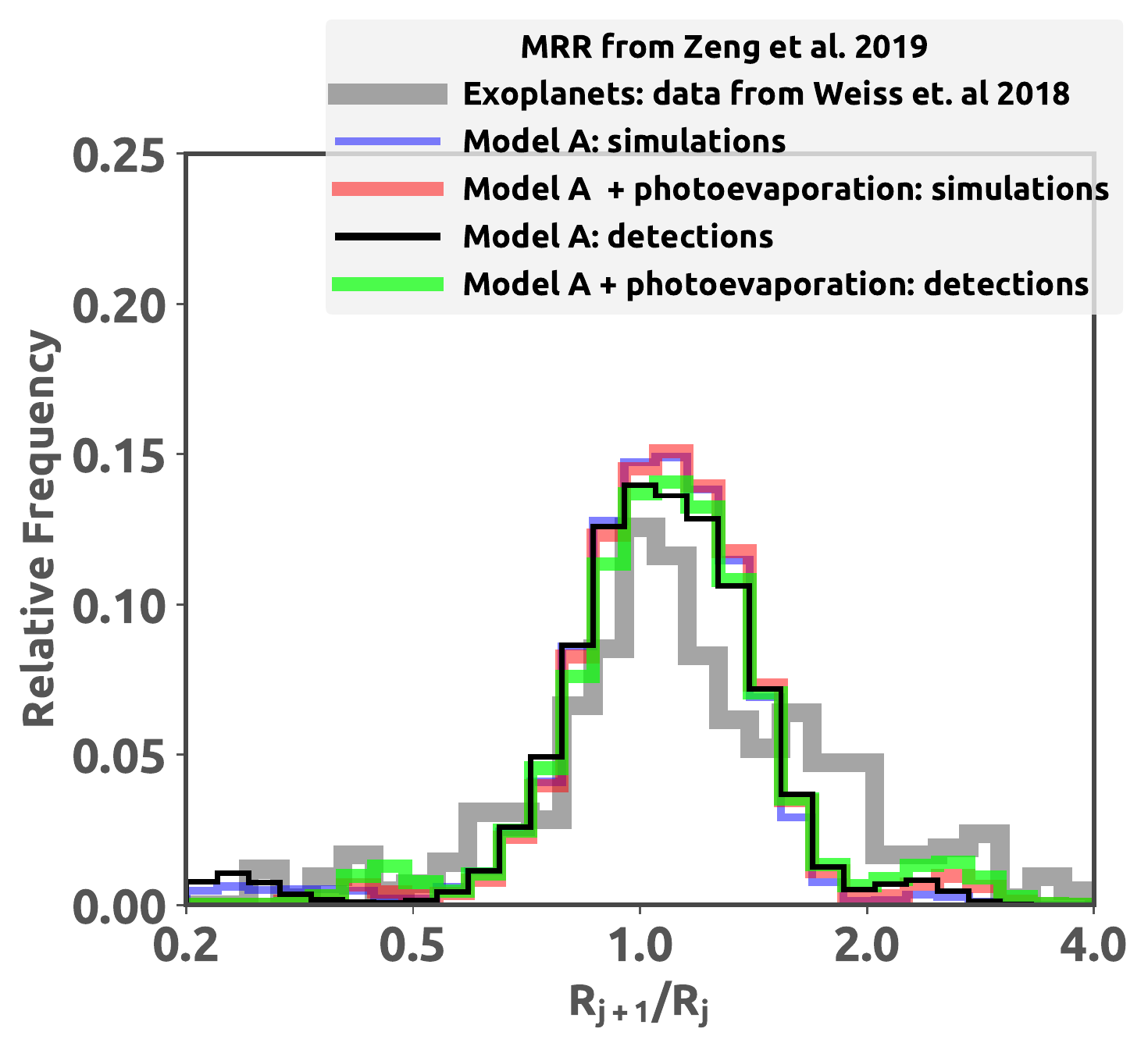}
    \includegraphics[scale=.45]{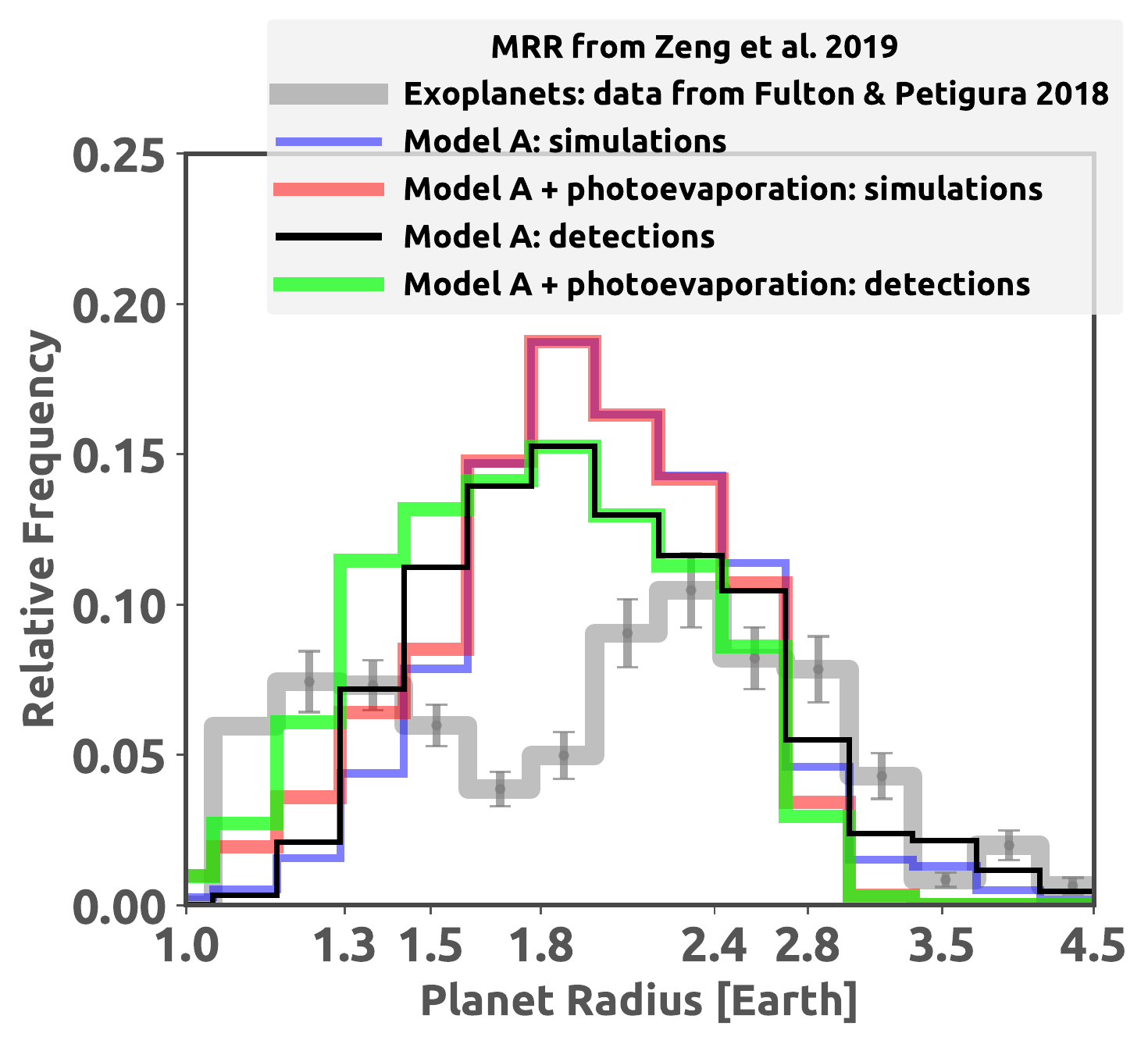}
        \includegraphics[scale=.45]{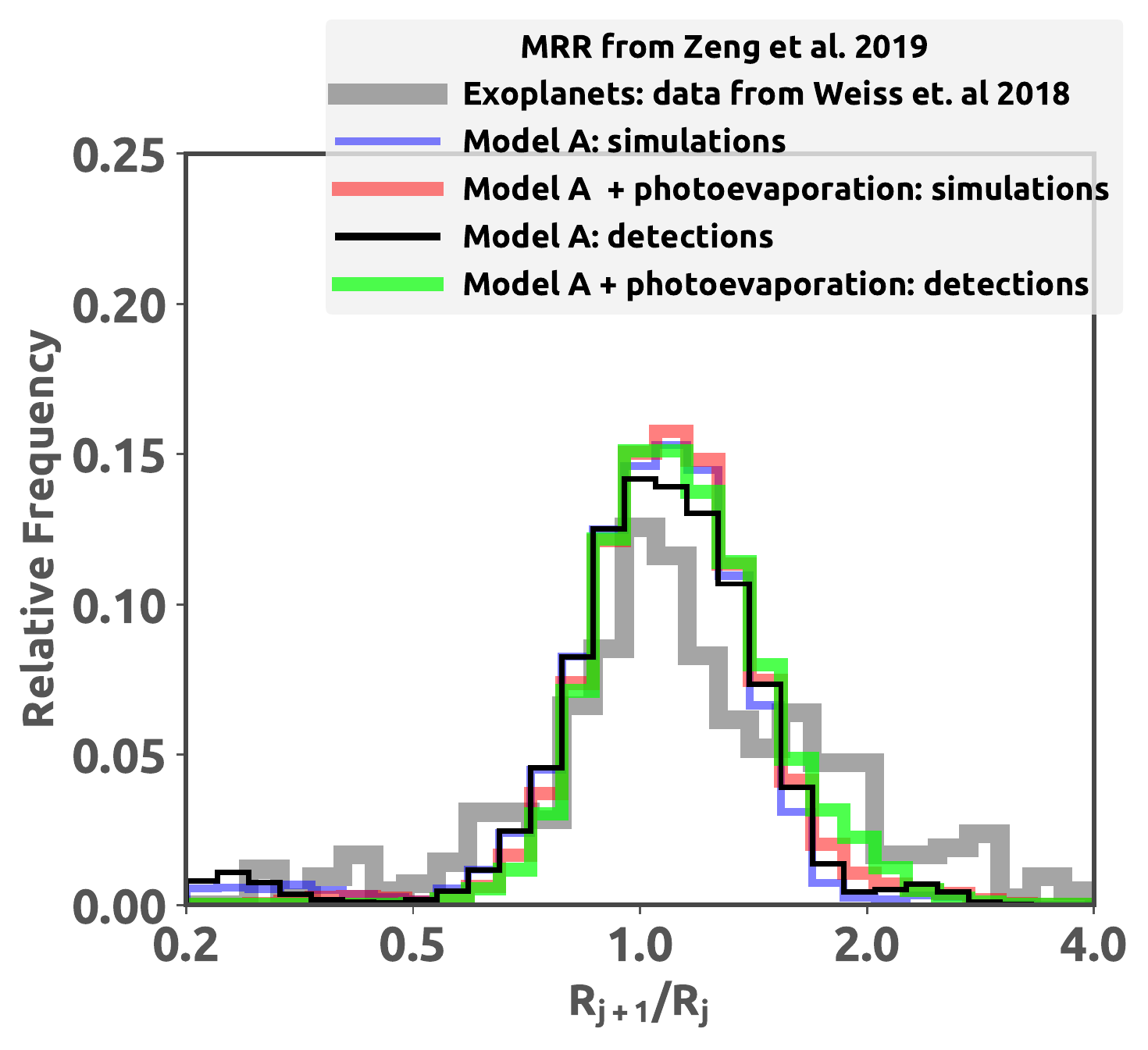}            
    \caption{Planet radius distribution (left) and planet size-ratio distribution (right) of model A assuming an initial atmospheric mass fractions of 0.1\% (top panels), 1\% (middle panels), and 5\% (bottom panels). Observations are shown in gray. Blue shows the outcome of our planet formation simulations. Red shows the outcome of our planet formation simulations modeling the effects of photo-evaporation. Black shows the synthetic transit observations of our simulations. Green shows the synthetic transit observations of our simulations including the effects of photo-evaporation.}
     \label{fig:ModelA_all}  
\end{figure}

\begin{figure}
\centering
    \includegraphics[scale=.45]{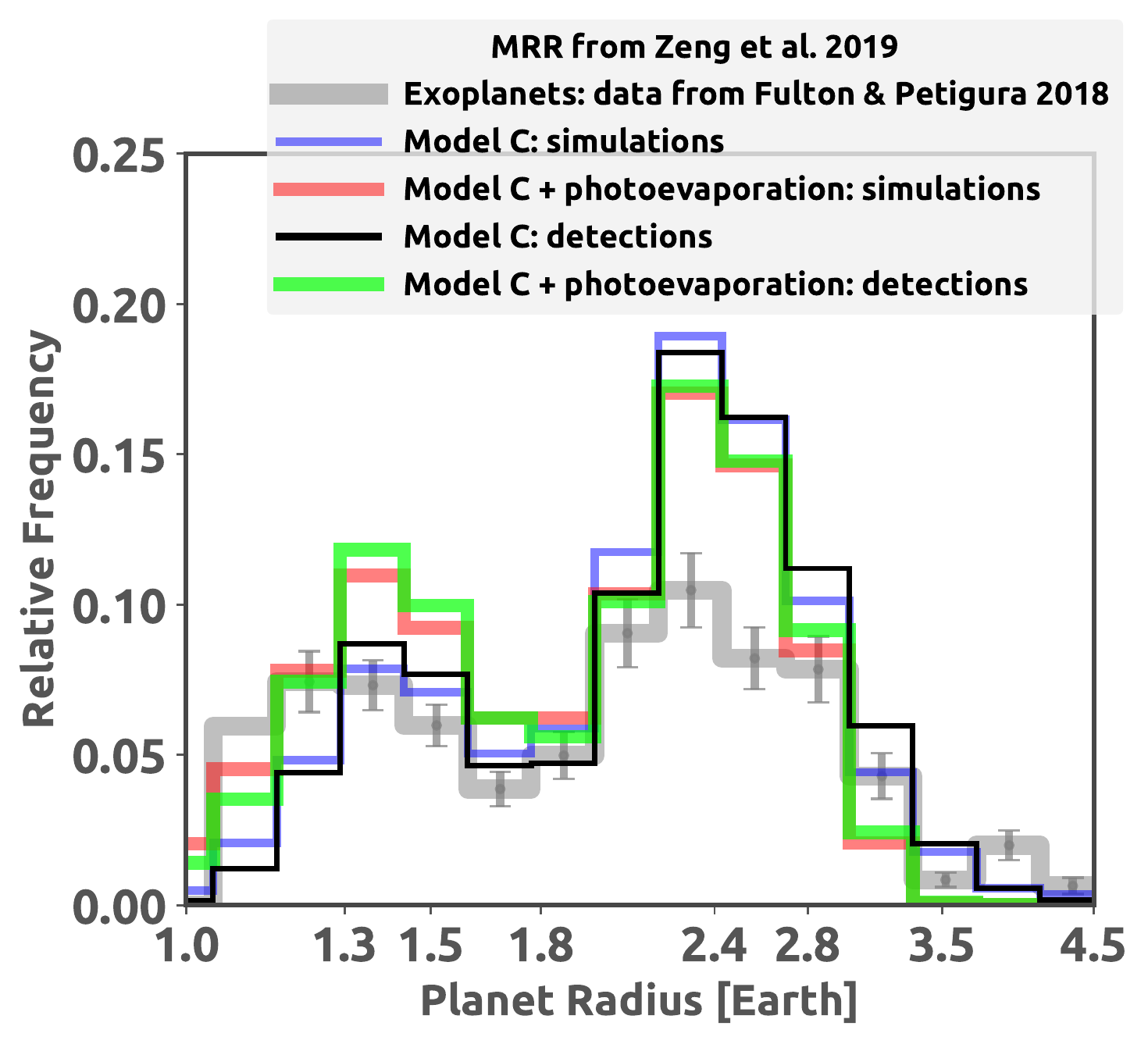}
        \includegraphics[scale=.45]{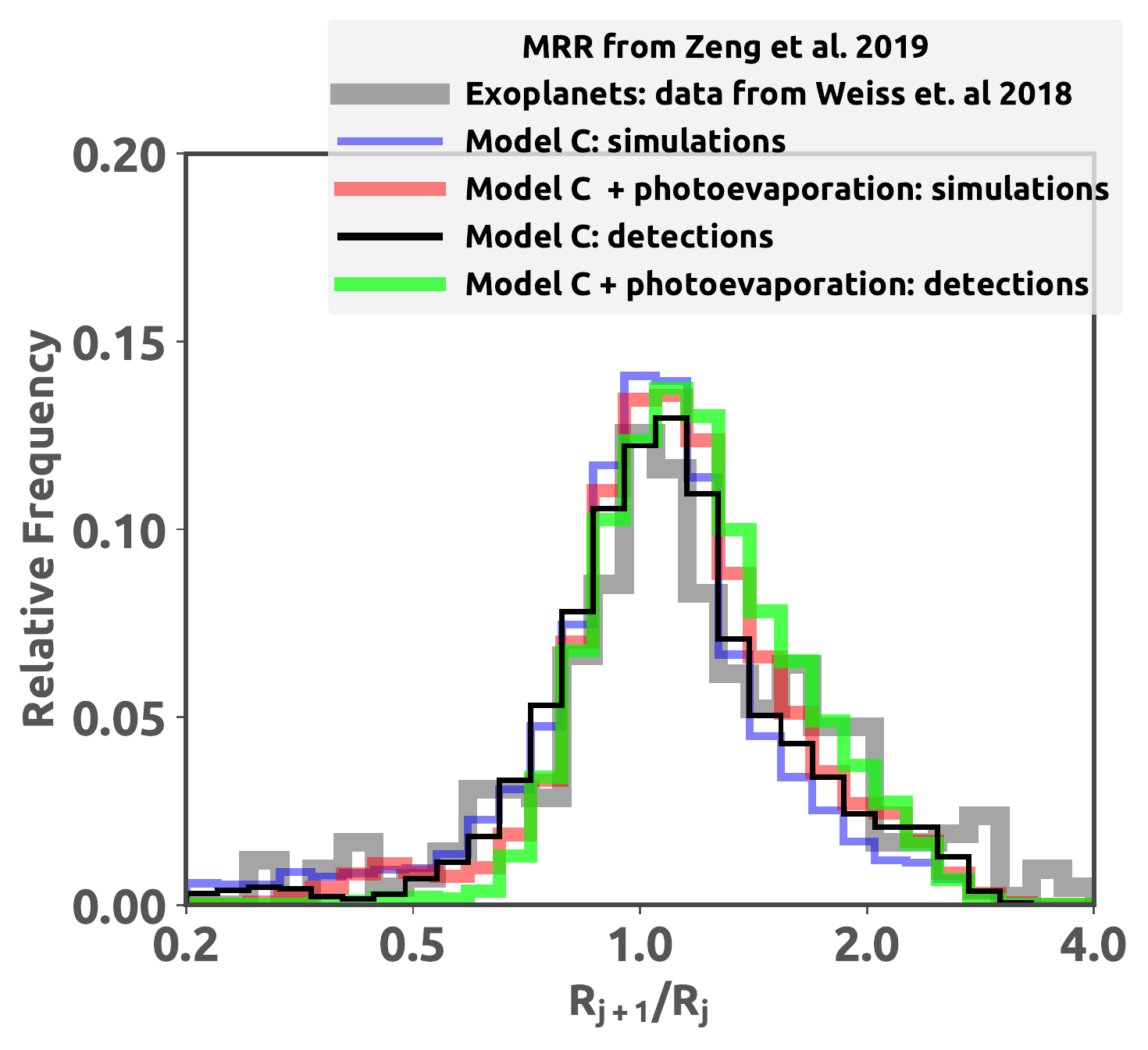}
    \includegraphics[scale=.45]{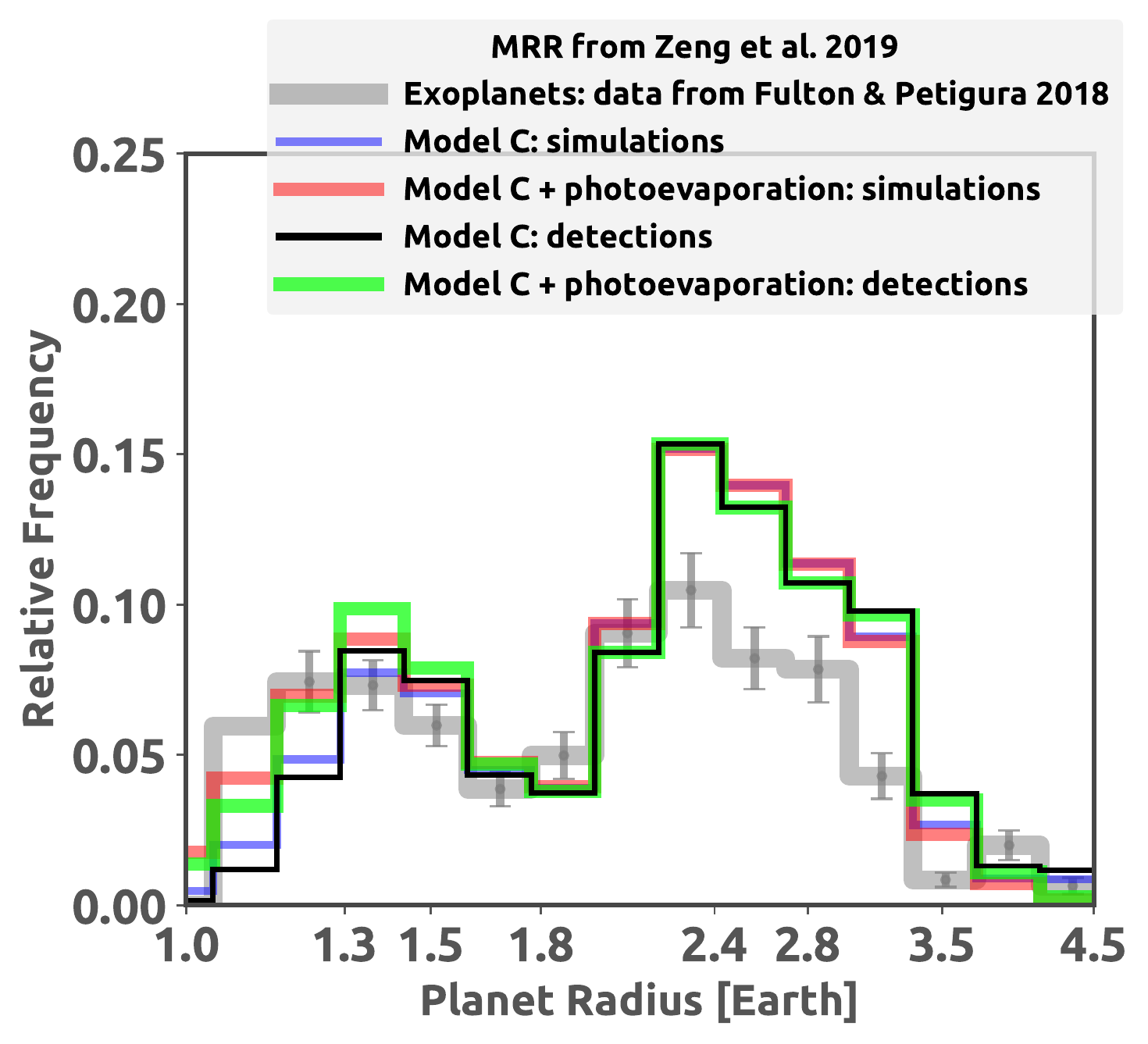}
        \includegraphics[scale=.45]{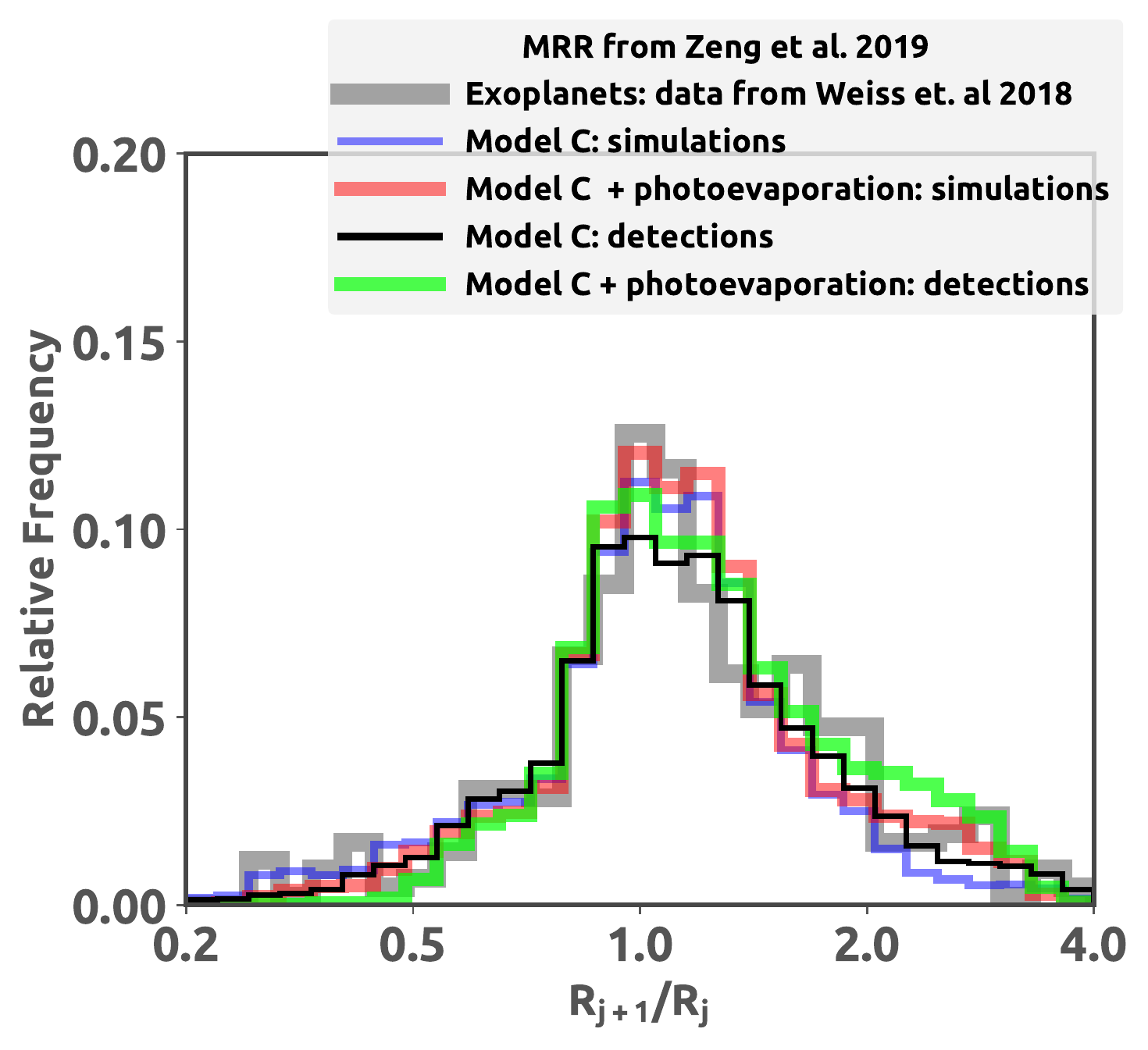}
    \includegraphics[scale=.45]{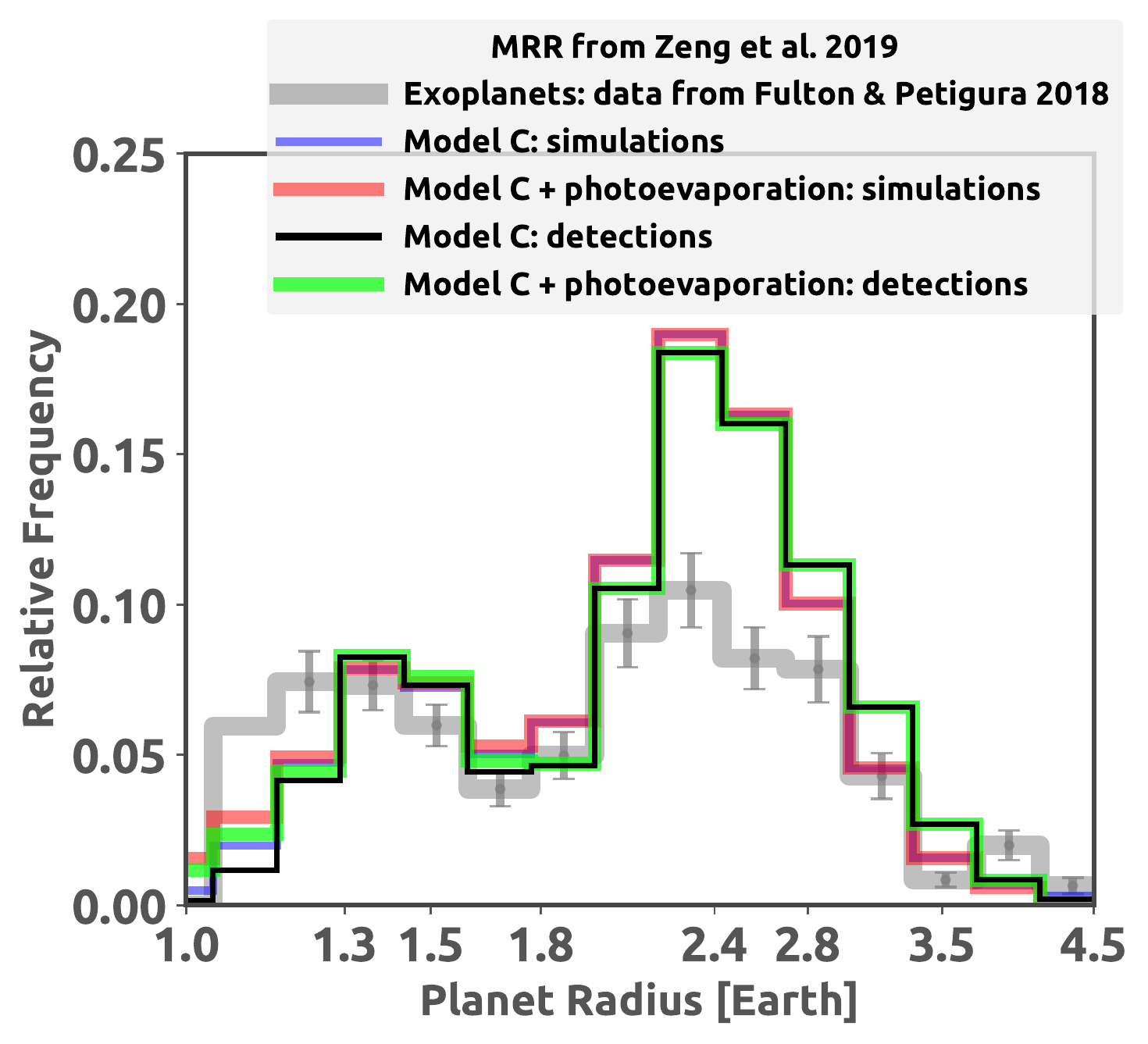}
        \includegraphics[scale=.45]{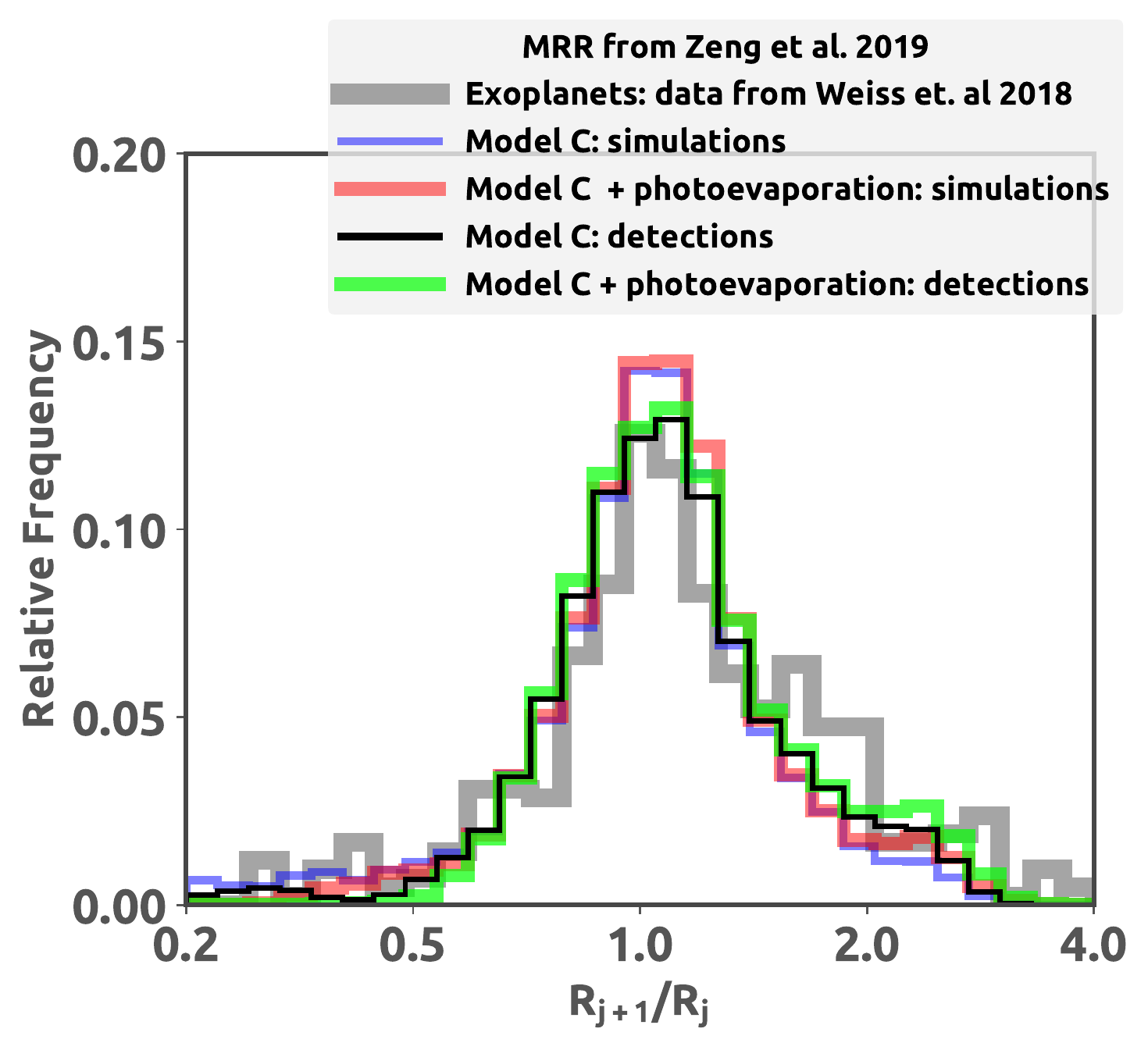}            
    \caption{Planet radius distribution (left) and planet size-ratio distribution (right) of model C assuming an initial atmospheric mass fractions of 0.1\% (top panels), 1\% (middle panels), and 5\% (bottom panels). Observations are shown in gray. Blue shows the outcome of our planet formation simulations. Red shows the outcome of our planet formation simulations modeling the effects of photo-evaporation. Black shows the synthetic transit observations of our simulations. Green shows the synthetic transit observations of our simulations including the effects of photo-evaporation.}
     \label{fig:ModelC_all}   
\end{figure}

Figure \ref{fig:ModelA_all} and \ref{fig:ModelC_all} show the planet radius distribution and planet size-ratio distribution of adjacent planets when we assume that primordial atmospheres correspond to 0.1\%, 1\% and 5\% of the planet/core mass. The radius valley and peas-in-a-pod feature are matched well for model C but not for model A, regardless of the assumed atmosphere mass.

\begin{figure}
\centering
    \includegraphics[scale=.6]{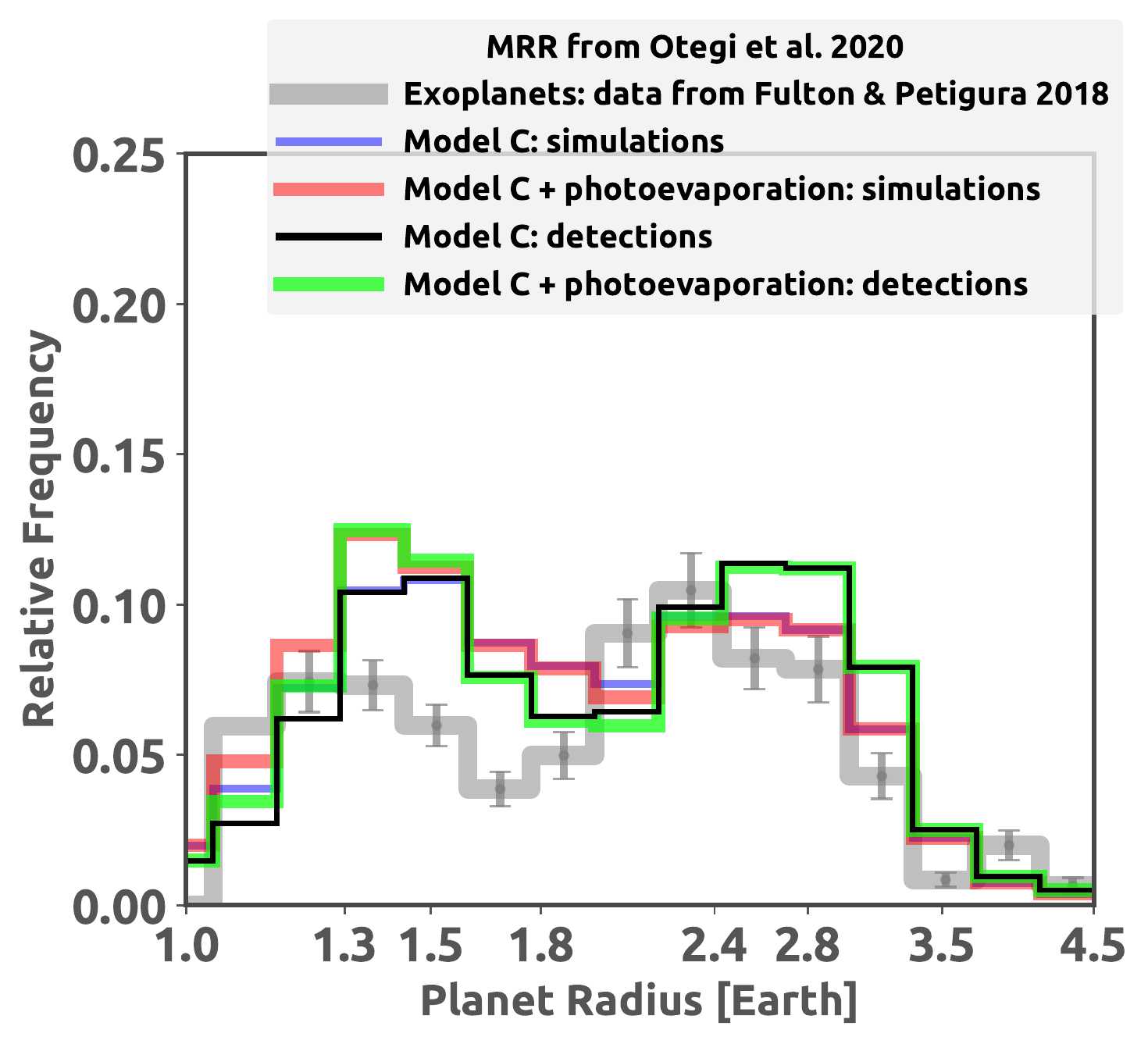}
        \includegraphics[scale=.6]{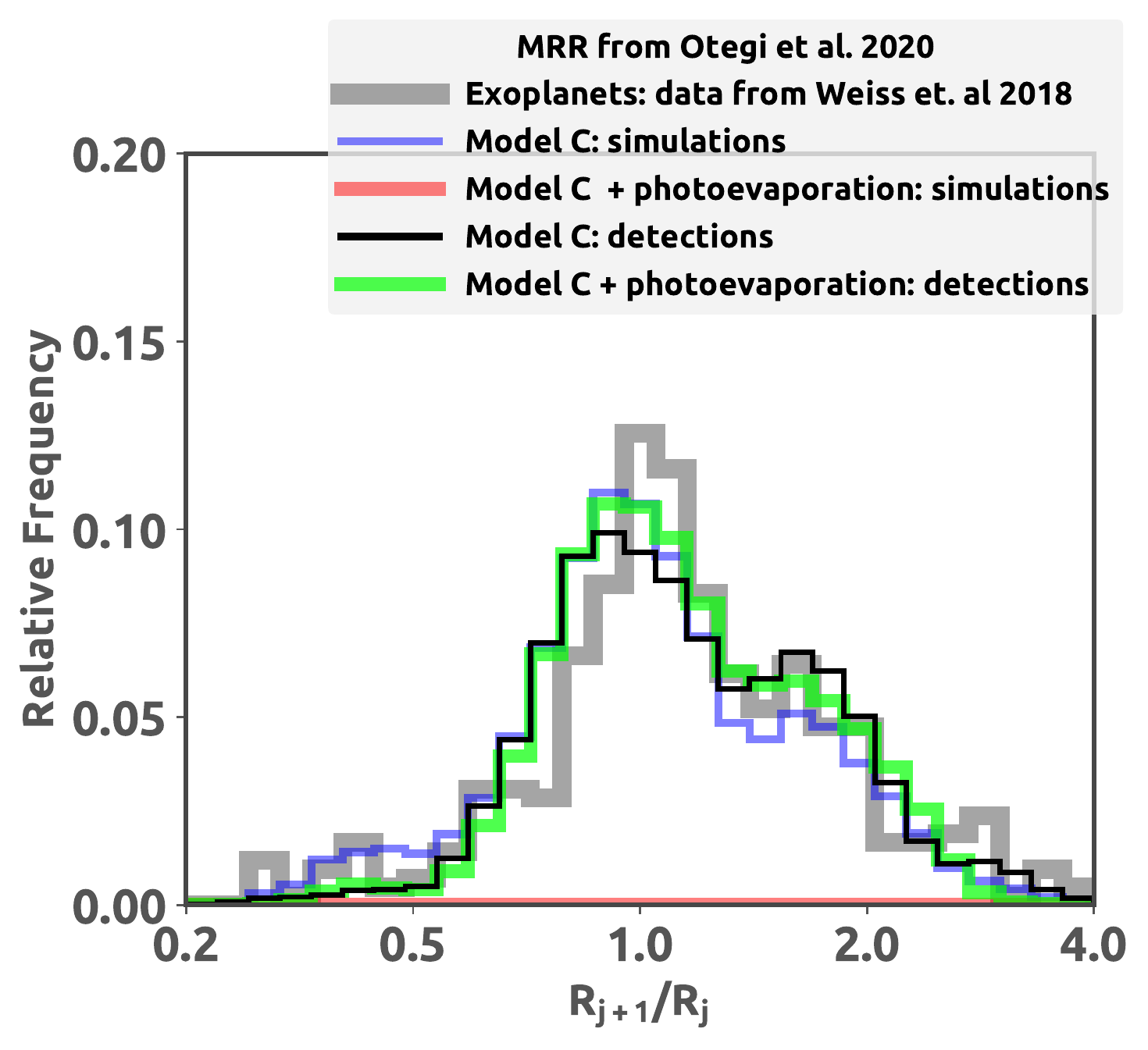}
    \caption{Planet radius distribution (left) and planet size-ratio distribution (right) of model C assuming an initial atmospheric mass fraction of 0.3\%. Planet radii are computed using the empirical mass radius relationship of ~\cite{otegietal20}. We use the high density mass-radius relationship equation of \cite{otegietal20} to compute the sizes of rocky planets that experienced late impacts, and the low-density equation for planets with water-rich compositions or rocky-cores with atmospheres (avoided impacts).  Observations are shown in gray. Blue shows the outcome of our planet formation simulations. Red shows the outcome of our planet formation simulations modeling the effects of photo-evaporation. Black shows the synthetic transit observations of our simulations. Green shows the synthetic transit observations of our simulations including the effects of photo-evaporation.}
    \label{fig:otegi}    
\end{figure}

Figure \ref{fig:otegi} shows the planet radius distribution and planet size-ratio distribution of adjacent planets when we assume the empirical mass radius relationship of \cite{otegietal20}. We assume that giant impacts make ``high density''  rocky planets  and we used the mass radius relationship for high density planets from ~\cite{otegietal20}. We use the low density mass-radius relationship to compute the radius of water-rich planets or water-rich planets with primordial atmospheres regardless of the occurrence of late giant impacts (compare with bottom panel of Figure \ref{fig:distributions} and Figure \ref{fig:ModelC_all}). Compared to the results in the main paper, these two mass-radius determinations give very similar results in broadly matching the radius valley.

\begin{figure}
\centering
	\includegraphics[scale=.5]{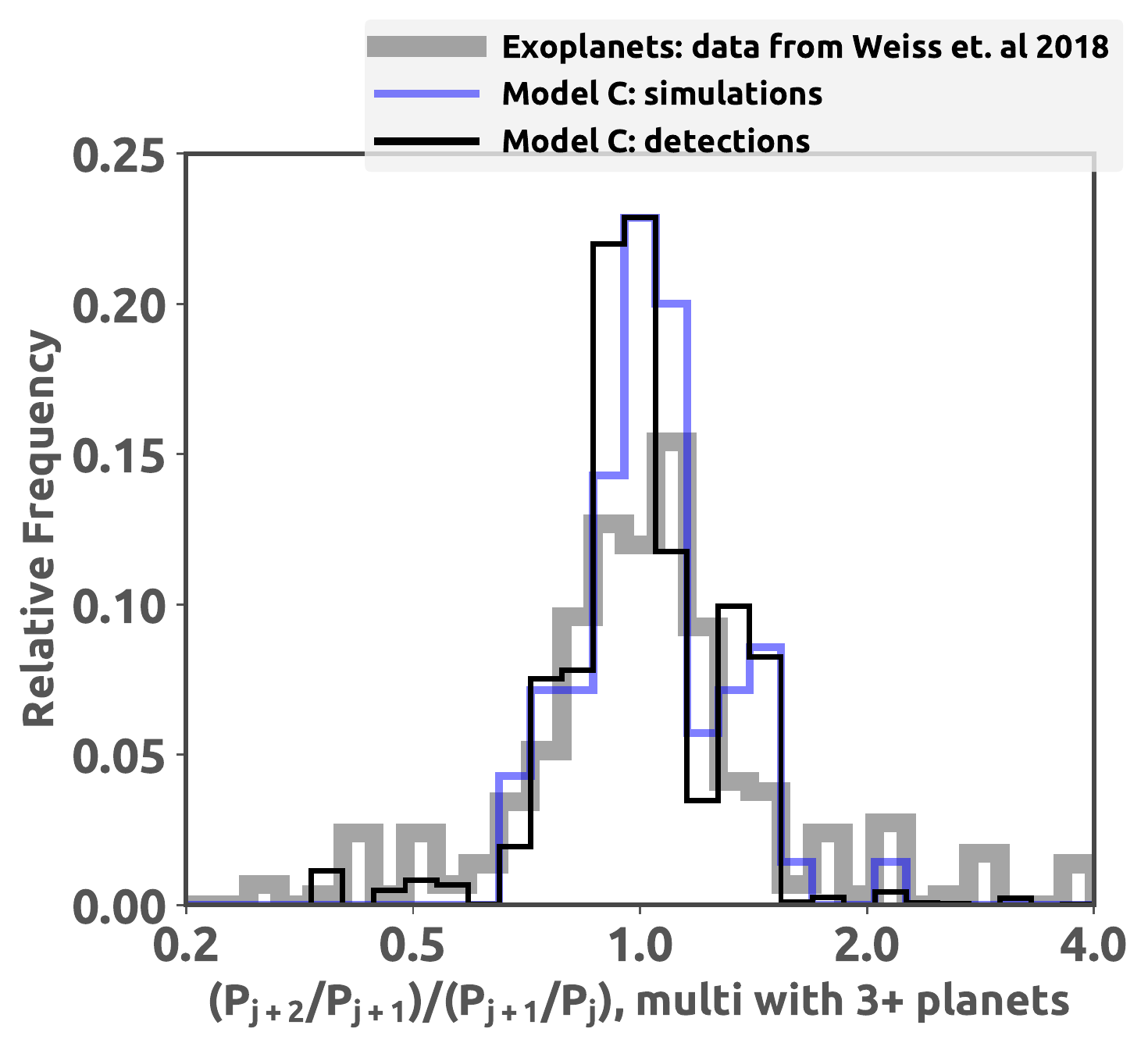}
 \caption{Ratio of orbital period ratios for adjacent triple of planets in the same planetary systems. Gray shows exoplanet data from \cite{weissetal18}. Blue and black show the results of our model C.}
 \label{fig:periodpeas}
\end{figure}

 Figure \ref{fig:periodpeas} shows the ratio of orbital period ratios of  adjacent triple of planets in a same planetary system. This figure shows that our model C is also broadly consistent with the regular orbital spacing of observed adjacent planet pairs~\citep{weissetal18}.

\bibliography{sample631}{}

\begin{thebibliography}{}
\expandafter\ifx\csname natexlab\endcsname\relax\def\natexlab#1{#1}\fi
\providecommand{\url}[1]{\href{#1}{#1}}
\providecommand{\dodoi}[1]{doi:~\href{http://doi.org/#1}{\nolinkurl{#1}}}
\providecommand{\doeprint}[1]{\href{http://ascl.net/#1}{\nolinkurl{http://ascl.net/#1}}}
\providecommand{\doarXiv}[1]{\href{https://arxiv.org/abs/#1}{\nolinkurl{https://arxiv.org/abs/#1}}}

\bibitem[{{Adams} {et~al.}(2008){Adams}, {Seager}, \&
  {Elkins-Tanton}}]{adamsseageretal08}
{Adams}, E.~R., {Seager}, S., \& {Elkins-Tanton}, L. 2008, \apj, 673, 1160,
  \dodoi{10.1086/524925}

\bibitem[{{Baruteau} {et~al.}(2014){Baruteau}, {Crida}, {Paardekooper},
  {Masset}, {Guilet}, {Bitsch}, {Nelson}, {Kley}, \&
  {Papaloizou}}]{baruteauetal14}
{Baruteau}, C., {Crida}, A., {Paardekooper}, S.-J., {et~al.} 2014, Protostars
  and Planets VI, 667, \dodoi{10.2458/azu_uapress_9780816531240-ch029}

\bibitem[{{Bashi} {et~al.}(2017){Bashi}, {Helled}, {Zucker}, \&
  {Mordasini}}]{bashietal17}
{Bashi}, D., {Helled}, R., {Zucker}, S., \& {Mordasini}, C. 2017, \aap, 604,
  A83, \dodoi{10.1051/0004-6361/201629922}

\bibitem[{{Batalha} {et~al.}(2013){Batalha}, {Rowe}, {Bryson}, {Barclay},
  {Burke}, {Caldwell}, {Christiansen}, {Mullally}, {Thompson}, {Brown},
  {Dupree}, {Fabrycky}, {Ford}, {Fortney}, {Gilliland}, {Isaacson}, {Latham},
  {Marcy}, {Quinn}, {Ragozzine}, {Shporer}, {Borucki}, {Ciardi}, {Gautier},
  {Haas}, {Jenkins}, {Koch}, {Lissauer}, {Rapin}, {Basri}, {Boss}, {Buchhave},
  {Carter}, {Charbonneau}, {Christensen-Dalsgaard}, {Clarke}, {Cochran},
  {Demory}, {Desert}, {Devore}, {Doyle}, {Esquerdo}, {Everett}, {Fressin},
  {Geary}, {Girouard}, {Gould}, {Hall}, {Holman}, {Howard}, {Howell},
  {Ibrahim}, {Kinemuchi}, {Kjeldsen}, {Klaus}, {Li}, {Lucas}, {Meibom},
  {Morris}, {Pr{\v s}a}, {Quintana}, {Sanderfer}, {Sasselov}, {Seader},
  {Smith}, {Steffen}, {Still}, {Stumpe}, {Tarter}, {Tenenbaum}, {Torres},
  {Twicken}, {Uddin}, {Van Cleve}, {Walkowicz}, \& {Welsh}}]{batalhaetal13}
{Batalha}, N.~M., {Rowe}, J.~F., {Bryson}, S.~T., {et~al.} 2013, \apjs, 204,
  24, \dodoi{10.1088/0067-0049/204/2/24}

\bibitem[{{Bean} {et~al.}(2021){Bean}, {Raymond}, \& {Owen}}]{beanetal21}
{Bean}, J.~L., {Raymond}, S.~N., \& {Owen}, J.~E. 2021, Journal of Geophysical
  Research (Planets), 126, e06639, \dodoi{10.1029/2020JE006639}

\bibitem[{{B{\'e}thune} \& {Rafikov}(2019)}]{bethunerafikov19}
{B{\'e}thune}, W., \& {Rafikov}, R.~R. 2019, \mnras, 488, 2365,
  \dodoi{10.1093/mnras/stz1870}

\bibitem[{{Biersteker} \& {Schlichting}(2019)}]{bierstekeretal19}
{Biersteker}, J.~B., \& {Schlichting}, H.~E. 2019, \mnras, 485, 4454,
  \dodoi{10.1093/mnras/stz738}

\bibitem[{{Biersteker} \& {Schlichting}(2021)}]{bierstekeretal20}
---. 2021, \mnras, 501, 587, \dodoi{10.1093/mnras/staa3614}

\bibitem[{{Bitsch}(2019)}]{bitsch19}
{Bitsch}, B. 2019, \aap, 630, A51, \dodoi{10.1051/0004-6361/201935877}

\bibitem[{{Bitsch} {et~al.}(2019{\natexlab{a}}){Bitsch}, {Izidoro}, {Johansen},
  {Raymond}, {Morbidelli}, {Lambrechts}, \& {Jacobson}}]{bitschetal19}
{Bitsch}, B., {Izidoro}, A., {Johansen}, A., {et~al.} 2019{\natexlab{a}}, \aap,
  623, A88, \dodoi{10.1051/0004-6361/201834489}

\bibitem[{{Bitsch} {et~al.}(2018){Bitsch}, {Morbidelli}, {Johansen}, {Lega},
  {Lambrechts}, \& {Crida}}]{bitschetal18b}
{Bitsch}, B., {Morbidelli}, A., {Johansen}, A., {et~al.} 2018, \aap, 612, A30,
  \dodoi{10.1051/0004-6361/201731931}

\bibitem[{{Bitsch} {et~al.}(2019{\natexlab{b}}){Bitsch}, {Raymond}, \&
  {Izidoro}}]{bitschetal19b}
{Bitsch}, B., {Raymond}, S.~N., \& {Izidoro}, A. 2019{\natexlab{b}}, \aap, 624,
  A109, \dodoi{10.1051/0004-6361/201935007}

\bibitem[{{Bitsch} {et~al.}(2020){Bitsch}, {Trifonov}, \&
  {Izidoro}}]{bitschetal20}
{Bitsch}, B., {Trifonov}, T., \& {Izidoro}, A. 2020, \aap, 643, A66,
  \dodoi{10.1051/0004-6361/202038856}

\bibitem[{{Bolmont} \& {Mathis}(2016)}]{bolmontmathis16}
{Bolmont}, E., \& {Mathis}, S. 2016, Celestial Mechanics and Dynamical
  Astronomy, 126, 275, \dodoi{10.1007/s10569-016-9690-3}

\bibitem[{{Carrera} {et~al.}(2019){Carrera}, {Ford}, \&
  {Izidoro}}]{carreraetal19}
{Carrera}, D., {Ford}, E.~B., \& {Izidoro}, A. 2019, \mnras, 486, 3874,
  \dodoi{10.1093/mnras/stz974}

\bibitem[{{Carrera} {et~al.}(2018){Carrera}, {Ford}, {Izidoro}, {Jontof-
  Hutter}, {Raymond}, \& {Wolfgang}}]{carreraetal18}
{Carrera}, D., {Ford}, E.~B., {Izidoro}, A., {et~al.} 2018, \apj, 866, 104,
  \dodoi{10.3847/1538-4357/aadf8a}

\bibitem[{{Carter} {et~al.}(2012){Carter}, {Agol}, {Chaplin}, {Basu},
  {Bedding}, {Buchhave}, {Christensen-Dalsgaard}, {Deck}, {Elsworth},
  {Fabrycky}, {Ford}, {Fortney}, {Hale}, {Handberg}, {Hekker}, {Holman},
  {Huber}, {Karoff}, {Kawaler}, {Kjeldsen}, {Lissauer}, {Lopez}, {Lund},
  {Lundkvist}, {Metcalfe}, {Miglio}, {Rogers}, {Stello}, {Borucki}, {Bryson},
  {Christiansen}, {Cochran}, {Geary}, {Gilliland}, {Haas}, {Hall}, {Howard},
  {Jenkins}, {Klaus}, {Koch}, {Latham}, {MacQueen}, {Sasselov}, {Steffen},
  {Twicken}, \& {Winn}}]{carter12}
{Carter}, J.~A., {Agol}, E., {Chaplin}, W.~J., {et~al.} 2012, Science, 337,
  556, \dodoi{10.1126/science.1223269}

\bibitem[{{Chance} {et~al.}(2022){Chance}, {Ballard}, \&
  {Stassun}}]{chanceetal22}
{Chance}, Q., {Ballard}, S., \& {Stassun}, K. 2022, arXiv e-prints,
  arXiv:2208.05989.
\newblock \doarXiv{2208.05989}

\bibitem[{{Chatterjee} \& {Ford}(2015)}]{chatterjeeford15}
{Chatterjee}, S., \& {Ford}, E.~B. 2015, \apj, 803, 33,
  \dodoi{10.1088/0004-637X/803/1/33}

\bibitem[{{Chen} \& {Kipping}(2017)}]{chenkipping17}
{Chen}, J., \& {Kipping}, D. 2017, \apj, 834, 17,
  \dodoi{10.3847/1538-4357/834/1/17}

\bibitem[{{Christiansen} {et~al.}(2012){Christiansen}, {Jenkins}, {Caldwell},
  {Burke}, {Tenenbaum}, {Seader}, {Thompson}, {Barclay}, {Clarke}, {Li},
  {Smith}, {Stumpe}, {Twicken}, \& {Van Cleve}}]{christiansenetal12}
{Christiansen}, J.~L., {Jenkins}, J.~M., {Caldwell}, D.~A., {et~al.} 2012,
  \pasp, 124, 1279, \dodoi{10.1086/668847}

\bibitem[{{Cimerman} {et~al.}(2017){Cimerman}, {Kuiper}, \&
  {Ormel}}]{cimermanetal17}
{Cimerman}, N.~P., {Kuiper}, R., \& {Ormel}, C.~W. 2017, \mnras, 471, 4662,
  \dodoi{10.1093/mnras/stx1924}

\bibitem[{{Coleman} \& {Nelson}(2014)}]{colemannelson14}
{Coleman}, G.~A.~L., \& {Nelson}, R.~P. 2014, \mnras, 445, 479,
  \dodoi{10.1093/mnras/stu1715}

\bibitem[{{Coleman} \& {Nelson}(2016)}]{colemannelson16}
---. 2016, \mnras, 457, 2480, \dodoi{10.1093/mnras/stw149}

\bibitem[{{Cossou} {et~al.}(2014){Cossou}, {Raymond}, {Hersant}, \&
  {Pierens}}]{cossouetal14}
{Cossou}, C., {Raymond}, S.~N., {Hersant}, F., \& {Pierens}, A. 2014, A\&A,
  569, A56, \dodoi{10.1051/0004-6361/201424157}

\bibitem[{Cresswell \& Nelson(2008)}]{cresswellnelson08}
Cresswell, \& Nelson, R.~P. 2008, Astronomy \& Astrophysics, 482, 677,
  \dodoi{10.1051/0004-6361:20079178}

\bibitem[{Cresswell \& Nelson(2006)}]{cresswellnelson06}
Cresswell, P., \& Nelson, R.~P. 2006, Astronomy \& Astrophysics, 450, 833,
  \dodoi{10.1051/0004-6361:20054551}

\bibitem[{{Dorn} {et~al.}(2015){Dorn}, {Khan}, {Heng}, {Connolly}, {Alibert},
  {Benz}, \& {Tackley}}]{dornetal15}
{Dorn}, C., {Khan}, A., {Heng}, K., {et~al.} 2015, \aap, 577, A83,
  \dodoi{10.1051/0004-6361/201424915}

\bibitem[{{Dr{\c a}{\.z}kowska} \& {Alibert}(2017)}]{drazkowskaalibert17}
{Dr{\c a}{\.z}kowska}, J., \& {Alibert}, Y. 2017, \aap, 608, A92,
  \dodoi{10.1051/0004-6361/201731491}

\bibitem[{{Esteves} {et~al.}(2022){Esteves}, {Izidoro}, {Bitsch}, {Jacobson},
  {Raymond}, {Deienno}, \& {Winter}}]{estevesetal2022}
{Esteves}, L., {Izidoro}, A., {Bitsch}, B., {et~al.} 2022, \mnras, 509, 2856,
  \dodoi{10.1093/mnras/stab3203}

\bibitem[{{Esteves} {et~al.}(2020){Esteves}, {Izidoro}, {Raymond}, \&
  {Bitsch}}]{estevesetal20}
{Esteves}, L., {Izidoro}, A., {Raymond}, S.~N., \& {Bitsch}, B. 2020, \mnras,
  497, 2493, \dodoi{10.1093/mnras/staa2112}

\bibitem[{{Fabrycky} {et~al.}(2014){Fabrycky}, {Lissauer}, {Ragozzine}, {Rowe},
  {Steffen}, {Agol}, {Barclay}, {Batalha}, {Borucki}, {Ciardi}, {Ford},
  {Gautier}, {Geary}, {Holman}, {Jenkins}, {Li}, {Morehead}, {Morris},
  {Shporer}, {Smith}, {Still}, \& {Van Cleve}}]{fabryckyetal14}
{Fabrycky}, D.~C., {Lissauer}, J.~J., {Ragozzine}, D., {et~al.} 2014, \apj,
  790, 146, \dodoi{10.1088/0004-637X/790/2/146}

\bibitem[{{Fortney} {et~al.}(2007){Fortney}, {Marley}, \&
  {Barnes}}]{fortneyetal07}
{Fortney}, J.~J., {Marley}, M.~S., \& {Barnes}, J.~W. 2007, \apj, 659, 1661,
  \dodoi{10.1086/512120}

\bibitem[{{Fressin} {et~al.}(2013){Fressin}, {Torres}, {Charbonneau}, {Bryson},
  {Christiansen}, {Dressing}, {Jenkins}, {Walkowicz}, \&
  {Batalha}}]{fressinetal13}
{Fressin}, F., {Torres}, G., {Charbonneau}, D., {et~al.} 2013, \apj, 766, 81,
  \dodoi{10.1088/0004-637X/766/2/81}

\bibitem[{{Fulton} \& {Petigura}(2018)}]{fultonpetigura18}
{Fulton}, B.~J., \& {Petigura}, E.~A. 2018, \aj, 156, 264,
  \dodoi{10.3847/1538-3881/aae828}

\bibitem[{{Fulton} {et~al.}(2017){Fulton}, {Petigura}, {Howard}, {Isaacson},
  {Marcy}, {Cargile}, {Hebb}, {Weiss}, {Johnson}, {Morton}, {Sinukoff},
  {Crossfield}, \& {Hirsch}}]{fultonetal17}
{Fulton}, B.~J., {Petigura}, E.~A., {Howard}, A.~W., {et~al.} 2017, \aj, 154,
  109, \dodoi{10.3847/1538-3881/aa80eb}

\bibitem[{{Gillon} {et~al.}(2017){Gillon}, {Triaud}, {Demory}, {Jehin}, {Agol},
  {Deck}, {Lederer}, {de Wit}, {Burdanov}, {Ingalls}, {Bolmont}, {Leconte},
  {Raymond}, {Selsis}, {Turbet}, {Barkaoui}, {Burgasser}, {Burleigh}, {Carey},
  {Chaushev}, {Copperwheat}, {Delrez}, {Fernandes}, {Holdsworth}, {Kotze}, {Van
  Grootel}, {Almleaky}, {Benkhaldoun}, {Magain}, \& {Queloz}}]{gillonetal17}
{Gillon}, M., {Triaud}, A.~H.~M.~J., {Demory}, B.-O., {et~al.} 2017, \nat, 542,
  456, \dodoi{10.1038/nature21360}

\bibitem[{{Ginzburg} {et~al.}(2016){Ginzburg}, {Schlichting}, \&
  {Sari}}]{ginzburgetal16}
{Ginzburg}, S., {Schlichting}, H.~E., \& {Sari}, R. 2016, \apj, 825, 29,
  \dodoi{10.3847/0004-637X/825/1/29}

\bibitem[{{Ginzburg} {et~al.}(2018){Ginzburg}, {Schlichting}, \&
  {Sari}}]{ginzburgetal18}
---. 2018, \mnras, 476, 759, \dodoi{10.1093/mnras/sty290}

\bibitem[{{Goldberg} \& {Batygin}(2022)}]{goldbergbatygin22}
{Goldberg}, M., \& {Batygin}, K. 2022, \aj, 163, 201,
  \dodoi{10.3847/1538-3881/ac5961}

\bibitem[{{Gupta} {et~al.}(2022){Gupta}, {Nicholson}, \&
  {Schlichting}}]{guptaetal22}
{Gupta}, A., {Nicholson}, L., \& {Schlichting}, H.~E. 2022, arXiv e-prints,
  arXiv:2205.14020.
\newblock \doarXiv{2205.14020}

\bibitem[{{Gupta} \& {Schlichting}(2019)}]{guptaetal19}
{Gupta}, A., \& {Schlichting}, H.~E. 2019, \mnras, 487, 24,
  \dodoi{10.1093/mnras/stz1230}

\bibitem[{{Gupta} \& {Schlichting}(2020)}]{guptaetal20}
---. 2020, \mnras, 493, 792, \dodoi{10.1093/mnras/staa315}

\bibitem[{{Hawthorn} {et~al.}(2022){Hawthorn}, {Bayliss}, {Wilson}, {Bonfanti},
  {Adibekyan}, {Alibert}, {Sousa}, {Collins}, {Bryant}, {Osborn}, {Armstrong},
  {Abe}, {Acton}, {Addison}, {Agabi}, {Alonso}, {Alves}, {Anglada-Escud{\'e}},
  {B{\'a}rczy}, {Barclay}, {Barrado}, {Barros}, {Baumjohann}, {Bendjoya},
  {Benz}, {Bieryla}, {Bonfils}, {Bouchy}, {Brandeker}, {Broeg}, {Brown},
  {Burleigh}, {Buttu}, {Cabrera}, {Caldwell}, {Casewell}, {Charbonneau},
  {Charnoz}, {Cloutier}, {Collier Cameron}, {Collins}, {Conti}, {Crouzet},
  {Czismadia}, {Davies}, {Deleuil}, {Delgado-Mena}, {Delrez}, {Demangeon},
  {Demory}, {Dransfield}, {Dumusque}, {Egger}, {Ehrenreich}, {Eigm{\"u}ller},
  {Erickson}, {Essack}, {Fortier}, {Fossati}, {Fridlund}, {G{\"u}nther},
  {G{\"u}del}, {Gandolfi}, {Gillard}, {Gillon}, {Gnilka}, {Goad}, {Goeke},
  {Guillot}, {Hadjigeorghiou}, {Hellier}, {Henderson}, {Heng}, {Hooton},
  {Horne}, {Howell}, {Hoyer}, {Irwin}, {Jenkins}, {Jenkins}, {Jensen}, {Kane},
  {Kendall}, {Kielkopf}, {Kiss}, {Lacedelli}, {Laskar}, {Latham}, {Lecavalier
  des Etangs}, {Leleu}, {Lendl}, {Lillo-Box}, {Lovis}, {M{\'e}karnia},
  {Massey}, {Masters}, {Maxted}, {Nascimbeni}, {Nielsen}, {O'Brien},
  {Olofsson}, {Osborn}, {Pagano}, {Pall{\'e}}, {Persson}, {Piotto}, {Plavchan},
  {Pollacco}, {Queloz}, {Ragazzoni}, {Rauer}, {Ribas}, {Ricker},
  {S{\'e}gransan}, {Salmon}, {Santerne}, {Santos}, {Scandariato}, {Schmider},
  {Schwarz}, {Seager}, {Shporer}, {Simon}, {Smith}, {Srdoc}, {Steller},
  {Suarez}, {Szab{\'o}}, {Teske}, {Thomas}, {Tilbrook}, {Triaud}, {Udry}, {Van
  Grootel}, {Walton}, {Wang}, {Wheatley}, {Winn}, {Wittenmyer}, \&
  {Zhang}}]{hawthornetal22}
{Hawthorn}, F., {Bayliss}, D., {Wilson}, T.~G., {et~al.} 2022, arXiv e-prints,
  arXiv:2208.07328.
\newblock \doarXiv{2208.07328}

\bibitem[{{He} {et~al.}(2019){He}, {Ford}, \& {Ragozzine}}]{heetal19}
{He}, M.~Y., {Ford}, E.~B., \& {Ragozzine}, D. 2019, \mnras, 490, 4575,
  \dodoi{10.1093/mnras/stz2869}

\bibitem[{{He} {et~al.}(2021){He}, {Ford}, \& {Ragozzine}}]{heetal2021}
---. 2021, \aj, 162, 216, \dodoi{10.3847/1538-3881/ac1db8}

\bibitem[{{Hellary} \& {Nelson}(2012)}]{hellarynelson12}
{Hellary}, P., \& {Nelson}, R.~P. 2012, \mnras, 419, 2737,
  \dodoi{10.1111/j.1365-2966.2011.19815.x}

\bibitem[{{Howard}(2013)}]{howardetal13}
{Howard}, A.~W. 2013, Science, 340, 572, \dodoi{10.1126/science.1233545}

\bibitem[{{Howard} {et~al.}(2012){Howard}, {Marcy}, {Bryson}, {Jenkins},
  {Rowe}, {Batalha}, {Borucki}, {Koch}, {Dunham}, {Gautier}, {Van Cleve},
  {Cochran}, {Latham}, {Lissauer}, {Torres}, {Brown}, {Gilliland}, {Buchhave},
  {Caldwell}, {Christensen-Dalsgaard}, {Ciardi}, {Fressin}, {Haas}, {Howell},
  {Kjeldsen}, {Seager}, {Rogers}, {Sasselov}, {Steffen}, {Basri},
  {Charbonneau}, {Christiansen}, {Clarke}, {Dupree}, {Fabrycky}, {Fischer},
  {Ford}, {Fortney}, {Tarter}, {Girouard}, {Holman}, {Johnson}, {Klaus},
  {Machalek}, {Moorhead}, {Morehead}, {Ragozzine}, {Tenenbaum}, {Twicken},
  {Quinn}, {Isaacson}, {Shporer}, {Lucas}, {Walkowicz}, {Welsh}, {Boss},
  {Devore}, {Gould}, {Smith}, {Morris}, {Prsa}, {Morton}, {Still}, {Thompson},
  {Mullally}, {Endl}, \& {MacQueen}}]{howardetal12}
{Howard}, A.~W., {Marcy}, G.~W., {Bryson}, S.~T., {et~al.} 2012, \apjs, 201,
  15, \dodoi{10.1088/0067-0049/201/2/15}

\bibitem[{{Ida} \& {Lin}(2008)}]{idalin08}
{Ida}, S., \& {Lin}, D.~N.~C. 2008, \apj, 685, 584, \dodoi{10.1086/590401}

\bibitem[{{Ida} \& {Lin}(2010)}]{idalin10}
---. 2010, \apj, 719, 810, \dodoi{10.1088/0004-637X/719/1/810}

\bibitem[{{Inamdar} \& {Schlichting}(2016)}]{inamdarschlichting16}
{Inamdar}, N.~K., \& {Schlichting}, H.~E. 2016, \apjl, 817, L13,
  \dodoi{10.3847/2041-8205/817/2/L13}

\bibitem[{{Izidoro} {et~al.}(2021{\natexlab{a}}){Izidoro}, {Bitsch}, {Raymond},
  {Johansen}, {Morbidelli}, {Lambrechts}, \& {Jacobson}}]{izidoroetal21a}
{Izidoro}, A., {Bitsch}, B., {Raymond}, S.~N., {et~al.} 2021{\natexlab{a}},
  \aap, 650, A152, \dodoi{10.1051/0004-6361/201935336}

\bibitem[{{Izidoro} {et~al.}(2021{\natexlab{b}}){Izidoro}, {Dasgupta},
  {Raymond}, {Deienno}, {Bitsch}, \& {Isella}}]{izidoroetal21b}
{Izidoro}, A., {Dasgupta}, R., {Raymond}, S.~N., {et~al.} 2021{\natexlab{b}},
  Nature Astronomy, 6, 357, \dodoi{10.1038/s41550-021-01557-z}

\bibitem[{Izidoro {et~al.}(2014)Izidoro, Morbidelli, \&
  Raymond}]{izidoroetal14b}
Izidoro, A., Morbidelli, A., \& Raymond, S.~N. 2014, \apj, 794, 11,
  \dodoi{10.1088/0004-637X/794/1/11}

\bibitem[{{Izidoro} {et~al.}(2017){Izidoro}, {Ogihara}, {Raymond},
  {Morbidelli}, {Pierens}, {Bitsch}, {Cossou}, \& {Hersant}}]{izidoroetal17}
{Izidoro}, A., {Ogihara}, M., {Raymond}, S.~N., {et~al.} 2017, \mnras, 470,
  1750, \dodoi{10.1093/mnras/stx1232}

\bibitem[{{Jin} \& {Mordasini}(2018)}]{jinmordasinietal18}
{Jin}, S., \& {Mordasini}, C. 2018, \apj, 853, 163,
  \dodoi{10.3847/1538-4357/aa9f1e}

\bibitem[{{Johansen} \& {Lambrechts}(2017)}]{johansenlambrechts17}
{Johansen}, A., \& {Lambrechts}, M. 2017, Annual Review of Earth and Planetary
  Sciences, 45, 359, \dodoi{10.1146/annurev-earth-063016-020226}

\bibitem[{{Johnson} {et~al.}(2017){Johnson}, {Petigura}, {Fulton}, {Marcy},
  {Howard}, {Isaacson}, {Hebb}, {Cargile}, {Morton}, {Weiss}, {Winn}, {Rogers},
  {Sinukoff}, \& {Hirsch}}]{johnsonetal17}
{Johnson}, J.~A., {Petigura}, E.~A., {Fulton}, B.~J., {et~al.} 2017, \aj, 154,
  108, \dodoi{10.3847/1538-3881/aa80e7}

\bibitem[{Kruijer {et~al.}(2017)Kruijer, Burkhardt, Budde, \&
  Kleine}]{Kruijeretal17}
Kruijer, T.~S., Burkhardt, C., Budde, G., \& Kleine, T. 2017, Proceedings of
  the National Academy of Sciences, 114, 6712, \dodoi{10.1073/pnas.1704461114}

\bibitem[{{Kuchner}(2003)}]{kuchner03}
{Kuchner}, M.~J. 2003, \apjl, 596, L105, \dodoi{10.1086/378397}

\bibitem[{{Kurosaki} {et~al.}(2014){Kurosaki}, {Ikoma}, \&
  {Hori}}]{kurosakietal13}
{Kurosaki}, K., {Ikoma}, M., \& {Hori}, Y. 2014, \aap, 562, A80,
  \dodoi{10.1051/0004-6361/201322258}

\bibitem[{{Lai} \& {Pu}(2016)}]{laipu16}
{Lai}, D., \& {Pu}, B. 2016, ArXiv e-prints.
\newblock \doarXiv{1606.08855}

\bibitem[{{Lambrechts} \& {Johansen}(2012)}]{lambrechtsjohansen12}
{Lambrechts}, M., \& {Johansen}, A. 2012, \aap, 544, A32,
  \dodoi{10.1051/0004-6361/201219127}

\bibitem[{Lambrechts {et~al.}(2014)Lambrechts, Johansen, \&
  Morbidelli}]{lambrechtsetal14}
Lambrechts, M., Johansen, A., \& Morbidelli, A. 2014, Astronomy \&
  Astrophysics, 572, A35, \dodoi{10.1051/0004-6361/201423814}

\bibitem[{{Lambrechts} \& {Lega}(2017)}]{lambrechtslega17}
{Lambrechts}, M., \& {Lega}, E. 2017, \aap, 606, A146,
  \dodoi{10.1051/0004-6361/201731014}

\bibitem[{{Lambrechts} {et~al.}(2019){Lambrechts}, {Morbidelli}, {Jacobson},
  {Johansen}, {Bitsch}, {Izidoro}, \& {Raymond}}]{lambrechtsetal19}
{Lambrechts}, M., {Morbidelli}, A., {Jacobson}, S.~A., {et~al.} 2019, \aap,
  627, A83, \dodoi{10.1051/0004-6361/201834229}

\bibitem[{{Lee} \& {Chiang}(2015)}]{leechiang15}
{Lee}, E.~J., \& {Chiang}, E. 2015, \apj, 811, 41,
  \dodoi{10.1088/0004-637X/811/1/41}

\bibitem[{Leinhardt \& Stewart(2012)}]{leinhardtstewart12}
Leinhardt, Z.~M., \& Stewart, S.~T. 2012, \apj, 745, 79,
  \dodoi{10.1088/0004-637X/745/1/79}

\bibitem[{{Leleu} {et~al.}(2021){Leleu}, {Alibert}, {Hara}, {Hooton}, {Wilson},
  {Robutel}, {Delisle}, {Laskar}, {Hoyer}, {Lovis}, {Bryant}, {Ducrot},
  {Cabrera}, {Delrez}, {Acton}, {Adibekyan}, {Allart}, {Allende Prieto},
  {Alonso}, {Alves}, {Anderson}, {Angerhausen}, {Anglada Escud{\'e}},
  {Asquier}, {Barrado}, {Barros}, {Baumjohann}, {Bayliss}, {Beck}, {Beck},
  {Bekkelien}, {Benz}, {Billot}, {Bonfanti}, {Bonfils}, {Bouchy}, {Bourrier},
  {Bou{\'e}}, {Brandeker}, {Broeg}, {Buder}, {Burdanov}, {Burleigh},
  {B{\'a}rczy}, {Cameron}, {Chamberlain}, {Charnoz}, {Cooke}, {Corral Van
  Damme}, {Correia}, {Cristiani}, {Damasso}, {Davies}, {Deleuil}, {Demangeon},
  {Demory}, {Di Marcantonio}, {Di Persio}, {Dumusque}, {Ehrenreich}, {Erikson},
  {Figueira}, {Fortier}, {Fossati}, {Fridlund}, {Futyan}, {Gandolfi},
  {Garc{\'\i}a Mu{\~n}oz}, {Garcia}, {Gill}, {Gillen}, {Gillon}, {Goad},
  {Gonz{\'a}lez Hern{\'a}ndez}, {Guedel}, {G{\"u}nther}, {Haldemann},
  {Henderson}, {Heng}, {Hogan}, {Isaak}, {Jehin}, {Jenkins}, {Jord{\'a}n},
  {Kiss}, {Kristiansen}, {Lam}, {Lavie}, {Lecavelier des Etangs}, {Lendl},
  {Lillo-Box}, {Lo Curto}, {Magrin}, {Martins}, {Maxted}, {McCormac}, {Mehner},
  {Micela}, {Molaro}, {Moyano}, {Murray}, {Nascimbeni}, {Nunes}, {Olofsson},
  {Osborn}, {Oshagh}, {Ottensamer}, {Pagano}, {Pall{\'e}}, {Pedersen}, {Pepe},
  {Persson}, {Peter}, {Piotto}, {Polenta}, {Pollacco}, {Poretti}, {Pozuelos},
  {Queloz}, {Ragazzoni}, {Rando}, {Ratti}, {Rauer}, {Raynard}, {Rebolo},
  {Reimers}, {Ribas}, {Santos}, {Scandariato}, {Schneider}, {Sebastian},
  {Sestovic}, {Simon}, {Smith}, {Sousa}, {Sozzetti}, {Steller}, {Su{\'a}rez
  Mascare{\~n}o}, {Szab{\'o}}, {S{\'e}gransan}, {Thomas}, {Thompson},
  {Tilbrook}, {Triaud}, {Turner}, {Udry}, {Van Grootel}, {Venus}, {Verrecchia},
  {Vines}, {Walton}, {West}, {Wheatley}, {Wolter}, \& {Zapatero
  Osorio}}]{leleuetal21}
{Leleu}, A., {Alibert}, Y., {Hara}, N.~C., {et~al.} 2021, \aap, 649, A26,
  \dodoi{10.1051/0004-6361/202039767}

\bibitem[{Levison {et~al.}(2011)Levison, Morbidelli, Tsiganis, Nesvorn{\'{y}},
  \& Gomes}]{levisonetal11}
Levison, H.~F., Morbidelli, A., Tsiganis, K., Nesvorn{\'{y}}, D., \& Gomes, R.
  2011, Astron. J., 142, 152, \dodoi{10.1088/0004-6256/142/5/152}

\bibitem[{{Lissauer} {et~al.}(2011){Lissauer}, {Ragozzine}, {Fabrycky},
  {Steffen}, {Ford}, {Jenkins}, {Shporer}, {Holman}, {Rowe}, {Quintana},
  {Batalha}, {Borucki}, {Bryson}, {Caldwell}, {Carter}, {Ciardi}, {Dunham},
  {Fortney}, {Gautier}, {Howell}, {Koch}, {Latham}, {Marcy}, {Morehead}, \&
  {Sasselov}}]{lissauertal11a}
{Lissauer}, J.~J., {Ragozzine}, D., {Fabrycky}, D.~C., {et~al.} 2011, \apjs,
  197, 8, \dodoi{10.1088/0067-0049/197/1/8}

\bibitem[{{Liu} {et~al.}(2015){Liu}, {Hori}, {Lin}, \& {Asphaug}}]{liuetal15}
{Liu}, S.-F., {Hori}, Y., {Lin}, D.~N.~C., \& {Asphaug}, E. 2015, \apj, 812,
  164, \dodoi{10.1088/0004-637X/812/2/164}

\bibitem[{{Lopez} \& {Fortney}(2013)}]{lopezfortney13}
{Lopez}, E.~D., \& {Fortney}, J.~J. 2013, \apj, 776, 2,
  \dodoi{10.1088/0004-637X/776/1/2}

\bibitem[{{Lopez} \& {Fortney}(2014)}]{lopezfortney14}
---. 2014, \apj, 792, 1, \dodoi{10.1088/0004-637X/792/1/1}

\bibitem[{{Lopez} {et~al.}(2012){Lopez}, {Fortney}, \&
  {Miller}}]{lopezfortney12}
{Lopez}, E.~D., {Fortney}, J.~J., \& {Miller}, N. 2012, \apj, 761, 59,
  \dodoi{10.1088/0004-637X/761/1/59}

\bibitem[{{Luger} {et~al.}(2015){Luger}, {Barnes}, {Lopez}, {Fortney},
  {Jackson}, \& {Meadows}}]{lugeretal15}
{Luger}, R., {Barnes}, R., {Lopez}, E., {et~al.} 2015, Astrobiology, 15, 57,
  \dodoi{10.1089/ast.2014.1215}

\bibitem[{{Luger} {et~al.}(2017){Luger}, {Sestovic}, {Kruse}, {Grimm},
  {Demory}, {Agol}, {Bolmont}, {Fabrycky}, {Fernandes}, {Van Grootel},
  {Burgasser}, {Gillon}, {Ingalls}, {Jehin}, {Raymond}, {Selsis}, {Triaud},
  {Barclay}, {Barentsen}, {Howell}, {Delrez}, {de Wit}, {Foreman-Mackey},
  {Holdsworth}, {Leconte}, {Lederer}, {Turbet}, {Almleaky}, {Benkhaldoun},
  {Magain}, {Morris}, {Heng}, \& {Queloz}}]{lugeretal17}
{Luger}, R., {Sestovic}, M., {Kruse}, E., {et~al.} 2017, Nature Astronomy, 1,
  0129, \dodoi{10.1038/s41550-017-0129}

\bibitem[{{Marcus} {et~al.}(2010){Marcus}, {Sasselov}, {Hernquist}, \&
  {Stewart}}]{marcusetal10}
{Marcus}, R.~A., {Sasselov}, D., {Hernquist}, L., \& {Stewart}, S.~T. 2010,
  \apj, 712, L73, \dodoi{10.1088/2041-8205/712/1/L73}

\bibitem[{{Marcy} {et~al.}(2014){Marcy}, {Weiss}, {Petigura}, {Isaacson},
  {Howard}, \& {Buchhave}}]{marcyetal14}
{Marcy}, G.~W., {Weiss}, L.~M., {Petigura}, E.~A., {et~al.} 2014, Proceedings
  of the National Academy of Science, 111, 12655,
  \dodoi{10.1073/pnas.1304197111}

\bibitem[{{Matsumoto} \& {Ogihara}(2020)}]{matsumotoogihara20}
{Matsumoto}, Y., \& {Ogihara}, M. 2020, \apj, 893, 43,
  \dodoi{10.3847/1538-4357/ab7cd7}

\bibitem[{{Mayor} {et~al.}(2011){Mayor}, {Marmier}, {Lovis}, {Udry},
  {S{\'e}gransan}, {Pepe}, {Benz}, {Bertaux}, {Bouchy}, {Dumusque}, {Lo Curto},
  {Mordasini}, {Queloz}, \& {Santos}}]{mayoretal11}
{Mayor}, M., {Marmier}, M., {Lovis}, C., {et~al.} 2011, ArXiv e-prints.
\newblock \doarXiv{1109.2497}

\bibitem[{{McNeil} \& {Nelson}(2010)}]{mcneilnelson10}
{McNeil}, D.~S., \& {Nelson}, R.~P. 2010, \mnras, 401, 1691,
  \dodoi{10.1111/j.1365-2966.2009.15805.x}

\bibitem[{{Millholland} \& {Winn}(2021)}]{millhollandwinn21}
{Millholland}, S.~C., \& {Winn}, J.~N. 2021, \apjl, 920, L34,
  \dodoi{10.3847/2041-8213/ac2c77}

\bibitem[{{Mills} {et~al.}(2016){Mills}, {Fabrycky}, {Migaszewski}, {Ford},
  {Petigura}, \& {Isaacson}}]{millsetal16}
{Mills}, S.~M., {Fabrycky}, D.~C., {Migaszewski}, C., {et~al.} 2016, Nature,
  533, 509, \dodoi{10.1038/nature17445}

\bibitem[{{Misener} \& {Schlichting}(2021)}]{misenerschlichting21}
{Misener}, W., \& {Schlichting}, H.~E. 2021, \mnras, 503, 5658,
  \dodoi{10.1093/mnras/stab895}

\bibitem[{{Moldenhauer} {et~al.}(2022){Moldenhauer}, {Kuiper}, {Kley}, \&
  {Ormel}}]{moldenhaueretal22}
{Moldenhauer}, T.~W., {Kuiper}, R., {Kley}, W., \& {Ormel}, C.~W. 2022, \aap,
  661, A142, \dodoi{10.1051/0004-6361/202141955}

\bibitem[{{Morbidelli} {et~al.}(2022){Morbidelli}, {Bailli{\'e}}, {Batygin},
  {Charnoz}, {Guillot}, {Rubie}, \& {Kleine}}]{morbidellietal22}
{Morbidelli}, A., {Bailli{\'e}}, K., {Batygin}, K., {et~al.} 2022, Nature
  Astronomy, 6, 72, \dodoi{10.1038/s41550-021-01517-7}

\bibitem[{{Morbidelli} {et~al.}(2015){Morbidelli}, {Lambrechts}, {Jacobson}, \&
  {Bitsch}}]{morbidellietal15b}
{Morbidelli}, A., {Lambrechts}, M., {Jacobson}, S., \& {Bitsch}, B. 2015,
  \icarus, 258, 418, \dodoi{10.1016/j.icarus.2015.06.003}

\bibitem[{{Mordasini}(2014)}]{mordasini14}
{Mordasini}, C. 2014, \aap, 572, A118, \dodoi{10.1051/0004-6361/201423702}

\bibitem[{{Mordasini}(2020)}]{mordasini20}
---. 2020, \aap, 638, A52, \dodoi{10.1051/0004-6361/201935541}

\bibitem[{{Mousis} {et~al.}(2020){Mousis}, {Deleuil}, {Aguichine}, {Marcq},
  {Naar}, {Aguirre}, {Brugger}, \& {Gon{\c{c}}alves}}]{mousisetal20}
{Mousis}, O., {Deleuil}, M., {Aguichine}, A., {et~al.} 2020, \apjl, 896, L22,
  \dodoi{10.3847/2041-8213/ab9530}

\bibitem[{{Mulders}(2018)}]{mulders18}
{Mulders}, G.~D. 2018, {Planet Populations as a Function of Stellar Properties}
  (Springer), 153, \dodoi{10.1007/978-3-319-55333-7_153}

\bibitem[{{Mulders} {et~al.}(2018){Mulders}, {Pascucci}, {Apai}, \&
  {Ciesla}}]{muldersetal18}
{Mulders}, G.~D., {Pascucci}, I., {Apai}, D., \& {Ciesla}, F.~J. 2018, \aj,
  156, 24, \dodoi{10.3847/1538-3881/aac5ea}

\bibitem[{Nesvorný \& Morbidelli(2012)}]{nesvornymorbidelli12}
Nesvorný, D., \& Morbidelli, A. 2012, The Astronomical Journal, 144, 117.
\newblock \url{http://stacks.iop.org/1538-3881/144/i=4/a=117}

\bibitem[{{Ogihara} {et~al.}(2018){Ogihara}, {Kokubo}, {Suzuki}, \&
  {Morbidelli}}]{ogiharaetal18}
{Ogihara}, M., {Kokubo}, E., {Suzuki}, T.~K., \& {Morbidelli}, A. 2018, \aap,
  615, A63, \dodoi{10.1051/0004-6361/201832720}

\bibitem[{{Ogihara} {et~al.}(2015){Ogihara}, {Morbidelli}, \&
  {Guillot}}]{ogiharaetal15a}
{Ogihara}, M., {Morbidelli}, A., \& {Guillot}, T. 2015, \aap, 578, A36,
  \dodoi{10.1051/0004-6361/201525884}

\bibitem[{{Ormel} \& {Klahr}(2010)}]{ormelklahr10}
{Ormel}, C.~W., \& {Klahr}, H.~H. 2010, \aap, 520, A43,
  \dodoi{10.1051/0004-6361/201014903}

\bibitem[{{Otegi} {et~al.}(2020){Otegi}, {Bouchy}, \& {Helled}}]{otegietal20}
{Otegi}, J.~F., {Bouchy}, F., \& {Helled}, R. 2020, \aap, 634, A43,
  \dodoi{10.1051/0004-6361/201936482}

\bibitem[{{Owen} \& {Campos Estrada}(2020)}]{owencampos20}
{Owen}, J.~E., \& {Campos Estrada}, B. 2020, \mnras, 491, 5287,
  \dodoi{10.1093/mnras/stz3435}

\bibitem[{{Owen} \& {Wu}(2013{\natexlab{a}})}]{owenyanqin13}
{Owen}, J.~E., \& {Wu}, Y. 2013{\natexlab{a}}, \apj, 775, 105,
  \dodoi{10.1088/0004-637X/775/2/105}

\bibitem[{{Owen} \& {Wu}(2013{\natexlab{b}})}]{owenwu13}
---. 2013{\natexlab{b}}, \apj, 775, 105, \dodoi{10.1088/0004-637X/775/2/105}

\bibitem[{{Owen} \& {Wu}(2017)}]{owenwu17}
---. 2017, \apj, 847, 29, \dodoi{10.3847/1538-4357/aa890a}

\bibitem[{{Paardekooper} {et~al.}(2011){Paardekooper}, {Baruteau}, \&
  {Kley}}]{paardekooperetal11}
{Paardekooper}, S.-J., {Baruteau}, C., \& {Kley}, W. 2011, \mnras, 410, 293,
  \dodoi{10.1111/j.1365-2966.2010.17442.x}

\bibitem[{{Petigura}(2020)}]{petigura20}
{Petigura}, E.~A. 2020, \aj, 160, 89, \dodoi{10.3847/1538-3881/ab9fff}

\bibitem[{{Petigura} {et~al.}(2013){Petigura}, {Howard}, \&
  {Marcy}}]{petiguraetal13}
{Petigura}, E.~A., {Howard}, A.~W., \& {Marcy}, G.~W. 2013, Proceedings of the
  National Academy of Science, 110, 19273.
\newblock \doarXiv{1311.6806}

\bibitem[{{Petigura} {et~al.}(2017){Petigura}, {Howard}, {Marcy}, {Johnson},
  {Isaacson}, {Cargile}, {Hebb}, {Fulton}, {Weiss}, {Morton}, {Winn}, {Rogers},
  {Sinukoff}, {Hirsch}, \& {Crossfield}}]{petiguraetal17}
{Petigura}, E.~A., {Howard}, A.~W., {Marcy}, G.~W., {et~al.} 2017, \aj, 154,
  107, \dodoi{10.3847/1538-3881/aa80de}

\bibitem[{{Petigura} {et~al.}(2022){Petigura}, {Rogers}, {Isaacson}, {Owen},
  {Kraus}, {Winn}, {MacDougall}, {Howard}, {Fulton}, {Kosiarek}, {Weiss},
  {Behmard}, \& {Blunt}}]{petiguraetal2022}
{Petigura}, E.~A., {Rogers}, J.~G., {Isaacson}, H., {et~al.} 2022, \aj, 163,
  179, \dodoi{10.3847/1538-3881/ac51e3}

\bibitem[{{Poon} {et~al.}(2020){Poon}, {Nelson}, {Jacobson}, \&
  {Morbidelli}}]{poonetal20}
{Poon}, S. T.~S., {Nelson}, R.~P., {Jacobson}, S.~A., \& {Morbidelli}, A. 2020,
  \mnras, 491, 5595, \dodoi{10.1093/mnras/stz3296}

\bibitem[{{Pu} \& {Wu}(2015)}]{puwu15}
{Pu}, B., \& {Wu}, Y. 2015, \apj, 807, 44, \dodoi{10.1088/0004-637X/807/1/44}

\bibitem[{{Raymond} {et~al.}(2018){Raymond}, {Boulet}, {Izidoro}, {Esteves}, \&
  {Bitsch}}]{raymondetal18}
{Raymond}, S.~N., {Boulet}, T., {Izidoro}, A., {Esteves}, L., \& {Bitsch}, B.
  2018, \mnras, 479, L81, \dodoi{10.1093/mnrasl/sly100}

\bibitem[{{Raymond} {et~al.}(2020){Raymond}, {Izidoro}, \&
  {Morbidelli}}]{raymondetal20}
{Raymond}, S.~N., {Izidoro}, A., \& {Morbidelli}, A. 2020, in Planetary
  Astrobiology, ed. V.~S. {Meadows}, G.~N. {Arney}, B.~E. {Schmidt}, \& D.~J.
  {Des Marais}, 287, \dodoi{10.2458/azu_uapress_9780816540068}

\bibitem[{{Raymond} {et~al.}(2022){Raymond}, {Izidoro}, {Bolmont}, {Dorn},
  {Selsis}, {Turbet}, {Agol}, {Barth}, {Carone}, {Dasgupta}, {Gillon}, \&
  {Grimm}}]{raymondetal22}
{Raymond}, S.~N., {Izidoro}, A., {Bolmont}, E., {et~al.} 2022, Nature
  Astronomy, 6, 80, \dodoi{10.1038/s41550-021-01518-6}

\bibitem[{{Rogers} {et~al.}(2021){Rogers}, {Gupta}, {Owen}, \&
  {Schlichting}}]{rogersetal21}
{Rogers}, J.~G., {Gupta}, A., {Owen}, J.~E., \& {Schlichting}, H.~E. 2021,
  \mnras, 508, 5886, \dodoi{10.1093/mnras/stab2897}

\bibitem[{{Rogers} \& {Seager}(2010)}]{rogersseager10}
{Rogers}, L.~A., \& {Seager}, S. 2010, \apj, 712, 974,
  \dodoi{10.1088/0004-637X/712/2/974}

\bibitem[{{Spalding} \& {Batygin}(2016)}]{spaldingbatygin16}
{Spalding}, C., \& {Batygin}, K. 2016, \apj, 830, 5,
  \dodoi{10.3847/0004-637X/830/1/5}

\bibitem[{{Terquem} \& {Papaloizou}(2007)}]{terquempapaloizou07}
{Terquem}, C., \& {Papaloizou}, J.~C.~B. 2007, \apj, 654, 1110,
  \dodoi{10.1086/509497}

\bibitem[{{Van Eylen} {et~al.}(2018){Van Eylen}, {Agentoft}, {Lundkvist},
  {Kjeldsen}, {Owen}, {Fulton}, {Petigura}, \& {Snellen}}]{vaneylenetal18b}
{Van Eylen}, V., {Agentoft}, C., {Lundkvist}, M.~S., {et~al.} 2018, \mnras,
  479, 4786, \dodoi{10.1093/mnras/sty1783}

\bibitem[{{Van Eylen} {et~al.}(2019){Van Eylen}, {Albrecht}, {Huang},
  {MacDonald}, {Dawson}, {Cai}, {Foreman-Mackey}, {Lundkvist}, {Silva Aguirre},
  {Snellen}, \& {Winn}}]{vaneylenetal18}
{Van Eylen}, V., {Albrecht}, S., {Huang}, X., {et~al.} 2019, \aj, 157, 61,
  \dodoi{10.3847/1538-3881/aaf22f}

\bibitem[{{Venturini} {et~al.}(2020){Venturini}, {Guilera}, {Haldemann},
  {Ronco}, \& {Mordasini}}]{venturinietal20}
{Venturini}, J., {Guilera}, O.~M., {Haldemann}, J., {Ronco}, M.~P., \&
  {Mordasini}, C. 2020, \aap, 643, L1, \dodoi{10.1051/0004-6361/202039141}

\bibitem[{{Weiss} \& {Marcy}(2014)}]{weissmarcy14}
{Weiss}, L.~M., \& {Marcy}, G.~W. 2014, \apjl, 783, L6,
  \dodoi{10.1088/2041-8205/783/1/L6}

\bibitem[{{Weiss} {et~al.}(2022){Weiss}, {Millholland}, {Petigura}, {Adams},
  {Batygin}, {Bloch}, \& {Mordasini}}]{weissetal2022}
{Weiss}, L.~M., {Millholland}, S.~C., {Petigura}, E.~A., {et~al.} 2022, arXiv
  e-prints, arXiv:2203.10076.
\newblock \doarXiv{2203.10076}

\bibitem[{{Weiss} \& {Petigura}(2020)}]{weisspetigura20}
{Weiss}, L.~M., \& {Petigura}, E.~A. 2020, \apjl, 893, L1,
  \dodoi{10.3847/2041-8213/ab7c69}

\bibitem[{{Weiss} {et~al.}(2018){Weiss}, {Marcy}, {Petigura}, {Fulton},
  {Howard}, {Winn}, {Isaacson}, {Morton}, {Hirsch}, {Sinukoff}, {Cumming},
  {Hebb}, \& {Cargile}}]{weissetal18}
{Weiss}, L.~M., {Marcy}, G.~W., {Petigura}, E.~A., {et~al.} 2018, \aj, 155, 48,
  \dodoi{10.3847/1538-3881/aa9ff6}

\bibitem[{{Wolfgang} {et~al.}(2016){Wolfgang}, {Rogers}, \&
  {Ford}}]{wolfgangetal16}
{Wolfgang}, A., {Rogers}, L.~A., \& {Ford}, E.~B. 2016, \apj, 825, 19,
  \dodoi{10.3847/0004-637X/825/1/19}

\bibitem[{{Zeng} {et~al.}(2016){Zeng}, {Sasselov}, \& {Jacobsen}}]{zengetal16}
{Zeng}, L., {Sasselov}, D.~D., \& {Jacobsen}, S.~B. 2016, \apj, 819, 127,
  \dodoi{10.3847/0004-637X/819/2/127}

\bibitem[{{Zeng} {et~al.}(2019){Zeng}, {Jacobsen}, {Sasselov}, {Petaev},
  {Vanderburg}, {Lopez-Morales}, {Perez-Mercader}, {Mattsson}, {Li}, {Heising},
  {Bonomo}, {Damasso}, {Berger}, {Cao}, {Levi}, \& {Wordsworth}}]{zengetal19}
{Zeng}, L., {Jacobsen}, S.~B., {Sasselov}, D.~D., {et~al.} 2019, Proceedings of
  the National Academy of Science, 116, 9723, \dodoi{10.1073/pnas.1812905116}

\bibitem[{{Zeng} {et~al.}(2021){Zeng}, {Jacobsen}, {Hyung}, {Levi}, {Nava},
  {Kirk}, {Piaulet}, {Lacedelli}, {Sasselov}, {Petaev}, {Stewart}, {Alam},
  {L{\'o}pez-Morales}, {Damasso}, \& {Latham}}]{zengetal21}
{Zeng}, L., {Jacobsen}, S.~B., {Hyung}, E., {et~al.} 2021, \apj, 923, 247,
  \dodoi{10.3847/1538-4357/ac3137}

\bibitem[{{Zhang} {et~al.}(2022){Zhang}, {Knutson}, {Dai}, {Wang}, {Ricker},
  {Schwarz}, {Mann}, \& {Collins}}]{zhangetal22}
{Zhang}, M., {Knutson}, H.~A., {Dai}, F., {et~al.} 2022, arXiv e-prints,
  arXiv:2207.13099.
\newblock \doarXiv{2207.13099}

\bibitem[{{Zhu} {et~al.}(2018){Zhu}, {Petrovich}, {Wu}, {Dong}, \&
  {Xie}}]{zhuetal18}
{Zhu}, W., {Petrovich}, C., {Wu}, Y., {Dong}, S., \& {Xie}, J. 2018, \apj, 860,
  101, \dodoi{10.3847/1538-4357/aac6d5}

\end{thebibliography}
\bibliographystyle{aasjournal}



\end{document}